%% file: thesis.tex
\documentclass{dmathesis}
\usepackage[colon]{natbib}
\input{format}

\begin{document}



\input{frontpage}

\pagenumbering{arabic}
\setcounter{page}{1}


\input{./resu/resu}

\newpage{\
\thispagestyle{empty}
}

\input{./intro/intro}

\newpage{\
\thispagestyle{empty}
}

\input{./jstat/jstat}
\newpage{\
\thispagestyle{empty}
}

\input{./epl/epl}

\newpage{\
\thispagestyle{empty}
}

\input{./prl/prl}



\input{./pre/pre}



\input{./cr/cr}



\input{./con/con}

\newpage{\
\thispagestyle{empty}
}

\input{./con_es/con_es}

\newpage{\
\thispagestyle{empty}
}

\appendix

\input{./rap/rap}


\newpage{\
\thispagestyle{empty}
}

\input{./cr/appendix_cr}


\input{./nest/nest}



\input{./pub/pub}


\newpage{\
\thispagestyle{empty}
}

\cleardoublepage
\addcontentsline{toc}{chapter}{\numberline{}References}
\renewcommand\bibname{References}
\bibliographystyle{plain}
\bibliography{references}

\newpage{\
\thispagestyle{empty}
}

\end{document}

%% file: format.tex
\usepackage{fancyhdr}
\usepackage{epsfig}
\usepackage{cite}
\usepackage{graphicx}
\usepackage{amsmath}
\usepackage{theorem}
\usepackage{amssymb}
\usepackage{latexsym}
\usepackage{breakcites}
\usepackage{epic}

\newcounter{ind}

{\theorembodyfont{\rmfamily}}
{\theorembodyfont{\rmfamily}}
{\theorembodyfont{\rmfamily}}
{\theorembodyfont{\rmfamily}}
{\theorembodyfont{\rmfamily}}
{\theorembodyfont{\rmfamily}}



%% file: frontpage.tex

\pagenumbering{roman}
\setcounter{page}{1}


\newpage
\thispagestyle{empty}
\begin{center}
  \vspace*{0cm}

   \begin{flushleft}
   \includegraphics[width=3cm]{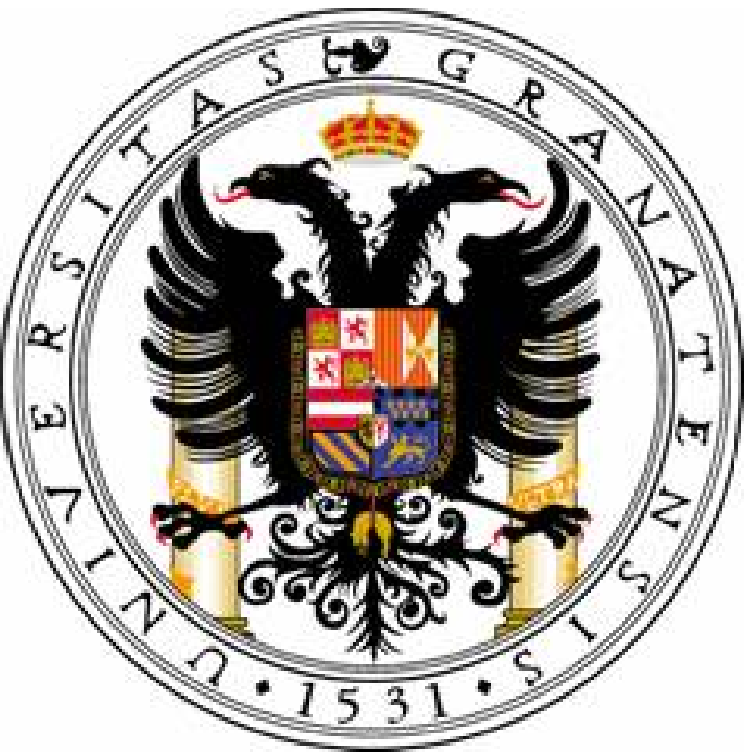}
   \end{flushleft}

   \vspace*{-3.3cm}
   \begin{flushright}
   \includegraphics{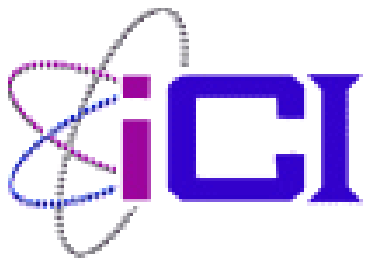}
   \end{flushright}

  {\large Department of Electromagnetism and Physics of Matter \& \\
          [-2mm] Institute {\em Carlos I} for Theoretical and Computational Physics,\\
          [-0mm] University of Granada, Spain.}

  \vspace{3cm}
  {\Huge \bf Interplay between Network Topology and Dynamics in Neural Systems}

  \vspace*{2cm}
  {\LARGE\bf Samuel Johnson}

  \vfill

  {\LARGE Ph.D. Thesis\\}
  \vspace*{0.6cm}
  {\LARGE Advisors: Joaqu\'{\i}n J. Torres Agudo\\ \& Joaqu\'in Marro Borau}
  \vspace*{0.4cm}

  {\LARGE Granada, April 2011}

\end{center}

\newpage{\
\thispagestyle{empty}

\newpage
\thispagestyle{empty}
\begin{flushright}
 \vspace*{6cm}
  \textit{\LARGE {Dedicated to my family}}\\ 
\end{flushright}

\newpage{\
\thispagestyle{empty}

\chapter*{Acknowledgements}
\addcontentsline{toc}{chapter}{\numberline{}Acknowledgements}
Aunque hay muchas m\'as personas a las que estoy agradecido, por tantas cosas, de las que puedo enumerar en un espacio razonable, tratar\'e de poner el umbral en alg\'un sitio y mencionar expl\'icitamente s\'olo a aquellas que me han ayudado de alguna manera directa  con esta tesis. En primer lugar quiero hacer constar mi sincero agradecimiento a mis directores: Joaqu\'in Torres no s\'olo es quien me introdujo al caos, los fractales, el SOC, las redes, la simulaci\'on... sino el que me mostr\'o que la neurociencia es algo tan concreto y estudiable como la f\'isica estad\'istica; Joaqu\'in Marro, por su parte, al compartir conmigo su genial perspectiva sobre la ciencia y la complejidad, me ha abierto los ojos a toda una manera de ver el mundo. Miguel \'Angel Mu\~noz ha sido tambi\'en una influencia cientf\'ifica important\'isima, con su maravillosa combinaci\'on de talento y buen humor. Agradezco tambi\'en a tod@s l@s dem\'as compa\~ner@s con quienes he trabajado y convivido estos a\~nos, tanto por lo que he aprendido de ellos como por los muchos buenos ratos: Pablo Hurtado, Pedro Garrido, Paco de los Santos, Antonio Lacomba, Elvira Romera, Ram\'on Contreras, Paco Ramos, Jes\'us Cort\'es, Juan Antonio Bonachela, Luca Donetti, Jos\'e Manuel Mart\'in, Omar al-Hammal, Jes\'us del Pozo, Carlos Espigares, Clara Guglieri, Pablo Sartori, Marina Manrique, Jordi Hidalgo, Virginia Dom\'inguez, Jorge Mej\'ias, Sebastiano de Franciscis, Alejandro Pinto, Leticia Rubio, Jordi Garces, Luca Sabino, Simone Pigolotti, Lu\'is Seoane, Daniele Vilone y Miguel Ib\'a\~nez. Como bien sab\'eis, sois compa\~ner@s mucho m\'as que de trabajo, si\'endolo tambi\'en, seg\'un el caso, de piso, de grupo musical, de decrepitud, de juegos diversos, del alma... Por supuesto, hay otras personas con esta multiplicidad de roles a quienes sin embargo no voy a tratar de nombrar aqu\'i: por favor, no pens\'eis que es por falta de reconocimiento, sino por ahorrar papel y tinta, y porque de todas maneras seguramente no vay\'ais a leer esto.

I'm grateful to Mike Ramsey, Mohammed Boudjada and Helmut Rucker for treating me so well in Graz as well as 
introducing me to the world of research.
Doy gracias a Marcelo del Castillo por invitarme a la UNAM y por ser, junto con su familia y amig@s, tan buen anfitri\'on en M\'exico; a Ezequiel Albano, Gabriel Baglietto, Bel\'en Moglia, Nara Guisoni y Luis Diambra por tratarme tan bien en La Plata; and to Nick Jones and Sumeet Agarwal for making me so welcome in Oxford. I am also grateful to all the people who have helped in one way or another with my research, be it reading manuscripts, providing data, suggesting ideas, or simply with stimulating conversations. I'm probably leaving many deserving people out when I mention Dante Chialvo, \'Alex Arenas, Ginestra Bianconi, Yamir Moreno, Jennifer Dunne, Alberto Pascual, V\'ictor Egu\'iluz, Sasha Goltsev, Gorka Zamora, Lars Rudolf, Sabine Hilfiker, Tiago Peixoto, Ole Paulsen and Peter Latham.
Tambi\'en estoy agradecido a Juan Soler, Alberto Prieto, Juan Calvo, Pilar Guerrero, Irene Mendoza, Estrella Ryan, Luna \'Alvarez, Javier (Chancly) Pascual, Nikolina Dimitrov, Felisa Torralba y Caroline de Cannart por su diversa ayuda. Three members of my family who have been particularly influential on my scientific interests and helpful in various ways are my uncle Dave Jones, my grandfather Tony Jones, and my aunt Sue Ziebland.

C\'ecile Poirier, qui as \'et\'e avec moi pendant la plus part de ces derni\`eres quatre ans, tu m'as fait surmonter tant d'obstacles et m'as aid\'e tellement que je n'ai vraiment pas de mots pour te remercier assez.
And the mental freedom and emotional basis I need to undertake this or any other project is grounded in the unconditional support and love from my parents, Jenni and Alan, and my sisters, Jazz and Abby. Thanks so much to all of you.

Finally, thanks are also due
to the government of the United States for its perseverance in attempting to crush
the news organization Wikileaks, whose continual publication of enlightening documents 
so often kept me from working on this thesis; and to myself, for at all times resisting the temptation
of perfectionism when writing this up -- as the reader will no doubt observe.



\newpage{\
\thispagestyle{empty}

\null\vfill
\textit{``To say that a man is made up of certain chemical elements is a satisfactory description only for those who intend to use him as a fertilizer.''}

\begin{flushright}
Hermann Joseph Muller
\end{flushright}

\null\vfill
\textit{``Je n'avais pas besoin de cette hypoth\`ese-l\`a.''}

\begin{flushright}
Pierre-Simon Laplace
\end{flushright}

\null\vfill
\textit{``Research is what I'm doing when I don't know what I'm doing.''}

\begin{flushright}
Werner von Braun
\end{flushright}

\vfill\vfill\vfill\vfill\vfill\vfill\null
\clearpage  



\newpage{\
\thispagestyle{empty}

\chapter*{Abstract}
\addcontentsline{toc}{chapter}{\numberline{}Abstract}

This thesis is a compendium of research which brings together ideas from the fields of Complex Networks 
and Computational Neuroscience to address two questions regarding neural systems: 
\\
1) How the activity of neurons, via synaptic changes, can shape the topology of the network they form part of, and
\\
2) How the resulting network structure, in its turn, might condition aspects of brain behaviour.
\\
Although the emphasis is on neural networks, several theoretical findings which are relevant for complex networks in general are presented -- such as a method for studying network evolution as a stochastic process, or a theory that allows for ensembles of correlated networks, and sets of dynamical elements thereon, to be treated mathematically and computationally in a model-independent manner. Some of the results are used to explain experimental data -- certain properties of brain tissue, the spontaneous emergence of correlations in all kinds of networks... -- and predictions regarding statistical aspects of the central nervous system are made. The mechanism of Cluster Reverberation is proposed to account for the near-instant storage of novel information the brain is capable of.

\clearpage  

\newpage{\
\thispagestyle{empty}

\chapter*{Preamble: The Ant, the Grasshopper and Complexity}
\addcontentsline{toc}{chapter}{\numberline{}Preamble: The Ant, the Grasshopper and Complexity}

Once upon a time, in a charming and peaceful little valley, a grasshopper sat under the shade
of a sunflower, idly strumming up a tune, when a young worker ant came into view. The
grasshopper watched as she trundled her way laboriously up an incline under the weight of a
large piece of leaf. When she was close enough, he hailed her:
\\
`Ahoy there, friend. I hope I won't seem tactless if I point out what a singularly cumbersome
bit of leaf you have there. Would you not rather put it down for a while and join me for a
quick jam session? You could bang along on some twigs or something.'
\\
`Thank you for the offer, but I must continue on my way,' replied the ant, glancing up in
slight surprise at being thus addressed.
\\
`Oh, what a pity,' the grasshopper rejoined. `And where, if I may be so bold as to inquire,
would you be taking your rather unappetising ration of cellulose?'
\\
`Well, I can't say I really know... I just follow this trail of pheromones I've come across. I'm sure
it's for some noble purpose though.'
\\
`Ah, that must be reassuring. And I suppose when you get to wherever it is you don't know
you're going you intend to eat your bit of leaf...'
\\
`Oh no, I can't digest something like this -- who do you take me for?'
\\
`You can't? Well, how strange...'
\\
`What's strange?'
\\
`However did an animal evolve which, instead of engaging in biologically reasonable (not to
mention enjoyable) activities, such as playing music to attract sexual partners, prefers to lug
useless bits of leaf about? How on earth can that serve to spread your genes?'
\\
`I'm not interested in music or sex, whatever those are. I just follow simple rules, like all my identical
sisters. You could say we're automata.'
\\
`Thanks, I was going to but wasn't sure whether you'd be offended. Well, let me wish you an
agreeable day of toil, you frigid little automaton.'
\\

With that, the grasshopper gave a big leap into the air, slightly exasperated by the folly so
often displayed by his fellow insects. Looking down, he spotted a few more ants, all carrying
leaves in the same direction as the one he had just met. Intrigued, he fluttered slightly higher
(since grasshoppers can, actually, fly, if not all that well). He realised the ants were all heading
for a nest some way off. In fact, there were many ant trails leading to various sources of food.
It dawned on the grasshopper that although the individual ants were just boring little morons
idiotically following rules, the nest as a whole was managing to find the closest leaves, bring
them back along optimal routes, and feed them to its plantations of fungi. The colony was
behaving like an intelligent organism, in some respects not so different from he himself, who
functioned thanks to the cells of his body -- each with the same genome, like the ants --
cooperating through the obedience to relatively simple rules.
\\

This thought impressed the grasshopper very much, driving him to flutter even higher so as to
see things in greater perspective. From there he considered the apparently fragile web of
trophic, parasitical and symbiotic interactions linking all the living beings in the valley -- a
network which nonetheless must have evolved a particularly robust structure not to shatter at
the first environmental fluctuation. He became so enthralled by the idea of such complexity on
one scale emerging from simplicity on another that he didn't even pay any attention to an
attractive young grasshopperess making her wanton way just below him. Instead, he couldn't
help fearing that a butterfly he noticed gently flapping his wings would probably set off a
hurricane somewhere. As he flew ever higher, he began to see snowflakes glide by,
overwhelmingly intricate and beautiful patterns self-organised out of the simplest little water
molecules. Finally he was so high that he began to reflect on how the very stellar systems,
galaxies,
clusters, superclusters, filaments of galaxies...
-- of which his whole world was but an
infinitesimal component -- also interacted with each other via the simple rules of gravity and
pressure to form objects marvellous beyond conception.
\\

What he didn't notice until it was too
late, as he left behind the cosy protection of the atmosphere, was how ultraviolet sunlight and
ionising cosmic rays were steadily burning his wings each to a crisp. Beginning to fall, he only
hoped he would have time to consider the several morals to his tragic tale.
\\

After a while spent plummeting to his doom he realised that, the freefall terminal velocity and life expectancy of a grasshopper being what they respectively were, he would most likely die peacefully of old age somewhere along his way down -- never again contemplating his Edenic valley except, like some prophetic locust, from afar.

\clearpage

\tableofcontents

\newpage{\
\thispagestyle{empty}

\listoffigures
\addcontentsline{toc}{chapter}{\numberline{}List of figures}
\clearpage
\newpage{\
\thispagestyle{empty}
\listoftables
\addcontentsline{toc}{chapter}{\numberline{}List of tables}
\clearpage
\newpage{\
\thispagestyle{empty}
\clearpage

%% file: resu/resu.tex
\chapter{Resumen en espa\~nol}

\label{Chapter_RESU}

Paradigma de sistema complejo y el peor comprendido de nuestros \'organos, el cerebro es, esencialmente, una inmensa red de c\'elulas que se comunican entre s\'i mediante se\~nales electro-qu\'imicas. Este trabajo recoge y desarrolla ideas del joven campo de las Redes Complejas para tratar de mejorar nuestro entendimiento acerca del comportamiento colectivo complejo que puede emerger en las redes de neuronas a partir de din\'amicas individuales relativamente sencillas.

El Cap\'itulo \ref{Chapter_INTRO} es una breve introducci\'on a las Redes Complejas y a la Neurociencia Computacional. Se describe, entre otras cosas, el modelo de Hopfield de red neuronal atractora, en que cada nodo representa una neurona y las sinapsis son representadas por los enlaces. Este sistema puede almacenar informaci\'on, en forma de patrones o configuraciones concretas de neuronas activas e inactivas, en los {\it pesos sin\'apticos}; es decir, en la intensidad con la que la actividad de una neurona influye sobre sus vecinas. Si, para representar un patr\'on dado, dos neuronas vecinas han de adoptar el mismo estado (activo o inactivo), se refuerza la interacci\'on entre ambas, mientras que se debilita en caso contrario. Repitiendo esta operaci\'on para cada pareja conectada de neuronas y para cada patr\'on, estos patrones se convierten en los estados que minimizan la energ\'ia total (atractores de la din\'amica), y el sistema evoluciona siempre hacia el patr\'on que m\'as se parezca a su estado inicial. Este mecanismo, llamado de {\it memoria asociativa}, es la responsable del almacenaje y la recuperaci\'on de informaci\'on tanto en modelos m\'as realistas de medios neuronales, como en muchos aparatos artificiales que desempe\~nan tareas tales como la identificaci\'on y la clasificaci\'on de im\'agenes. Adem\'as, hoy en d\'ia existen evidencias experimentales suficientes para asegurar que algo parecido ocurre en el cerebro: mediante los procesos bioqu\'imicos de potenciaci\'on de largo plazo (LTP, por sus siglas en ingl\'es) y depresi\'on de largo plazo (LTD), las sinapsis modifican gradualmente sus conductancias durante el aprendizaje.

El Cap\'itulo \ref{Chapter_JSTAT} aborda el problema de c\'omo puede desarrollarse una red con el tipo de estructuras que se observa en el cerebro. Para ello se formaliza como un proceso estoc\'astico una red que evoluciona mediante cambios probabil\'isticos que dependen de cualquier manera de informaci\'on local y global de los grados (n\'umeros de vecinos) de los nodos, tal como se hace en la Ref. \citep{Johnson_JSTAT}. Se considera que estas suposiciones son relevantes para el caso del cerebro ya que la arborizaci\'on y la atrofia sin\'apticas dependen de la actividad el\'ectrica de la neurona en cuesti\'on, que a su vez puede estar relacionada con el n\'umero de vecinos que tenga, y con la densidad sin\'aptica media en la red. Se demuestra c\'omo esta situaci\'on viene descrita por una ecuaci\'on de Fokker-Planck, y se aplica a dos conjuntos de datos reales neurofisiol\'ogicos: por una parte, la curvas de {\it poda sin\'aptica} (fuerte reducci\'on de la densidad sin\'aptica que sufre el c\'ortex durante la infancia) de autopsias humanas pueden explicarse con unas suposiciones m\'inimas; por otra, varias magnitudes estad\'isticas de la red del an\'elido C. Elegans (distribuci\'on de grados, perfil de correlaciones, {\it clustering} o agrupamiento y camino m\'inimo medio) emergen con cierta precisi\'on y de manera natural justo en la transici\'on de fase que presenta el modelo. Esto da fuerza a la idea de que el sistema nervioso optimiza su rendimiento coloc\'andose cerca de un punto cr\'itico. Un caso parecido, en que los enlaces de la red, en vez de desaparecer o aparecer, son redirigidos estoc\'asticamente, presentado en la Ref. \citep{Johnson_PRE}, se describe en el Ap\'endice \ref{Appendix_RAP}.

El resto de la tesis se centra en los efectos que pueden tener sobre el comportamiento colectivo de sistemas de neuronas las caracter\'isticas topol\'ogicas descritas en el Cap\'itulo \ref{Chapter_JSTAT}. Se sabe que la heterogeneidad de la distribuci\'on de grados de la red suele tener una influencia significativa en la din\'amica de elementos conectados mediante sus enlaces. En el caso de redes neuronales de Hopfield, Torres {\it et al.} \citep{Torres_influence} demostraron que, en redes {\it libres de escala} (que son altamente heterog\'eneas), el rendimiento aumenta con la heterogeneidad. El Cap\'itulo \ref{Chapter_EPL} examina el mismo efecto en una red neuronal que incluye otro ingrediente biol\'ogico: la {\it depresi\'on sin\'aptica}, gracias a la cual se observa una transici\'on entre una fase de memoria est\'atica a otra en que el sistema salta ca\'oticamente entre los patrones guardados. Resulta que cerca de este punto cr\'itico (el famoso Borde del Caos) la red es capaz de realizar una tarea din\'amica necesaria para los seres vivos: reconocer, de entre varios patrones que tenga almacenados, uno dado que se le ``ense\~ne'' y retenerlo indefinidamente despu\'es. Como demostramos en la Ref. \citep{Johnson_EPL}, la heterogeneidad de la distribuci\'on de grados de la red acerca el punto cr\'itico a una regi\'on del espacio de par\'ametros en que apenas hay depresi\'on sin\'aptica. Teniendo en cuenta que esta depresi\'on empeora la capacidad de memoria del sistema, se concluye que una red altamente heterog\'enea es la \'optima para realizar este tipo de tarea din\'amica. Las redes funcionales medidas en el c\'ortex humano durante tareas del estilo adopta la red libre de escala m\'as heterog\'ena posible, por lo que cabe la hip\'otesis de que el cerebro est\'e maximizando as\'i su rendimiento.

Otra propiedad altamente estudiada de las redes complejas es la existencia de correlaciones entre los grados de nodos vecinos. Cuando dichas correlaciones son positivas (nodos muy conectados se suelen conectar con otros que tambi\'en tienen muchos vecinos, y los que tienen pocos con otros parecidos) se dice que la red es {\it asortativa}; mientras que es {\it disasortativa} si las correlaciones son negativas (los que tienen muchas conexiones se conectan, preponderantemente, con los que tienen pocas). Curiosamente, se hab\'ia observado que por lo general las redes sociales (por ejemplo, redes de colaboraciones profesionales o de contactos sexuales) son asortativas, mientras que pr\'acticamente todas las dem\'as (gen\'eticas, tr\'oficas, proteicas, de transportes, de palabras, Internet, la Web...) son significativamente disasortativas. Aunque se hab\'ia estudiado los efectos de estas correlaciones en varios sistemas, las t\'ecnicas matem\'aticas y computacionales para ello padec\'ian de inconvenientes que restaban generalidad a los resultados. Para solventar esto, en el Cap\'itulo \ref{Chapter_PRL} se describe un nuevo m\'etodo para particionar el espacio de las fases de redes en regiones de correlaciones iguales, una t\'ecnica que permite tanto an\'alisis te\'orico como computacional de este tipo de sistemas. Utilizando este m\'etodo junto con ideas de Teor\'ia de la Informaci\'on se demuestra tambi\'en el resultado principal de la Ref. \citep{Johnson_PRL}: que la disasortatividad es el estado ``natural'' (en cuanto a situaci\'on de equlibrio) de las redes heterog\'eneas, lo cual explica la preponderancia en la realidad de este tipo de configuraciones. La preferencia de los humanos por agregarse en funci\'on de propiedades similares ser\'ia la explicaci\'on de que las redes sociales se encuentren fuera del equilibrio, en regiones asortativas del espacio de fases.

En el Cap\'itulo \ref{Chapter_PRE} se aplica el m\'etodo del Cap\'itulo \ref{Chapter_PRL} al caso de una red neuronal de Hopfield que no s\'olo presenta heterogeneidad, sino tambi\'en correlaciones nodo-nodo. Se encuentra, como ya fue descrito en la Ref. \citep{Sebas}, que estos sistemas pueden aumentar de manera notable su robustez frente a ruido gracias a las correlaciones positivas. De nuevo, esto parece encajar, al menos cualitativamente, con resultados experimentales que han encontrado redes funcionales en el c\'ortex humano altamente asortativas.

Hemos dicho que las redes neuronales pueden aprender gracias a una apropiada modificaci\'on de los pesos sin\'apticos mediante LTP y LTD, lo que explica la memoria de largo plazo. Pero estos procesos bioqu\'imicos ocurren en un tiempo caracter\'istico de al menos minutos. Los modelos de memoria de corto plazo, que ocurren en escalas de tiempo menores, suelen dar por hecho que la informaci\'on que se utiliza est\'a de antemano almacenada en el cerebro, y que el sujeto realizando la tarea s\'olo ha de activar y mantener de alguna manera la configuraci\'on correcta (como en el Cap\'itulo \ref{Chapter_EPL}). Pero es f\'acil darse cuenta de que esto no puede ser el caso para cualquier tarea: basta mirar algo totalmente nuevo por un instante, cerrar los ojos, y pensar en lo que se ha visto. Los \'unicos modelos de memoria de corto plazo existentes que no requieren aprendizaje sin\'aptico se basan en que cada neurona mantenga de alguna manera la informaci\'on que le corresponde (t\'ipicamente gracias a una serie de procesos sub-celulares). Pero al no emerger la memoria como propiedad colectiva del sistema, sino como suma de memorias individuales, estos modelos padecen de una gran falta de robustez frente a ruido. Y, lejos de presentar un comportamiento individual fiable, las neuronas se caracterizan justamente por ser c\'elulas de una alta variabilidad, con tendencia a disparar de manera m\'as o menos aleatoria. En el Cap\'itulo \ref{Chapter_CR} se propone un mecanismo, llamado {\it Cluster Reverberation} (CR), o Reverberaci\'on de Grupo, gracias al cual incluso sistemas  como redes con unidades simples, binarias, como en el modelo de Hopfield pueden almacenar informaci\'on instant\'aneamente sin necesidad de aprendizaje sin\'aptico, y de una manera que puede ser todo lo robusto frente a ruido como se quiera \citep{Johnson_CR}. Para ello el sistema aprovecha la existencia de estados metastables (situaciones que minimizan la energ\'ia del sistema localmente, sin corresponder al m\'inimo global) y como consecuencia aparecen transitorios en la dinámica de la actividad neuronal cuyas propiedades son consecuencia inmediata de las caracter\'{\i}sticas de la topolog\'ia subyacente y que es del tipo  de las descritas anteriormente en el Cap\'itulo \ref{Chapter_JSTAT} y en experimentos, esto es, el grado de agrupamiento o la modularidad de la red. B\'asicamente, grupitos densamente interconectados de neuronas pueden mantener un estado conjunto de alta o baja actividad, en promedio. Considerando cada grupito como un elemento funcional elemental, en vez de cada neurona, se consigue la aparici\'on de las propiedades requeridas. Es m\'as, algunas otras  caracter\'isticas de la memoria de corto plazo emergen de manera natural de este mecanismo. En particular, se demuestra que la informaci\'on se pierde gradualmente con el tiempo seg\'un una ley aproximadamente potencial, como ha sido descrito en experimentos psicof\'isicos. 

En conclusi\'on, las principales aportaciones originales de esta Tesis son:
\begin{itemize}
 \item 
M\'etodos anal\'iticos y computacionales para estudiar redes evolutivas \citep{Johnson_PRE,Johnson_JSTAT} y redes con correlaciones nodo-nodo \citep{Johnson_PRL, Sebas}.

\item
Una respuesta a la pregunta de por qu\'e la mayor\'ia de las redes reales son disasortativas \citep{Johnson_PRL}.

\item
Propiedades topol\'ogicas que pueden optimizar el rendimiento de redes neuronales \citep{Johnson_EPL,Sebas}.

\item
Un mecanismo que pudiera estar detr\'as de la memoria de corto plazo \citep{Johnson_CR}.

\end{itemize}

%% file: intro/intro.tex


\chapter{Where we are and where we'd like to go} 
\label{Chapter_INTRO}

\section{From bridges to brains}

\begin{figure}[t!]
\begin{center}
\hspace{-0.5cm}\epsfig{file=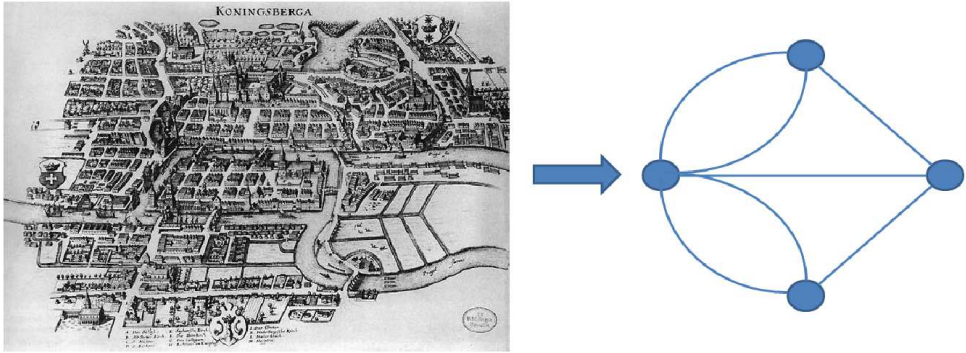,scale=0.3}
\caption{The problem of the Seven Bridges of K\"onigsberg can be reduced to a graph in which nodes and edges represent land masses and bridges, respectively.}
\label{fig_kon}%
\end{center}
\end{figure}

Strolling through the streets of K\"onigsberg, a young Immanuel Kant may have wondered whether, as some hoped, a path could be found that would take him once and only once over each of the city's celebrated seven bridges and back to where he started. 
In 1736 Leonard Euler pointed out that for this or any other problem of the kind all that mattered was which land masses were connected to each other, and by how many bridges. In other words, the situation could be captured by a graph, as in Fig. \ref{fig_kon}, in which each land mass is represented by a node (also called vertex) and each bridge by a link (or edge). He showed that in the case of K\"onigsberg no such walk could be found, since an ``Eulerian cycle'' in a connected graph exists if and only if the degrees of all nodes are even numbers -- the degree of a node being its number of edges \citep{Euler}. And thus was Graph Theory born.

For over two centuries, the graphs people were interested in were precisely defined objects, usually sufficiently small to be drawn on a piece of paper. But in the late nineteen fifties, mathematicians began to study {\it random graphs} -- i.e., defined only by some random generation process -- perhaps with a view to better dealing with ever-growing communications networks \citep{Bollobas}. E. N. Gilbert considered a situation in which there are $n$ nodes and each pair is connected by an edge with probability $q$ \citep{Gilbert}. For different values of these parameters, he was able to obtain the likelihood of the graph being connected (that is, of there being a path joining any two nodes). A similar model was proposed by Paul Erd\"os and Alfr\'ed R\'enyi: each of all the possible graphs with $n$ nodes and $m$ edges had an equal chance of being ``picked'' \citep{Erdos}. In fact, a given graph will be generated with equal probability in either scenario, so the descriptions are equivalent, and usually known as the Erd\"os-R\'enyi (ER) model. It is simple to see that if one were to average over many graphs generated by either of these processes, the degrees would follow a binomial distribution -- tending, for large $n$, to a Poisson distribution. That is, $p(k)$ is symmetrically centred around its mean value and drops off exponentially -- where $k_{i}$ is the degree of node $i$. An interesting phenomenon that can be observed using the ER model is that of percolation. If we measure the size $\Phi$ of the {\it largest connected component} (that is, of the highest number of nodes in the graph forming a connected subgraph) we obtain at different values of the probability $q$ (or, equivalently, of $m=\frac{1}{2}q n^{2}$), we see that there is a critical value, $q_{c}=1/n$, above which $\Phi/n$ does not vanish for high $n$ -- that is, there will usually exist a connected subgraph of a size comparable to that of the whole system. This passing from one situation (or phase) to another, each characterised by some qualitatively different characteristic, is known in physics as a {\it phase transition}. In this case, it is a second-order transition, since the control parameter $\Phi$ varies continuously (and not abruptly, as in first-order transitions), and has innumerable applications. For instance, the nodes might be people susceptible to some disease, trees which may be set on fire, or oil bubbles in a porous medium. The epidemic will spread, the forest will burn, or the oil will be extractable if the density of edges -- contagious contact, fire-conducting proximity, or links between pores -- is over the critical value.



In his 1929 short story {\it Chains} ({\it L\'ancszemek}, in the original Hungarian) Frigyes Karinthy suggested that the number of people in a chain of acquaintances grows exponentially with size, and thus that very few steps are needed to join anyone with any other person. This Small World idea was taken up in 1967 by Stanley Milgram, who performed a series of experiments that, while somewhat less controversial than his well-known obedience-to-authority explorations, have nonetheless been widely discussed \citep{Milgram}. He and his colleagues sent various letters to random people with the request to attempt to send them on to a particular individual many thousands of miles away, plus the constraint that this had to be done via people with whom the sender was on first-name terms. Many of the letters reached their destination, after having been sent on by a surprisingly small number of intervening people. This was later popularised as the Six Degrees of Separation -- the famous idea that any two people are linked by a path of only six acquaintances. Within the connected component of an ER random graph any two nodes are also joined by a short path -- of the order of the logarithm of the number of nodes. However, this is less surprising, since these networks are not {\it clustered}; that is, they do not have the property typical of social networks whereby ``the friends of my friends are (also likely to be) my friends.'' In 1998, Duncan Watts and Steven Strogatz put forward a network model which took this feature into account. They considered a ring of $n$ nodes, each connected to their $k$ nearest neighbours (they set $k=4$). Each edge was then broken from one of its nodes and rewired to some other random node with a probability $p$. Thus, $p=0$ leaves the ring intact, while $p=1$ changes it into an ER random graph.  Two magnitudes were measured for different values of $p$: the mean-minimum-path length, $l$, and the clustering coefficient, $C$. The first is simply an average over all pairs of nodes of the minimum-paths connecting them. The latter can be seen as the probability that two neighbours of a given node are directly connected to each other. For $p=0$, the clustering is high ($C=\frac{1}{2}$) and independent of the network size, while the mean minimum path scales with $n$ ($l\simeq n/8$). At the other extreme, $p=1$ yields a vanishingly small clustering ($C=k/n$) but short paths ($l\simeq \ln n/\ln k$). The most interesting case is found at intermediate values of $p$. As $p$ grows from zero, $l$ falls very rapidly to a value close to the random case, but $C$ does not present this drop until a much higher value is reached. Watts and Strogatz called this intermediate zone the Small-World region, since everyone is highly-interconnected while much of the local structure is conserved. They suggested that this is actually a property of many real networks (as has since turned out to be the case \citep{Newman_rev}), most especially of social networks -- in which $C$ is often several orders of magnitude greater than if the graph were random, while $l$ is not much larger than in such a case. As the authors point out, it is essential to take this feature into account for the study of, say, epidemics. 

Another feature of networks which is quite ubiquitous in the real world is that degree distributions are highly heterogeneous; in fact, they often follow power-laws, $p(k)\sim k^{-\gamma}$, with $\gamma$ a positive constant typically between 2 and 3. Such networks are nowadays referred to as {\it scale free}. In the nineteen fifties, Herbet Simon showed that these distributions come about when ``the rich get richer'' \citep{Simon}. Applying this idea to the case of scientific citations, Derek de Solla Price proposed the first known model of a scale-free network, in which nodes represent papers and edges are citations \citep{Price_1}. Each node has an in-degree (the number of papers citing it) and an out-degree (papers it cites). That is, the network is {\it directed}, since edges have a direction. Assuming that the probability a paper has of being cited by a new one is proportional to the number already citing it (its in-degree), the network is built up through the gradual addition of nodes, the neighbours of these being chosen according to their existing in-degrees. Price showed analytically that such a mechanism leads to an in-degree distribution $p(k)\sim k^{-(2+1/m)}$, with $m$ a parameter of the model equivalent to the mean degree\footnote{Note that in a directed network, the mean in-degree and mean out-degree coincide.}. He called this mechanism {\it cumulative advantage}. Somewhat ironically -- considering that Price, with a PhD in history from Cambridge, is best known as the father of scientometrics -- this work was mostly ignored by the scientific community. The model was rediscovered in 1999 by Albert-L\'aszl\'o Barab\'asi and Reka Albert, with the difference that they considered the network to be undirected
\citep{Barabasi}. They coined the term {\it preferential attachment} for the rich-get-richer mechanism, which is now generally assumed to be behind the formation of most scale-free networks (although other mechanisms exist \citep{Caldarelli_fitness, Krapivsky, Newman_power-laws}). Among many interesting consequences of such degree heterogeneity, Mark Newman showed that the clustering and mean-minimum-path length are respectively higher and lower than in homogeneous networks, making all scale-free networks to some extent small worlds \citep{Newman_random}. It also has important consequences for dynamical processes taking place among elements on the network, such as the synchronisation of coupled oscillators \citep{Barahona}.

As mentioned above, networks can be made up of separate components such that no path exists between nodes in different subgraphs. This is an extreme case of {\it community structure}. However, what is usually more interesting is the fact that communities may exist such that
there is a higher density of edges within them than without, 
even if the network is connected \citep{Girvan}. These communities are also at times called modules or clusters (although this can create some confusion with the related but distinct idea of clustering referred to above). Given a network, one can make a partition -- that is, divide the nodes up into groups -- and calculate what proportion of the edges fall within these, compared with the random expectation. This measure is called the modularity of this partition, and sometimes one speaks of the modularity of a graph referring to that of the partition for which this value is maximum. Determining the community structure of empirical networks can often provide useful insights into aspects of the systems. For instance, the communities may correspond to functional groups in a metabolic network, or groups of people who share some trait. However, there are many problems related to making this kind of measurement. For one thing, there are so many possible partitions that even an ER random graph can have a fairly high modularity due simply to statistical fluctuations \citep{Reichardt}. Then there is the fact that community structure can exist on may different levels -- that is, the groups considered can be of any size -- so one must usually consider a hierarchy of modules \citep{Arenas_scales}. Furthermore, finding an optimal partition is an NP-Complete problem \citep{Garey}, which makes comparing the modularity of each possible partition intractable in all but very small networks. For these and other reasons, in recent years much work has gone into finding efficient algorithms to determine the community structure of networks, albeit approximately \citep{Girvan,Luca_communities,Louvain} -- as well as into comparing the results offered by each approach \citep{Danon}.

Finally, another feature of networks worth mentioning arises when the nodes of a network are endowed with some property and this is reflected by the layout of the edges: the situation is called a {\it mixing pattern} \citep{Newman_mixing_PRL, Newman_mixing_PRE}. For instance, people might tend to choose sexual partners who share certain characteristics, such as mother-tongue or self-defined race. In these cases the network is {\it assortative}, since nodes of a kind assort, or group together. However, if the property in question were, say, gender, then the same graphs would be {\it disassortative} if most of the relations were heterosexual. In these cases the property can be considered discrete, but it can be continuous -- for instance, people might assort according to age. An interesting case is when the property in question is the degree of each node, since it is then an entirely topological issue. The extent to which the degrees of neighbouring nodes are correlated -- as given, for example, by Pearson's correlation coefficient \citep{Newman_mixing_PRL} -- is then a measure of the {\it assortativity} of the graph, being positive for assortative networks and negative for disassortative ones. It turns out that there is a striking universality in the nature of the degree-degree correlations displayed by real-world networks, whether natural or artificial: social networks, like the ones just described, tend to be assortative, while almost all other kinds of network are disassortative \citep{Pastor-Satorras,Newman_rev}. Often specific functional constraints can be found to justify correlations of one or other kind, but in Chapter \ref{Chapter_PRL} of this thesis the purely topological explanation put forward in Ref. \citep{Johnson_PRL} is described. In any case, the degree-degree correlations of a network can play a significant role in the behaviour of processes taking place thereon. For example, assortative networks have lower percolation thresholds and are more robust to targeted attack \citep{Newman_mixing_PRE}, while disassortative ones make for more stable ecosystems and seem to be more synchronizable \citep{Brede}.

All the aspects of networks mentioned in this brief overview, as well as many others, have been shown to be relevant for a wide range of complex systems \citep{Albert_rev,Newman_rev,Boccaletti}. Among these is the brain, a paradigm of complexity as well as the least understood of our organs.
However, research focusing on how the collective behaviour of neural systems, as observed in mathematical models, is influenced by the topology of the underlying network is relatively scarce. This is perhaps due in part to the attention that other biological properties of the nervous system have tended to draw from the Computational Neuroscience community. Thanks to the flurry of activity that the field of Complex Networks has been enjoying over the last decade, this is a particularly good moment to undertake a more systematic analysis of how dynamics and topology are related in this kind of systems.


\section{Neural networks in neuroscience}

Ever since the publication of Santiago Ram\'on y Cajal's drawings of neurons -- in his
words, those ``mysterious butterflies of the soul'' -- it has been clear that the nervous
system is composed of a large number of such cells connected to one another to form a
network \citep{Cajal}.
Long axons, ending in terminals which form synapses to the dendrites which
branch out from neighbouring neurons, transmit
{\it action potentials} (APs) -- changes in the cellular membrane voltage -- 
and enable
neurons somehow to cooperate and give rise the astonishing emergent phenomenon called thought.
One of these APs 
is formed each time the membrane electric potential of a neuron surpasses a threshold value, leading to the opening of a great many voltage-dependent ionic gates between the cell and the extra-cellular medium. In turn, the membrane potential of a given neuron is constantly affected by action potentials arriving from neighbouring neurons, and thus an extremely complex web of cellular signalling is achieved.

The first model neuron was proposed by
\citet{McCulloch}. This was simply an element that would return a Heaviside step function of the sum of its inputs. Sets of such ``artificial neurons'' could be used to implement any logical gate. Shortly after this, another important suggestion was made, this time by the psychologist Donald Hebb. Attempting to relate Pavlovian
conditioning experiments with cellular plasticity, he conjectured, in 1949, the existence of some biological mechanism that would lead to neurons which repeatedly fired (i.e., let off
action potentials) together becoming more strongly coupled \citep{Hebb}.
The initiation and propagation of action potentials in individual neurons was first modelled mathematically by Alan L. Hodgkin and Andrew Huxley in 1952 by means of a set of nonlinear ordinary differential equations which took into account the various ion fluxes \citep{Hodgkin}.

However, the concept of a neural network (as understood in theoretical and
computational neuroscience) was partly inspired by mathematical models of {\it spin systems}. The first of these was the Ising model, put forward in 1920 by Wilhelm Lenz and studied by Ernst Ising with a view with a view to understanding phase transitions and magnets \citep{Onsager,Brush}. It was known that the spin of electrons conferred a magnetic moment to individual atoms, but
it wasn't clear how exactly a very many such spins could self-organise into a large body
producing a net magnetic field. By considering an infinite set of entities (spins) with
possible values plus or minus one (up or down, say) which, when placed at the nodes of a lattice, interact
in such a way that energy is lowest when neighbours are aligned,
and a temperature parameter to
govern the extent of random fluctuations, it was eventually shown that, below a certain
critical temperature (in two or more dimensions), symmetry is spontaneously broken
and most of the spins end up aligned \citep{Baxter_exact}.
This {\it ferromagnetic} solution comes about and is then 
maintained because it has a lower energy than any other 
configuration of spins. Subsequent models, in particular that of
\citet{Sherrington}, incorporated inhomogeneities in the coupling strengths such that there was no longer a configuration which simultaneously minimized all interaction energies, leading to frustrated states (spin-glasses).

\begin{figure}[t!]
\begin{center}
\epsfig{file=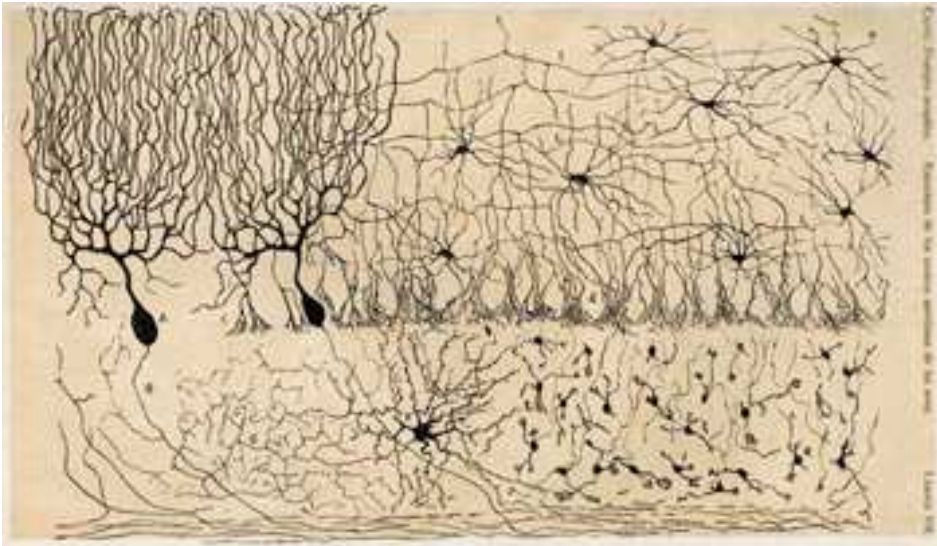,scale=1}
\caption{Drawing of the cells of the chick cerebellum, from ``Estructura de los centros nerviosos de las aves'', Madrid, 1905. Notice how the neurons make up a complex network of synaptic interactions.}
\label{fig_cajal}%
\end{center}
\end{figure}

\begin{figure}[t!]
\begin{center}
\epsfig{file=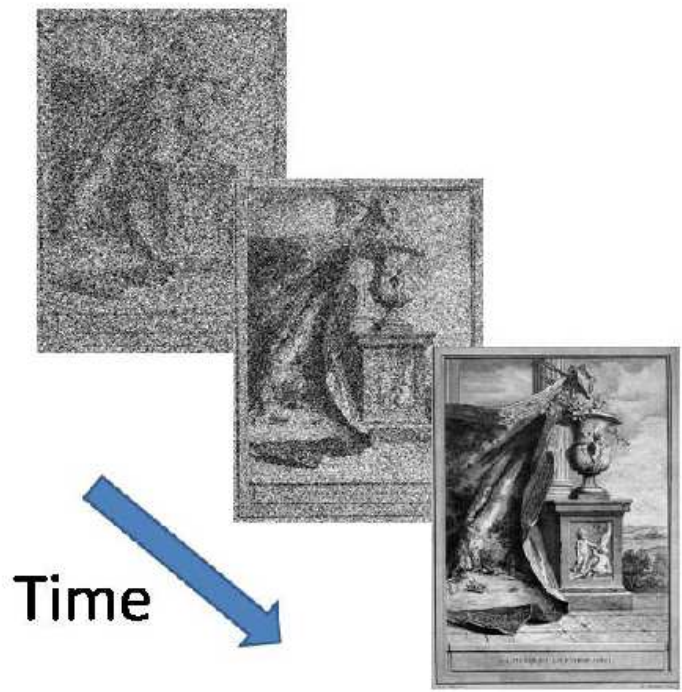,scale=0.65}%
\caption{In the Hopfield neural-network model, the interaction strengths (representing synaptic weights) store information in the form of particular patterns, or configurations of firing neurons, which become attractors of the dynamics. Whatever the initial state of the system, it will always evolve towards one of these patterns, thus allowing for the storage and retrieval of information. The mechanism, known as {\it associative memory}, is thought to be at the basis of memory in the brain. In this case, the network is ``remembering'' an illustration by Jean-Baptiste Oudry for Jean de la Fontaine's fable {\it La Cigale et la Fourmi}.
}%
\label{fig_ant_grasshopfield}%
\end{center}
\end{figure}

These ideas were put together, by
\citet{Amari} and then by
\citet{Hopfield}, in the first neural network models to exhibit the mechanism known as {\it associative memory}. Each model neuron was placed at the node of a network, originally assumed to be fully connected (all nodes connected to all the rest), and followed a dynamics which can be seen either as that of Ising spins or of McCulloch-Pitts neurons. However, a noise parameter usually referred to as ``temperature'' by analogy with spin systems could be included to allow for non-deterministic behaviour. By setting the interaction strengths (synaptic weights) not randomly, as in the Sherrington-Kirkpatrick model, but according to the Hebb rule referred to above, information could be stored and retrieved by the system. More specifically, a set of particular {\it patterns}, or configurations of positive and negative elements (firing and non-firing neurons), are recorded in the following way: for each pattern, one looks at each pair of neurons and adds a quantity to the weight of the synapse joining them if the pattern in question requires them to be in the same state, and subtracts it when they should be opposite. In this way, the minimum energy configurations correspond to the stored patterns, which therefore become attractors of the dynamics: if the temperature is not too heigh to destroy all order, the system will evolve towards whichever of these patterns most resembles the initial configuration it is placed in. Figure \ref{fig_ant_grasshopfield} illustrates how this mechanism works for a system such that the firing and non-firing neurons represent black and white pixels of a bitmap image.

Thanks to associative memory, if we were to store, say, a set of photos of various people and then ``show'' the network a different picture of one of the same subjects, it would be able to retrieve the correct identity. Not only is this mechanism used nowadays in technology capable of performing tasks such as pattern discrimination and classification, but it is widely considered to underlie our own capacity for learning and recalling information.
There is evidence from neuronal readouts that this is so \citep{Amit_Hebb}, and
not long ago, {\it in vivo} experiments finally established that learning is indeed related to the processes of long term potentiation (LTP) and depression (LTD) -- by which synapses between neurons
that fire nearly simultaneously gradually increase or decrease their conductance depending on the interval of time elapsed \citep{Gruart, DeRoo}. 

The neural network models
studied nowadays generally include more realistic dynamics both for the neurons and
for the synapses, taking into account a variety of cellular and subcellular processes \citep{Amit,Torres_rev}. For
example, the fact that the conductance of synapses in reality depends on their workload
has been found to enable a network to switch from one pattern to another -- either
spontaneously or as a reaction to sensory stimuli -- providing a means for the
performance of dynamic tasks \citep{Cortes_effects, Holcman}; this result also seems to agree well with physiological data \citep{Korn}. In fact, there is
evidence that the brain somehow maintains itself close to a boundary -- called, in physics, a critical point -- between an ordered and a chaotic regime \citep{Eguiluz,Chialvo_critical,Chialvo_AIP,Bonachela,Torres_rev}. This would be in line with research that shows how certain useful properties -- such as the computational capacity of some neural-network models \citep{Bert}, or the dynamic range of sensitivity to stimuli in sensory systems \citep{Kinouchi} -- are optimised at this ``edge of chaos \citep{Chialvo_psycho}''.

That these models should actually reflect, albeit in an enormously simplified way, what actually goes on in our brains tends to fit in quite well with intuitive expectations --
to the extent that so-called {\it connectionist} models seem to be gradually becoming the accepted paradigm in relevant areas of psychology and philosophy \citep{Marcus, Frank}.
For instance, from this point of view the way in which the recollection of a particular detail
often evokes, almost instantly, a whole landscape of sensations and emotions makes
sense, since these concepts will have been stored in some way as the same pattern. Also,
the fact that new memories are recorded in synapses which were already being used to
store previous information would seem to explain why memories tend to fade slowly with time, yet can still be recalled, at least to some extent, when a particular
thought in some sense overlaps with (reminds one of) one of them. When this happens,
the old memory springs to mind and, if held there for long enough, can be refreshed via
long term potentiation and depression -- although interaction with other patterns or with current stimuli may well modify
the refreshed information. Similarly, previous information influences the storing of new
memories, leading to the well known fact that we tend to ``see'' things the way we
expect them to be.

It seems, then, that the basic mechanisms behind the ability of our brains to remember things, at least when the information is stored slowly enough for the biochemical processes of LTP and LTD to be at work (long-term memory), are now understood. Not only are the implications of such knowledge far-reaching in themselves. The way in which it was developed
is also particularly notworthy.
More or less sketchy ideas from areas as diverse as behavioural psychology, neurophysiology and theoretical physics were brought together in order to come up with a minimal mathematical model capable of manifesting the sought-after phenomenon of information retrieval
as a consequence of the known properties of a great many simple elements.
This kind of research can at first seem more like a mathematical game than anything to do with nitty-gritty reality. But the fact the basic mechanism of associative memory has since borne up to decades of experimenting and theoretical probing reveals how insightful it can actually be. It is likely that other features of brain function -- short-term memory, information processing or emotional tagging, to name but the first few that spring to mind -- will eventually be thrown under a similar light. In fact, we can expect the nature of even such an elusive and intimate phenomenon as consciousness some day to become clear. After all, the explanations behind other emergent properties of matter which in their day seemed almost mystical, such as temperature or life, are now well understood.


\section{A declaration of intent}

As Zora Neale Hurston put it, ``Research is formalized curiosity. It is poking and prying with a purpose.'' But there are many possible purposes, and even more different ways of poking and prying. The motivation behind the work presented here is to understand how the phenomena we observe in certain systems on a macroscopic scale can come about from interactions of their many relatively simple constituent elements. In the case of
neural systems,
it seems reasonable to assume that these basic elements are neurons, and that it is thanks to the cooperation of a great many of these cells that such organs are able to think and feel. The human brain -- with about 100 billion neurons connected by 100 trillion synapses -- being the most complex system we know of, an enormous degree of simplification will be required for our description to be of any use to this purpose. (In fact, if we could somehow simulate a brain in all detail, the result would be just as unfathomable as the original object, however
exciting
the activity may prove for other reasons.) The physiology of the neuron is nowadays quite well understood.
However, just as the properties of atoms or transistors that are key to understanding phase transitions or the workings of a microchip are, respectively, magnetic interactions and voltage-dependent gating, we must try to ascertain exactly which neuronal features are necessary for the macroscopic behaviour we are interested in to occur. One way to do this is to start by considering only the most basic characteristics and explore what non-trivial phenomena emerge from these, allowing us then to add new ingredients one at a time to pinpoint the relevant ones. In this line, we  consider large sets of Hopfield's simple binary model neurons to study how network properties are related to collective behaviour.

This work is laid out as follows.
Chapter \ref{Chapter_JSTAT} deals with development. The appearance and disappearance of edges in a network (growth and death of synapses, in the case of the brain) is formalised as a stochastic process and studied in a general setting \citep{Johnson_PRE,Johnson_JSTAT}. It turns out that many of the topological features observed in experiments are well modelled in this way -- which to some extent justifies, {\it a posteriori}, our initial assumptions. The following chapters describe particular phenomena that emerge as a direct consequence of some of those topological features: {\it degree heterogeneity} in conjunction with synaptic depression improves the performance of dynamical tasks \citep{Johnson_EPL} (Chapter \ref{Chapter_EPL}); {\it assortativity} serves to enhance a neural network's robustness to noise \citep{Sebas} (Chapter \ref{Chapter_PRE}); and {\it clustering} or {\it modularity} can lead to metastable states with certain properties essential for some short-term memory abilities (properties hitherto lacking in previous models) \citep{Johnson_CR} (Chapter \ref{Chapter_CR}). Thanks to the extreme simplicity of the basic elements we are considering, we are able not only to simulate but also to understand mathematically how exactly the interesting phenomena emerge. This makes it possible to predict, to some extent, which extra ingredients {\it will not} invalidate the results if they are taken into account explicitly.

Some of the work has a more general scope than the study of neural networks. In particular, the equations obtained in Chapter \ref{Chapter_JSTAT} can be applied to any network that evolves under the influence of probabilistic addition and deletion of edges. And the method put forward in Chapter \ref{Chapter_PRL} for the study of correlated networks can be used not just for analysing particular models, as we go on to do in Chapter \ref{Chapter_PRE}, but to solve many other problems -- such as that of the ubiquity of {\it disassortative} networks in nature and technology \citep{Johnson_PRL},
or how the property of nestedness typical of ecosystems is related to other topological characteristics (c.f. Appendix \ref{Appendix_NEST}).

To sum up, the aim of the thesis is {\bf to shed light on how cellular dynamics can lead to the complex network structures of neural systems, and, in its turn, in what ways this topology can influence, optimise and determine the collective behaviour of such systems.}

The main contributions made are:

\begin{itemize}
 \item 
An analytical method to study the evolution of networks governed by a combination of local and global stochastic rules.

\item
A mathematical and computational technique for the study of correlated networks in a model-independent way.

\item
Possible biological justifications for two non-trivial features of the topology of the human cortex: heterogeneity of the degree distribution and high assortativity. 

\item
An answer to the long-standing question of why most networks are disassortative.

\item
Cluster Reverberation: the first mechanism proposed which would allow neural systems to store information instantaneously in a robust manner.

\end{itemize}

%% file: jstat/jstat.tex





\chapter{Evolving networks and the development of neural systems} 
\label{Chapter_JSTAT}


The highly heterogeneous degree distributions of most empirical networks is 
assumed in many cases to arise from some 
form of cummulative advantage, or preferential attachment.
However, the origin of various other topological features
is often not clear and attributed to specific functional requirements. 
We show how it is possible to analyse a very general scenario in which 
nodes gain or lose edges according to 
arbitrary
functions of local and/or global degree information. 
Applying our method to two rather different examples of 
brain development -- synaptic pruning in humans and the neural network 
of the worm {\it C. Elegans} -- we find that simple biologically motivated 
assumptions lead to very good agreement with experimental data. 
In particular, many nontrivial topological features of the worm's brain 
arise naturally at a critical point.




\section{Introduction}

The conceptual simplicity of a \textit{network} -- a set of nodes, some pairs of which connected by edges -- often suffices to capture the essence of cooperation in complex systems. Ever since Barab\'{a}si and Albert presented their evolving network model \citep{Barabasi}, in which linear preferential attachment leads asymptotically to a scale-free degree distribution (the degree, $k$, of a node being its number of neighbouring nodes), there have
been many variations or refinements to the original scenario \citep{Albert,Bianconi_bose,Krapivsky,Bianconi_fitness,Park_Li,Suhan}
(see also the review by \citet{Boccaletti}).
In Ref. \citep{Johnson_PRE}, we show how topological phase transitions and scale-free solutions can emerge in the case of nonlinear rewiring in fixed-size networks, and this work is summarized in Appendix \ref{Appendix_RAP}. In Ref. \citep{Johnson_JSTAT}
we extend our scope to more general and realistic situations, considering the evolution of networks
making only minimal assumptions about the attachment/detachment rules. In fact, all we assume is 
that these probabilities factorize into two parts: a local term that
depends on node degree, and a global term, which is a function of the
mean degree of the network. This is the work described in this chapter.

Our motivation can be found in the
mechanisms behind many real-world networks, but we focus, for the sake of
illustration, on the development of biological neural networks,
where nodes
represent neurons and edges play the part of synaptic interaction \citep{Amit,Sporns,Torres_rev}. Experimental neuroscience has shown that enhanced electric activity induces synaptic growth and dendritic arborization
\citep{Lee,Frank,Klintsova,DeRoo}. Since the activity of a
neuron depends on the net current received from its neighbours, which tends to be higher
the more neighbours it has, we can see node degree as a proxy for this
activity -- accounting for the local term alluded to above.
On the other hand, synaptic growth and death also depend on concentrations of various molecules, which can diffuse through large areas of tissue and therefore cannot in general be considered local. A feature of brain development in many animals is \textit{synaptic pruning} -- the large reduction in synaptic density undergone throughout infancy.
\citet{Chechik_2,Chechik} have shown that via an elimination of less needed synapses, this can reduce the energy consumed by the brain (which in a human at rest can account
for a quarter of total energy used) while maintaining near optimal memory performance. Going on this, we will take the mean degree of the network -- or mean synaptic density -- to reflect total energy consumption, hence the global terms in our attachment/detachment rules \citep{Johnson_LNCS}.

An alternative approach would be to consider some kind of model neurons explicitly and couple the probabilities of synaptic growth and death to neuronal dynamic variables, such as local and global fields. In an Amari-Hopfield network, for example, the expected value of the field (total incoming current) at node $i$ is proportional to $i$'s degree \citep{Torres_influence}, the total current (energy consumption) in the network therefore being proportional to the mean degree; qualitatively, these observations are likely to hold also in more realistic situations \citep{Magistretti}, although relations need not be linear. Co-evolving networks of this sort are currently attracting attention, with dynamics such as Prisoner's Dilemma \citep{Julia}, Voter Model \citep{Eguiluz_coevolution} or Random Walkers \citep{Antiqueira}. Although we consider this line of work particularly interesting, for generality and analytical tractability we opt here to use only topological information
for the attachment/detachment rules,
although our results can be applied to any situation in which the dynamical states of the elements at the nodes can be functionally related to degrees\footnote{For instance, the stationary distribution of walkers used for edge dynamics by \citet{Antiqueira} is actually obtained purely from topological information, although it can only be written in terms of local degrees for undirected networks.}.

Following a brief general analysis, we show how appropriate choices of functions induce the system to evolve towards heterogeneous (sometimes scale-free) networks while undergoing synaptic pruning in quantitative agreement with experiments. At the same time, degree-degree correlations emerge naturally, thus making the resulting networks {\it disassortative} -- as tends to be the case for most biological networks -- and leading to realistic small-world parameters.

\section{Basic considerations}
\label{sec_basic}
Consider a simple undirected network with $N$ nodes defined by the adjacency matrix $\hat{a}$, the element $\hat{a}_{ij}$ representing the existence or otherwise of an edge between nodes $i$ and $j$. Each node can be characterized by its degree, $k_{i}=\sum_{j}\hat{a}_{ij}$. Initially, the degrees follow some distribution $p(k,t=0)$ with mean $\kappa(t)$. 
We wish to study the evolution of networks in which nodes can gain or lose edges according to stochastic rules which only take into account local and global information on degrees. So as to implement this in the most general way, we will assume that at every
time step, each node has a probability of gaining a new edge, $P_{i}^{\mbox{gain}},$ to a random node; and a probability of losing a randomly chosen edge, $P_{i}^{\mbox{lose}}.$ We assume these factorize as $P_{i}^{\mbox{gain}}=u(\kappa)\pi(k_{i})$ and $P_{i}^{\mbox{lose}}=d(\kappa)\sigma(k_{i})$,
where $u$, $d$, $\pi$ and $\sigma$ can be arbitrary functions, 
but impose nothing else other than normalization.

For each edge that is withdrawn from the network, two nodes decrease in degree: $i$, chosen according to $\sigma(k_{i})$, and $j$, a random neighbour of $i$'s; so there is an added effective probability of loss $k_{j}/(\kappa N)$. Similarly, for each edge placed in the network, not only $l$ chosen according to $\pi(k_{l})$ increases its degree; a random node $m$ will also gain, with the consequent effective probability $N^{-1}$ (though see\footnote{We are ignoring the small corrections that arise because $j\neq i$ and $l\neq m$, which in any case would disappear if self-connections were allowed.}).
Let us introduce the notation $\tilde{\pi}(k)\equiv\pi(k)+N^{-1}$ and $\tilde{\sigma}(k)\equiv\sigma(k)+k/(\kappa N)$. Network evolution can now be seen as a \textit{one step process} \citep{vanKampen} with transition rates $u(\kappa)\tilde{\pi}(k)$ and $d(\kappa)\tilde{\sigma}(k)$. The expected value for the increment in a given $p(k,t) $ at each time step (which we equate with a temporal derivative) defines a master equation for the degree distribution \citep{Johnson_PRE}:
\begin{eqnarray}
\displaystyle\frac{\mbox{d} p(k,t)}{\mbox{d} t} =u\left(  \kappa\right)
\tilde{\pi}(k-1) p(k-1)
+d\left(  \kappa\right) \tilde{\sigma}(k+1)
p(k+1)\nonumber\\
\label{eq1}\\
\displaystyle-\left[  u\left(  \kappa\right)   \tilde{\pi}(k)
+d\left(  \kappa\right) \tilde{\sigma}(k)
\right] p(k,t).\nonumber
\end{eqnarray}
Assuming now that $p(k,t)$ evolves towards a stationary distribution,
$p_{\mbox{\small{st}}}(k)$,
then this must necessarily satisfy detailed balance
since it is a one step process \citep{vanKampen};
i.e., the flux of probability from $k$ to $k+1$ must equal the flux from $k+1$ to $k$, for all $k$ \citep{MarroBook}. This condition (sufficient for Eq. (\ref{eq1}) to be zero) can be written as 
\begin{equation}
\frac{\partial p_{\mbox{\small{st}}}(k)}{\partial k} =\left[ \frac{u(\kappa_{\mbox{\small{st}}})}{d(\kappa_{\mbox{\small{st}}})}   \frac{\tilde{\pi}(k)}{\tilde{\sigma}(k+1)}-1\right] p_{\mbox{\small{st}}}(k),\label{eq2}%
\end{equation}
where we have
substituted a difference for a partial derivative and
$\kappa_{\mbox{\small{st}}}\equiv\sum_{k}kp_{\mbox{\small{st}}}(k)$.
Setting $\pi$ and $\sigma$ so as to be normalized to one (i.e., $\sum_{k}p(k)\pi(k)=\sum_{k}p(k)\sigma(k)=1$, $\forall t$),
which is equivalent to saying that at each time step exactly $u(\kappa)$ nodes are chosen to gain edges and $d(\kappa)$ to lose them, 
then in the stationary state we will have $u(\kappa_{\mbox{\small{st}}})=d(\kappa_{\mbox{\small{st}}})$ since the total number of edges will be conserved.
From  Eq. (\ref{eq2}) we can see that $p_{\mbox{\small{st}}}(k)$ will have an extremum at some value $k_{e}$ if it satisfies $\tilde{\pi}(k_{e})=\tilde{\sigma}(k_{e}+1)$. $k_{e}$ will be a maximum (minimum) if the numerator
in  Eq. (\ref{eq2}) is smaller (greater) than the denominator for $k>k_{e}$,
and viceversa for $k<k_{e}$. Assuming, for example, that there is one and only one such
$k_{e}$, then a maximum implies a relatively homogeneous distribution, while a
minimum means $p_{\mbox{\small{st}}}(k)$ will be split in two, and therefore
highly heterogeneous.
More intuitively, if for nodes with large enough $k$ there is a higher probability of gaining edges than of losing them, the degrees of these nodes will grow indefinitely, leading to heterogeneity. If, on the other hand, highly connected nodes always lose more edges than they gain, we will obtain quite homogeneous networks. From this reasoning we can see that a particularly interesting case (which turns out to be critical) is that in which
$\pi(k)$ and $\sigma(k)$
are such that 
\begin{equation}
\tilde{\pi}(k)=\tilde{\sigma}(k)\equiv v(k),\quad\forall k.
\label{condition}
\end{equation}
According to Eq. (\ref{eq2}), Condition (\ref{condition}) means that for large $k$, $\partial p_{st}(k)/\partial k\rightarrow 0$, and $p_{st}(k)$ flattens out -- as for example a power-law does.

The standard Fokker-Planck approximation for the one step process defined by Eq. (\ref{eq1}) is \citep{vanKampen}:
\begin{eqnarray}
\displaystyle\frac{\partial p(k,t)}{\partial t}=\frac{1}{2}\frac{\partial^{2}}{\partial k^{2}}
\left\{
\left[d(\kappa)\tilde{\sigma}(k)+u(\kappa)\tilde{\pi}(k)  \right]
p(k,t)\right\}\nonumber
\\
\label{eq_FokkerPlanck}\\
\displaystyle+
\frac{\partial}{\partial k}
\left\{\left[d(\kappa)\tilde{\sigma}(k)-u(\kappa)\tilde{\pi}(k)  \right]
p(k,t)\right\}
\nonumber.
\end{eqnarray}
For transition rates which meet Condition (\ref{condition}), Eq. (\ref{eq_FokkerPlanck}) can be written as:
\begin{eqnarray}
\displaystyle\frac{\partial p(k,t)}{\partial t}   =\frac{1}{2}\left[ d\left(  \kappa\right)+u\left(  \kappa\right)  \right]  \frac{\partial^{2}}{\partial k^{2}}\left[
v(k)p(k,t)\right]  \nonumber\\
\label{diff_drift}\\
\displaystyle +\left[d\left(  \kappa\right)-  u\left(  \kappa\right)  \right]
\frac{\partial}{\partial k}\left[  v(k)p(k,t)\right].\nonumber
\end{eqnarray}
Ignoring boundary conditions, the stationary solution must satisfy, on the one
hand, $v(k)p_{\mbox{\small{st}}}(k)=Ak+B,$ so that the diffusion is
stationary, and, on the other, $u(\kappa_{\mbox{\small{st}}})=d(\kappa
_{\mbox{\small{st}}}),$ to cancel out the drift. For this situation to be
reachable from any initial condition, $u(\kappa)$ and $d(\kappa)$ must be
monotonous functions, decreasing and increasing respectively.

\section{Synaptic pruning}

\begin{figure}[t!]
\begin{center}
\epsfig{file=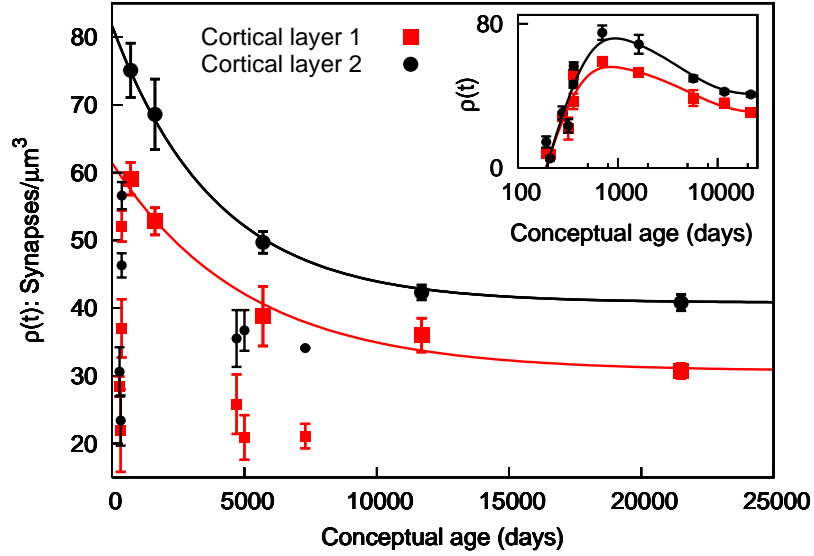,scale=0.6}%
\caption{
Synaptic densities in layers 1 (red squares) and 2 (black circles) of the human auditory cortex against time from conception. Data from \citet{Huttenlocher}, obtained by directly counting synapses in tissues from autopsies. Lines follow best fits to  Eq. (\ref{rho_t}), where the parameters were: for layer 1, $\tau_{\mbox{p}}=5041$ days; and for layer 2, $\tau_{\mbox{p}}=3898$ days (for $\rho_{\mbox{f}}$ we have used the last data pints: $30.7$ and $40.8$ synapses/$\mu m^{3}$, for layers 1 and 2 respectively). Data pertaining to the first year and to days $4700$, $5000$ $7300$, shown with smaller symbols, where omitted from the fit. Assuming the existence of transient growth factors, we can include the data points for the first year in the fit by using Eq. (\ref{rho_t2}). This is done in the inset (where time is displayed logarithmically). The best fits were: for layer 1, $\tau_{\mbox{g}}=151.0$ and $\tau_{\mbox{p}}=5221$; and for layer 2, $\tau_{\mbox{g}}=191.1$ and $\tau_{\mbox{p}}=4184$, all in days (we have approximated $t_{0}$ to the time of the first data points, $192$ days).}%
\label{fig_exp}%
\end{center}
\end{figure}

As a simple example,
we will first consider global probabilities which have the linear forms:
\begin{equation}
u[\kappa(t)]=\frac{n}{N}\left(  1-\frac{\kappa(t)}{\kappa_{\mbox{\small{max}}}%
}\right) \quad \mbox{ and }\quad d[\kappa(t)]=\frac{n}{N}\frac{\kappa(t)}{\kappa
_{\mbox{\small{max}}}},\label{u_d_linear}%
\end{equation}
where $n$ is the expected value of the number of additions and
deletions of edges per time step, and $\kappa_{\mbox{\small{max}}}$ is the maximum
value the mean degree can have. This choice describes a situation in which the
higher the density of synapses, the less likely new ones are to
sprout and the more likely existing ones
are to atrophy -- a situation that might arise, for instance, in the presence of a finite quantity of nutrients. Again taking into account that $\pi$ and $\sigma$
are normalized to one,
summing over $P_{i}^{\mbox{gain}}-P_{i}%
^{\mbox{lose}}$ we find that the expected increment in $\kappa(t)$ is 
$$
\langle\frac{\Delta\kappa(t)}{\Delta t}\rangle =2\lbrace u[\kappa(t)]-d[\kappa(t)]\rbrace=2\frac{n}{N}\left[  1-2\frac{\kappa(t)}{\kappa_{\mbox{\small{max}}}}%
\right] 
$$
(independently of the local probabilities). 
Therefore, the mean degree will increase
or decrease exponentially with time, from $\kappa(0)$ to $\frac{1}{2}\kappa_{\mbox{\small{max}}}$.
Assuming that the initial condition is, say, $\kappa(0)=\kappa
_{\mbox{\small{max}}}$, and expressing the solution in terms of the
\textit{mean synaptic density} -- i.e., $\rho(t)\equiv\kappa(t)N/(2V),$ with $V$
the total volume considered -- we have
\begin{equation}
\rho(t)=\rho_{\mbox{f}}\left(1+e^{-t/\tau_{\mbox{p}}}\right)  ,\label{rho_t}%
\end{equation}
where we have defined $\rho_{\mbox{f}}\equiv\rho(t\rightarrow\infty)$ and the time constant for pruning is $\tau_{\mbox{p}}=\rho_{\mbox{f}}N/n$. This equation was fitted in Fig. \ref{fig_exp}
to experimental data on layers 1 and 2 of the human auditory cortex\footnote{Data points for three particular days (smaller symbols) are omitted from the fit, since we believe these must be from subjects with inherently lower synaptic density.} obtained during autopsies by \citet{Huttenlocher}. 

It seems reasonable to assume that the initial overgrowth of synapses is due to the transient existence of some kind of growth factors. If we account for these by including a nonlinear, time-dependent term $g(t)\equiv
a\exp(-t/\tau_{\mbox{g}})$ in the probability of growth, i.e., $u[\kappa(t),t]=(n/N)[1-\kappa
(t)/\kappa_{\mbox{\small{max}}}+g(t)],$ leaving $d[\kappa(t)]$ as before, we find that $\rho(t)$ becomes
\begin{equation}
\rho(t)=\rho_{\mbox{f}}\left[1+e^{-t/\tau_{\mbox{p}}}-\left( 1+e^{-t_{0}/\tau_{\mbox{p}}} \right)e^{-\frac{t-t_{0}}{\tau_{\mbox{g}}}}  \right]  ,\label{rho_t2}%
\end{equation}
where $t_{0}$ is the time at which synapses begin to form ($t=0$ corresponds to the moment of conception) and $\tau_{\mbox{g}}$ is the time constant related to growth.
The inset in Fig. \ref{fig_exp} shows the best fit to the auditory cortex data.
Since the contour conditions $\rho_{\mbox{f}}$ and (for  Eq. (\ref{rho_t2})) $t_{0}$ are simply taken as the value of the last data point and the time of the first one, in each case, the time constants $\tau_{\mbox{p}}$ and $\tau_{\mbox{g}}$ are the only parameters needed for the fit.

\begin{figure}
[t!]
\begin{center}
\hspace{-0.6cm}\epsfig{file=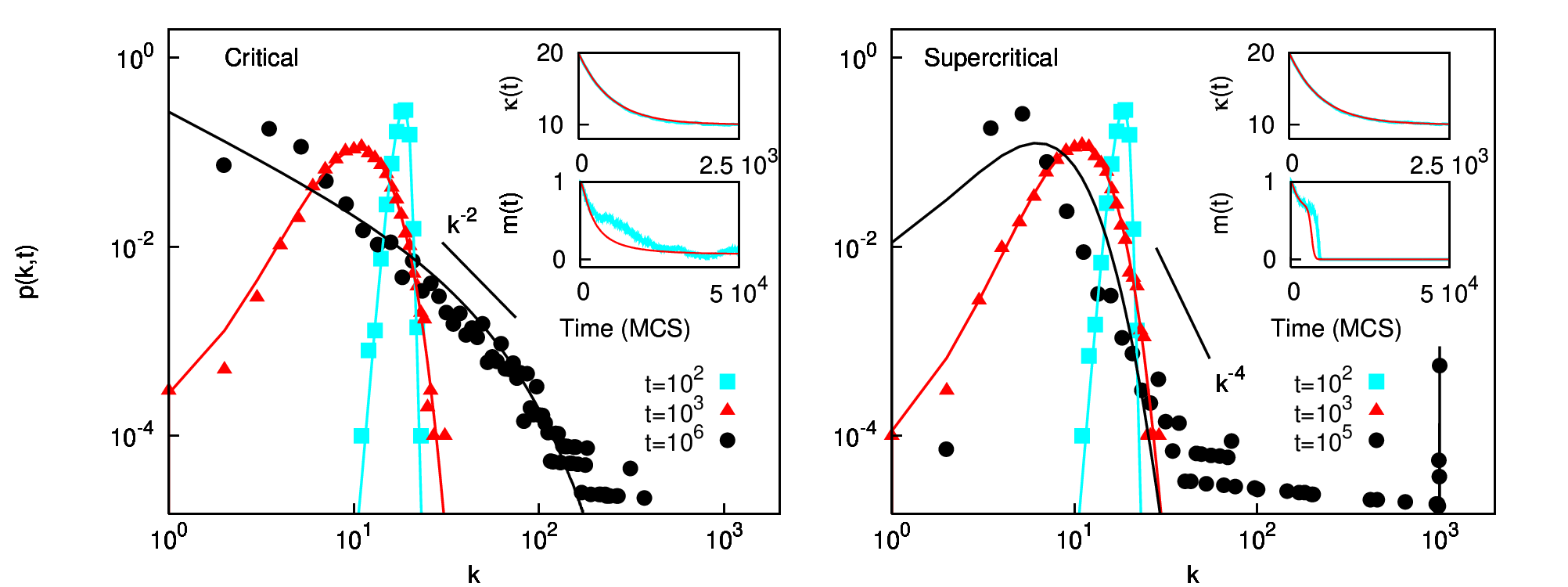,width=13cm}
\caption{
Evolution of the degree distributions of networks 
beginning as regular random graphs with $\kappa(0)=20$ in the critical (top)
and supercritical (bottom) regimes. Local probabilities are
$\sigma(k)=k/(\langle k\rangle N)$
in
both cases, and $\pi(k)=2\sigma(k)-N^{-1}$ and
$\pi(k)= k^{3/2}/(\langle k^{3/2}\rangle N)$
for the critical and supercritical ones, respectively. Global probabilities as in Eq. (\ref{u_d_linear}), with $n=10$ and $\kappa_{\mbox{\small{max}}}=20$. Symbols
in the main panels correspond to $p(k,t)$ at different times as obtained from
MC simulations. Lines result from numerical integration of Eq. (\ref{eq1}). 
Insets show typical time series of $\kappa$ and $m$. Light blue lines are from MC simulations and red lines are theoretical, 
given by  Eq. (\ref{rho_t}) and Eq. (\ref{eq1}), respectively. $N=1000$.}%
\label{fig_ab}%
\end{center}
\end{figure}

\section{Phase transitions}

The drift-like evolution of the mean degree we have just illustrated with
the example of synaptic pruning is independent of the local probabilities $\pi(k)$ and
$\sigma(k)$. The effect of these is rather in the diffusive behaviour which can
lead, as mentioned, either to homogeneous or to heterogeneous final states. A useful
bounded order parameter to characterize these phases is therefore $m\equiv\exp\left(
-\sigma^{2}/\kappa^{2}\right),$ where $\sigma^{2}=\langle k^{2}%
\rangle-\kappa^{2}$ is the variance of the degree distribution ($\langle\cdot\rangle\equiv N^{-1}\sum_{i}(\cdot)$ represents an average over nodes). We will use
$m_{\mbox{\small{st}}}\equiv\lim_{t\rightarrow\infty}m(t)$ to distinguish
between the different phases, since $m_{\mbox{\small{st}}}=1$ for a regular
network and $m_{\mbox{\small{st}}}\rightarrow0$ for one following a highly
heterogeneous distribution. Although there are particular choices of probabilities which lead to  Eq. (\ref{diff_drift}), these are not the only critical cases, since the transition from homogeneous to heterogeneous stationary states can come about also with functions which never meet Condition (\ref{condition}). Rather, this is a classic topological phase transition, the nature of which depends on the choice of functions \citep{Park_equilibrium,Burda,Derenyi} .

\begin{figure}
[t!]
\begin{center}
\epsfig{file=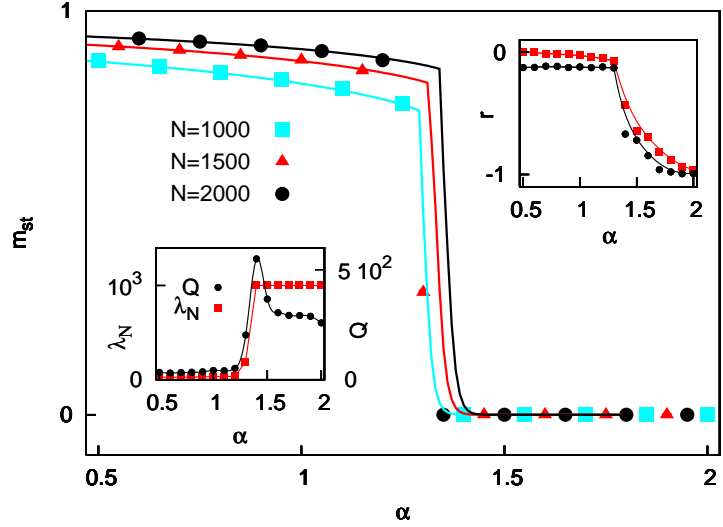,width=10cm}
\caption{
Phase transitions in $m_{\mbox{\small{st}}}$ for
$\pi(k)\sim k^{\alpha}$ and $\sigma(k)\sim k$, and $u(\kappa)$ and $d(\kappa)$
as in  Eq. (\ref{u_d_linear}). $N=1000$ (blue squares), $1500$ (red triangles)
and $2000$ (black circles); $\kappa(0)=\kappa_{\mbox{\small{max}}}=2n=N/50$.
Corresponding lines are from numerical integration of  Eq. (\ref{eq1}). 
The bottom left inset shows values of the highest eigenvalue of the Laplacian matrix (red squares) and of $Q=\lambda_{N}/\lambda_{2}$ (black circles), a measure of unsynchronizability; $N=1000$. The top right inset shows transitions for the same parameters in the final values of Pearson's correlation coefficient $r$ (see Section \ref{sec_correlations}), both for only one edge allowed per pair of nodes (red squares) and without this restriction (black circles).
}%
\label{fig_fss}%
\end{center}
\end{figure}

Evolution of the degree distribution is shown in Fig. \ref{fig_ab} for
critical and supercritical choices for the probabilities, as given
by MC simulations (starting from regular random graphs) and contrasted with theory. (The subcritical regime is not shown since the stationary state has a distribution similar to the ones at $t=10^{3}$ MCS in the other regimes.) The disparity between the theory and the
simulations for the final distributions is due to the build up of certain
correlations not taken into account in our analysis. This is because the
existence of some very highly connected nodes reduces the probability of there being very
low degree nodes. In particular, if there are, say, $x$ nodes connected to the
rest of the network, then a natural cutoff, $k_{min}=x$, emerges. Note that
this occurs only when we restrict ourselves to simple networks, i.e., with
only one edge allowed for each pair of nodes. 
This topological phase transition is
shown in Fig. \ref{fig_fss}, where $m_{\mbox{st}}$ is plotted against parameter $\alpha$ for global probabilities as in  Eq. (\ref{u_d_linear}) and local ones $\pi(k)\sim k^{\alpha}$ and $\sigma(k)\sim k$. This situation corresponds to one in which edges are eliminated randomly while nodes have a power-law probability of sprouting new ones (note that power-laws are good descriptions of a variety of monotonous response functions, yet only require one parameter).
Although, to our knowledge, there are not yet enough empirical data to ascertain what degree distribution the structural topology of the human brain follows, it is worth noting that its functional topology, at the level of brain areas, has been found to be scale-free with an exponent very close to $2$ \citep{Eguiluz}.

In general, most other measures can be expected to undergo a transition along with its variance. For instance, highly heterogeneous networks (such as scale-free ones) exhibit the small-world property, characterized by a high \textit{clustering coefficient,}
$C\gg \langle k\rangle /N$, and a low \textit{mean minimum path,} $l\sim \ln(N)$ \citep{Watts}. A particularly interesting topological feature of a network is its {\it synchronizability} -- i.e., 
given a set of oscillators placed at the nodes and coupled via the edges, how wide a range of coupling strengths will result in them all becoming synchronized.
Barahona and Pecora showed analytically that, for linear oscillators, a network is more
synchronizable the lower the relation $Q=\lambda_{N}/\lambda_{2}$ -- where $\lambda_{N}$ and $\lambda_{2}$ are the highest and lowest non-zero eigenvalues of the Laplacian matrix ($\hat{\Lambda}_{ij}\equiv \delta_{ij}k_{i}-\hat{a}_{ij}$), respectively \citep{Barahona}.
The bottom left inset in Fig. \ref{fig_fss} displays the values of $Q$ and $\lambda_{N}$ obtained for the different stationary states. There is a peak in $Q$
at the critical point.
It has been argued that this tendency of heterogeneous topologies to be particularly unsynchronizable poses a paradox given the wide prevalence of scale-free networks in nature, a problem that has been deftly got around by considering appropriate weighting schemes for the edges \citep{Adilson_paradox,Chavez} (see also\footnote{Using pacemaker nodes, scale-free networks have also been shown to emerge via rules which maximize synchrony \citep{Plos}.}, and the review by \citet{Arenas_rev}). However, there is no generic reason why high synchronizability should always be desirable. In fact, it has recently been shown that heterogeneity can improve the dynamical performance of model neural networks precisely because the fixed points are easily destabilised \citep{Johnson_EPL} (as well as
conferring robustness to thermal fluctuations and 
improving storage capacity \citep{Torres_influence}). This makes intuitive sense, since, presumably,
one would not usually want all the neurons in one's brain to be doing exactly the same thing. Therefore,
this point of maximum \textit{unsynchronizability} at the phase transition may be a particularly advantageous one.

On the whole, we find that three classes of network -- homogeneous, scale-free (at the critical point) and ones composed of starlike structures, with a great many small-degree nodes connected to a few hubs -- can emerge for any kind of attachment/detachment rules. It follows that a network subject to some sort of optimising mechanism, such as Natural Selection for the case of living systems, could thus evolve towards whichever topology best suits its requirements by tuning these microscopic actions.

\section{Correlations}
\label{sec_correlations}
Most real networks have been found to exhibit degree-degree correlations, also known as {\it mixing} by degree \citep{Pastor-Satorras,Newman_rev}. They can thus be classified as {\it assortative}, when the degree of a typical node is positively correlated with that of its neighbours, or {\it disassortative,} when the correlation is negative. This property has important implications for network characteristics such as connectedness and robustness \citep{Newman_mixing_PRL,Newman_mixing_PRE}. A useful measure of this phenomenon is Pearson's correlation coefficient applied to the edges \citep{Newman_rev, Newman_mixing_PRE, Boccaletti}: $
r= ([ k_{l}k'_{l}]-[ k_{l}]^{2})/([ k_{l}^{2}]-[ k_{l}]^{2}), $
where $k_{l}$ and $k'_{l}$ are the degrees of each of the two nodes
pertaining to edge $l$, and $[\cdot]\equiv(\langle k\rangle
N)^{-1}\sum_{l}(\cdot)$ represents an average over edges; $|r|\leq 1$. Writing
$\sum_{l}(\cdot)=\sum_{ij}\hat{a}_{ij}(\cdot)$, $r$ can be expressed in
terms of averages over nodes:
\begin{equation}
  r=\frac{\langle k\rangle \langle k^{2} k_{nn}(k)\rangle - 
    \langle k^{2}\rangle^{2} }{\langle k\rangle \langle k^{3}\rangle 
    - \langle k^{2}\rangle^{2}},
  \label{r_gen}
\end{equation}
where $k_{nn}(k)$ is the mean nearest-neighbour-degree function; i.e., if $k_{nn,i}\equiv k_{i}^{-1}\sum_{j}\hat{a}_{ij}k_{j}$ is the mean degree of the neighbours of node $i$, $k_{nn}(k)$ is its average over all nodes such that $k_{i}=k$. Whereas most social networks are assortative ($r>0$) -- due, probably, to mechanisms such as homophily \citep{Newman_rev} -- almost all other networks, whether biological, technological or information-related, seem to be generically disassortative. The top right inset in Fig. \ref{fig_fss} displays the stationary value of $r$ obtained in the same networks as in the main panel and lower inset. It turns out that the heterogeneous regime is disassortative, the more so the larger $\alpha$. (Note that a completely homogeneous network cannot have degree-degree correlations, since all degrees are the same.) It is known that the restriction of having at most one edge per pair of nodes induces disassortativity \citep{Park_correlations,Maslov}.
However, in our case this is not the sole origin of the correlations, as can also be seen in the same inset of Fig. \ref{fig_fss}, where we have plotted $r$ for networks in which we have lifted the restriction and allowed any number of edges per pair of nodes. In fact, when multiple edges are allowed, the correlations are slightly stronger. As we shall discuss in Chapter \ref{Chapter_PRL}, there is a general entropic reason for heterogeneous networks, in their equilibrium state (i.e., in the absence of correlating mechanisms), to become disassortative \citep{Johnson_PRL}. But neither is this here the case, since the networks generated are driven from equilibrium by the mechanisms of preferential attachment and detachment.

To understand how these specific correlations come about, consider a pair of nodes $(i,j)$, which, at a given moment, can either be occupied by an edge or unoccupied. We will call the expected times of permanence for occupied and unoccupied states $\tau_{ij}^{\mbox{o}}$ and $\tau_{ij}^{\mbox{u}}$, respectively. After sufficient evolution time (so that occupancy becomes independent of the initial state\footnote{Note that this will always happen eventually since the process is ergodic.}), the expected value of the corresponding element of the adjacency matrix, $E(\hat{a}_{ij})\equiv \hat{\epsilon}_{ij}$, will be
$$
\hat{\epsilon}_{ij}=\frac{\tau_{ij}^{\mbox{o}}}{\tau_{ij}^{\mbox{o}}+\tau_{ij}^{\mbox{u}}}.
$$
If $p_{ij}^{+}$ ($p_{ij}^{-}$) is the probability that $(i,j)$ will become occupied (unoccupied) given that it is unoccupied (occupied), then $\tau_{ij}^{\mbox{o}}\sim 1/p_{ij}^{-}$ and $\tau_{ij}^{\mbox{u}}\sim 1/p_{ij}^{+}$, yielding
$$
\hat{\epsilon}_{ij}=\left(1+\frac{p_{ij}^{-}}{p_{ij}^{+}}\right)^{-1}.
$$
Taking into account the probability that each node has of gaining or losing an edge, we obtain\footnote{Again, we are ignoring corrections due to the fact that $i$ is necessarily different from $j$.}:
$p_{ij}^{+}=u(\langle k\rangle)N^{-1}[\pi(k_{i})+\pi(k_{j})]$ and $p_{ij}^{-}=d(\langle k\rangle)[\sigma(k_{i})/k_{i}+\sigma(k_{j})/k_{j}]$.
Then, assuming that the network is sparse enough that $p_{ij}^{-}\gg p_{ij}^{+}$ (since the number of edges is much smaller than the number of pairs), and particularising for power-law local probabilities $\pi(k)\sim k^{\alpha}$ and $\sigma(k)\sim k^{\beta}$, the expected occupancy of the pair is
$$
\hat{\epsilon}_{ij}\simeq\frac{p_{ij}^{+}}{p_{ij}^{-}}=\frac{u(\langle k\rangle)}{d(\langle k\rangle)}\frac{\langle k^{\beta}\rangle}{\langle k^{\alpha}\rangle N}
\left(\frac{k_{i}^{\alpha}+k_{j}^{\alpha}}{k_{i}^{\beta-1}+k_{j}^{\beta-1}} \right).
$$
Considering the stationary state, when $u(\langle k\rangle)=d(\langle k\rangle)$, and for the case of random deletion of edges, $\beta=1$ (so that the only nonlinearity is due to $\alpha$), the previous expression reduces to
\begin{equation}
\hat{\epsilon}_{ij}\simeq\frac{\langle k\rangle}{2\langle k^{\alpha}\rangle N}
\left(k_{i}^{\alpha}+k_{j}^{\alpha} \right).
\label{epsi_b1}
\end{equation}
(Note that this matrix is not consistent term by term, since $\sum_{j}\hat{\epsilon}_{ij}\neq k_{i}$, although it is globally consistent: $\sum_{ij}\hat{\epsilon}_{ij}=\langle k\rangle N$.)
The nearest-neighbour-degree function is now
$$
k_{nn}(k_{i})=\frac{1}{k_{i}}\sum_{j}\hat{\epsilon}_{ij}k_{j}=
\frac{\langle k\rangle}{2\langle k^{\alpha}\rangle}
(\langle k\rangle k_{i}^{\alpha-1}+\langle k^{\alpha+1}\rangle k_{i}^{-1})
\label{knn}
$$
(a decreasing function for any $\alpha$),
with the result that Pearson's coefficient becomes
\begin{equation}
r=\frac{1}{\langle k^{\alpha}\rangle}\left(
\frac{\langle k\rangle^{3}\langle k^{\alpha+1}\rangle-\langle k^{2}\rangle^{2}\langle k^{\alpha}\rangle }{\langle k\rangle \langle k^{3}\rangle -\langle k^{2}\rangle^{2}}
\right).
\label{eq_r}
\end{equation}
More generally, one can understand the emergence of these correlations in the following way. For the network to become heterogeneous, we must have $\pi(k)+N^{-1}\geq\sigma(k)+k/(\langle k\rangle N)$ for large enough $k$, so that highly connected nodes do not lose more edges than they can acquire (see Section \ref{sec_basic}). This implies that $\pi(k)$ must be increasing and approximately linear or superlinear. The expected value of the degree of a node $i$, chosen according to $\pi(k_{i})$, is then $E(k_{i})=N^{-1}\sum_{k} \pi(k)k\gtrsim \langle k^{2}\rangle/\langle k\rangle$, while that of its new, randomly chosen neighbour, $j$, is only $E(k_{j})=\langle k\rangle$. This induces disassortative correlations which can never be compensated by the breaking of edges between nodes whose expected degree values are $N^{-1}\sum_{k} \sigma(k)k$ and $\langle k^{2}\rangle/\langle k\rangle$ if $\sigma(k)$ is an increasing function. It thus ensues that a scenario such as the one analysed in this paper will never lead to assortative networks except for some cases in which $\sigma(k)$ is a decreasing function -- meaning that less connected nodes should be more likely to lose edges. Assortativity could, however, arise if there were some bias also on the node chosen to be $i$'s neighbour, e.g. on the postsynaptic neuron -- which is precisely what happens in most social networks, where individuals do not generally choose their friends, partners, etc. randomly. Although there seem to be other reasons for the ubiquity of disassortative networks in nature \citep{Johnson_PRL}, it is possible that the generality of the scenario studied here may also play a part.

We can use the expected value matrix $\hat{\epsilon}$ to estimate other magnitudes. For example, the clustering coefficient, as defined by Watts and Strogatz \citep{Watts}, is an average over nodes of $C_{i}$, with $C_{i}$ the proportion of $i$'s neighbours which are connected to each other; so its expected value is $E(C_{i})=\hat{\epsilon}_{jl}$ conditioned to $j$ and $l$ being neighbours of $i$'s. This means that, on average, we can make the approximation that
$$
k_{j}=k_{l}=\langle k_{nn}\rangle=\frac{\langle k\rangle}{2\langle k^{\alpha}\rangle}[\langle k\rangle\langle k^{\alpha-1}\rangle+\langle k^{\alpha+1}\rangle\langle k^{-1}\rangle].
$$
Substituting this value in  Eq. (\ref{epsi_b1}), and taking into account that one edge of $j$'s and one of $l$'s are taken up by $i$, we have
\begin{equation}
C\simeq \frac{\langle k\rangle}{\langle k^{\alpha}\rangle N}(\langle k_{nn}\rangle-1)^{\alpha}.
\label{C}
\end{equation}
For a rough estimate of the mean minimum path (the minimum path between two nodes being the smallest number of edges one has to follow to get from one to the other), we can proceed as \citet{Albert_rev}. For a given node, let us define the number of nearest neighbours, $z_{1}$, next-nearest neighbours, $z_{2}$, and in general $m$th neighbours, $z_{m}$. Using the relation
$
z_{m}=z_{1}\left(z_{2}/z_{1}\right)^{m-1},
$
and assuming that the network is connected and can be obtained in $l$ steps, this yields
\begin{equation}
1+\sum_{1}^{l}z_{m}=N.
\label{lz}
\end{equation}
On average, $z_{1}=\langle k\rangle$ and $z_{2}=\langle k\rangle[(1-C)\langle k_{nn}\rangle-1]$ (since for each second nearest neighbour, one edge goes to the reference node and a proportion $C$ to mutual neighbours). Now, if $N\gg z_{1}$ and $z_{2}\gg z_{1}$,  Eq. (\ref{lz}) leads to
\begin{equation}
l\simeq 1+\frac{\ln(N/\langle k\rangle)}{\ln[(1-C)\langle k_{nn}\rangle-1]}.
\label{l}
\end{equation}

\section{The C. Elegans neural network}

There exists a biological neural network which has been entirely mapped (although not, to the best of our knowledge, at different stages of development) -- that of the much-investigated worm \textit{C. Elegans} \citep{White, Watts}. With a view to testing whether such a network could arise via simple stochastic rules of the kind we are here considering, we ran simulations for the same number of nodes, $N=307$, and (stationary) mean degree,
$\langle k\rangle =14.0$ (in the simple, undirected representation of the network).
Using the global probabilities given by  Eq. (\ref{u_d_linear}) and local ones $\pi(k)\sim k^{\alpha}$ and $\sigma(k)\sim k$ (as in Fig. \ref{fig_fss}), we obtain a surprising result. Precisely at the critical point, $\alpha=\alpha_{c}\simeq1.35$, there are some remarkable similarities between the biological network and the ones produced by the model.

\begin{figure}
[t!]
\begin{center}
\epsfig{file=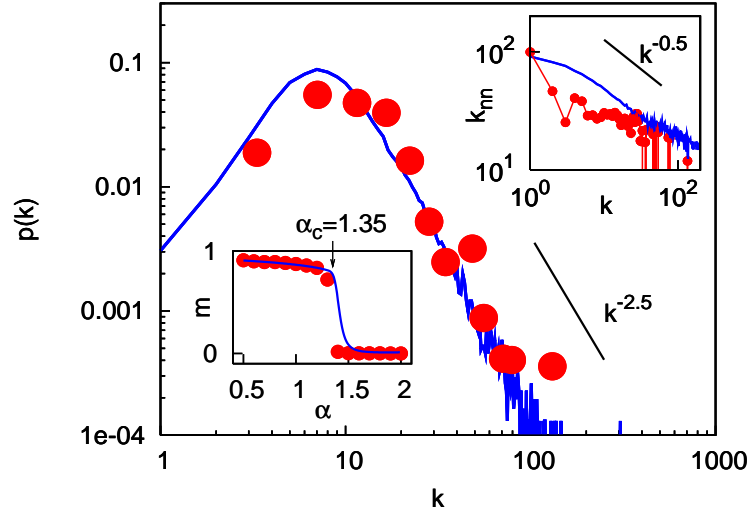,width=10cm}
\caption{
Degree distribution (binned) of the {\it C. Elegans} neural network (circles) \citep{White} and that obtained with MC simulations (line) in the stationary state ($t=10^{5}$ steps) for an equivalent network in which edges are removed randomly ($\beta =1$) at the critical point ($\alpha=1.35$). $N=307$, $\kappa_{\mbox{st}}=14.0$, averages over $100$ runs. Global probabilities as in Eq. (\ref{u_d_linear}). The slope is for $k^{-5/2}$. Top right inset: mean-neighbour-degree function $k_{nn}(k)$ as measured in the same empirical network (circles) and as given by the same simulations (line) as in the main panel. The slope is for $k^{-1/2}$. Bottom left inset: $m_{\mbox{st}}$ of equivalent network for a range of $\alpha$, both from simulations (circles) and as obtained with  Eq. (\ref{eq1}). (See also Table \ref{table_CElegans}.)
}%
\label{fig_CElegans}%
\end{center}
\end{figure}

Figure \ref{fig_CElegans} displays the degree distributions, both for the empirical network and for the average (stationary) simulated network corresponding to the critical point,
while the top inset shows the mean-nearest-neighbour degree function $k_{nn}(k)$ for the same networks.
Both $p(k)$ and $k_{nn}(k)$ of the simulated networks can be seen to be very similar to those measured in the biological one. 
Furthermore, as is displayed in Table \ref{table_CElegans}, the clustering coefficient obtained in simulation is almost the same as the empirical one. The mean minimum path is similar though slightly smaller in simulation, probably due to the worm's brain having modules related to functions \citep{Arenas_celegans}.
Finally,
Pearson's coefficient is
also in fairly good agreement, although the simulated networks are actually a bit more disassortative. It should, however, be stressed that the simulation results are for averages over $100$ runs, while the biological system is equivalent to a single run; given the small number of neurons, statistical fluctuations can be fairly large, so one should refrain from attributing too much importance to the precise values obtained -- at least until we can average over $100$ worms.
Table \ref{table_CElegans} also shows the values of $C$, $l$ and $r$ both as estimated form the theory laid out in Section \ref{sec_correlations}, and for the equivalent network in the \textit{configuration model} \citep{Newman_rev} -- generally taken as the null model for heterogeneous networks, where the probability of an edge existing between nodes $i$ and $j$ is $k_{i}k_{j}/(\langle k\rangle N)$. It is clear that whereas the configuration-model predictions deviate substantially from the
magnitudes measured in the {\it C. Elegans} neural network, the growth process we are here considering accounts for them quite well.
\begin{table}[t!]
\begin{center}
\begin{tabular}{l|llll}
\cr &Experiment &Simulation &Theory &Config.\\
\hline
\hline
\\
$C$ & 0.28 & 0.28 & 0.23 & 0.15 \\\\
$l$ & 2.46 & 2.19 &  1.86 & 1.96\\\\
$r$ & -0.163 & -0.207 & -0.305 & -0.101\\
\end{tabular}
\caption{\label{table_CElegans}Values of small-world parameters $C$ and $l$, and Pearson's correlation coefficient $r$, as measured in the neural network of the worm {\it C. Elegans} \citep{White}, and as obtained from simulations in the stationary state ($t=10^{5}$ steps) for an equivalent network at the critical point when edges are removed randomly -- i.e., for $\alpha=1.35$ and $\beta=1$. $N=307$, $\kappa_{\mbox{st}}=14.0$; averages over $100$ runs and global probabilities as in  Eq. (\ref{u_d_linear}). Theoretical estimates correspond to  Eqs. (\ref{C}), (\ref{l}) and (\ref{eq_r}) applied to the networks generated by the same simulations. The last column lists the respective \textit{configuration model} values: $C$ and $l$ are obtained theoretically as in \citep{Newman_rev}, while $r$, from MC simulations as in \citep{Maslov}, is the value expected due to the absence of multiple edges. (See also Fig. \ref{fig_CElegans}.)}
\end{center}
\end{table}
It is interesting that it should be at the critical point that a structural topology so similar to the empirical one emerges, since it seems that the brain's functional topology may also be related to a critical point \citep{Chialvo_critical,Chialvo_AIP}.

\section{Discussion}

With this work we have attempted, on the one hand, to extend our understanding of evolving networks so that any choice of transition probabilities dependent on local and/or global degrees can be treated analytically, thereby obtaining some model-independent results; and on the other, to illustrate how such a framework can be applied to realistic biological scenarios. For the latter, we have used two examples relating to two rather different nervous systems:
\\
{\bf i)} synaptic pruning in humans, for which the use of nonlinear global probabilities reproduces the initial increase and subsequent depletion in synaptic density in good accord with experiments -- to the extent that nonmonotonic data points spanning a lifetime can be very well fitted with only two parameters; and
\\
{\bf ii)} the structure of the {\it C. Elegans} neural network, for which it turns out that by only considering the numbers of nodes and edges, and imposing random deletion of edges and power-law probability of growth, the critical point leads to networks exhibiting many of the worm's nontrivial features -- such as the degree distribution, small-world parameters, and even level of disassortativity.

These examples indicate that it is not far-fetched to contemplate how many structural features of the brain or other networks -- and not just the degree distributions -- could arise by simple stochastic rules like the ones considered; although, undoubtedly, other ingredients such as natural modularity \citep{Arenas_celegans}, a metric \citep{Kaiser_spatial} or functional requirements \citep{Sporns} can also be expected to play a role in many instances.  We hope, therefore, that the framework laid out here -- in which for simplicity we have assumed the network to be undirected and to have a fixed size, although generalizations are straightforward -- may prove useful for interpreting data from a variety of fields. It would be particularly interesting to try to locate and quantify the biological mechanisms assumed to be behind this kind of network dynamics.




%% file: epl/epl.tex


\chapter{Bringing on the Edge of Chaos with heterogeneity} 
\label{Chapter_EPL}



The collective behaviour of systems of coupled excitable elements, such as neurons, 
has been shown to depend significantly on the heterogeneity of the degree distribution 
of the underlying network of interactions. For instance, broad -- in particular, 
scale-free -- distributions have been found to improve static memory performance 
in neural-network models. Here we look at the influence of degree heterogeneity 
in a neural network which, due to the effect of synaptic depression (a kind of 
fatigue of the interaction strengths), exhibits chaotic behaviour. Not only can 
the existence of a chaotic phase be related to neurophysiological experiments; 
it allows the system to perform a class of dynamic pattern-recognition tasks. 
We find first of all that, as has been described in a few other systems, 
optimal performance is achieved close to the phase transition -- i.e., at 
the so-called {\it Edge of Chaos}. Furthermore, we obtain a functional 
relationship between the level of synaptic depression required to bring 
on chaos and the heterogeneity of the degree distribution. This result 
points to a clear advantage of low-exponent scale-free networks, and suggests 
an explanation for their apparent ubiquity in certain biological systems.


\section{Exciting cooperation}

Excitable systems allow for the regeneration of waves propagating through
them, and may thus respond vigorously to weak stimulus. The brain and other
parts of the nervous system are well--studied paradigms, and forest fires
with constant ignition of trees and autocatalytic reactions in surfaces, for
instance, also share some of the basics \citep{bak,Meron,Lindner,Izhi,Arenas_rev}.
The fact that signals are not gradually damped by friction in these cases is
a consequence of
cooperation
among
many elements in a nonequilibrium
setting. These systems can be seen as large networks
of nodes that are \textquotedblleft excitable\textquotedblright . This
admits various realizations, but typically means that each element has a
threshold and a refractory time between consecutive responses -- a behaviour
that impedes thermal equilibrium.

Some brain tasks can be simulated with 
mathematical neural networks. As described in Chapter \ref{Chapter_INTRO}, these
consist of neurons -- often modelled as variables 
which are as simple as possible while still able to display the essence of the cooperative
behaviour of interest\footnote{Several studies have already shown that binary neurons can
capture the essence of cooperation in many more complex settings. See, for
instance, \citep{Pantic} in the case of integrate and fire neuron models of
pyramidal cells.} -- connected by edges representing synapses  
\citep{Amari,Hopfield,Amit,Torres_rev}. If the edges are weighted according to some prescription
-- such as the Hebb rule \citep{Hebb} -- which 
saves information from a set of given
patterns of activity (particular configurations of active and inactive neurons), 
these patterns become attractors of the phase-space
dynamics.
Therefore, the system is then able to retrieve 
the stored patterns; this mechanism is known as \textit{associative memory}. Actual neural
systems do much more than just recalling a memory and staying there,
however. That is, one should expect dynamic instabilities or some other
destabilizing mechanism. This expectation is reinforced by recent
experiments suggesting that synapses undergo rapid changes with time which
may both determine brain tasks \citep{Abbot_synaptic,Tsodyks,Sabine,Pantic}
and induce irregular and perhaps chaotic activity \citep{Barrie,Korn}.

One may argue that the observed rapid changes (which have been found to cause 
\textquotedblleft synaptic
depression\textquotedblright\ and/or \textquotedblleft
facilitation\textquotedblright\ on the time scale of milliseconds \citep{Tsodyks,Pantic} -- i.e.,
much faster than the plasticity processes whereby synapses \textit{store}
patterns \citep{Malenka}) may simply correspond to the characteristic behaviour of single
excitable elements. Furthermore, a fully-connected network which describes
cooperation among such excitable elements has recently been shown to
exhibit both attractors and chaotic instabilities \citep{Marro_complex}.
The work described here, first reported by \citet{Johnson_EPL},
extends and generalizes
this study to conclude on the influence of
the excitable network topology on dynamic behaviour.
We show, in particular,
an interesting correlation between certain wiring topology and optimal
functionality.

\section{The Fast-Noise model}

Consider $N$ binary nodes ($s_{i}=\pm 1)$
and the adjacency matrix, $\hat{a}
_{ij}=1,0,$ which indicates the existence or not of an edge between nodes $%
i,j=1,2,...,N.$ Let there be a set of \textit{M} patterns, $\xi _{i}^{\nu }=\pm 1,$ $%
\nu =1,...M$ (which we generate here at random), and assume that they are
\textquotedblleft stored\textquotedblright\ by giving each edge a base
weight $\overline{\omega _{ij}}=N^{-1}\sum_{\nu }\xi _{i}^{\nu }\xi
_{j}^{\nu }$. Actual weights are dynamic, however, namely, $\omega _{ij}=%
\overline{\omega _{ij}}x_{j}$ where $x_{j}$ is a stochastic variable.
Assuming the limit in which this varies in a time scale infinitely smaller
than the one for node dynamics, we can consider a stationary distribution
such as $P(x_{j}|S)=q\delta (x_{j}-\Xi _{j})+(1-q)\delta (x_{j}-1),$ $%
S=\left\{ s_{j}\right\} ,$ for instance. This amounts to assuming that, at
each time step, every connection has a probability \textit{q} of altering
its weight by a factor $\Xi _{j}$ which is a function (to be determined) of
the local \textit{field} at $j,$ defined as the net current arriving to $j$
from other nodes. This choice differs essentially from the one used by \citet{Marro_complex}, where $q$ depends on the global degree of order and $\Xi
_{j}$ is a constant independent of $j.$%

Assume independence of the noise at different edges, and that the transition
rate for the stochastic changes is%
\[
\frac{\bar{c}\left( S\rightarrow S^{i}\right) }{\bar{c}\left(
S^{i}\rightarrow S\right) }=\prod_{j/\hat{a} _{ij}=1}\frac{\int
dx_{j}P(x_{j}|S)\Psi (u_{ij})}{\int dx_{j}P(x_{j}|S^{i})\Psi (-u_{ij})}, 
\]%
where $u_{ij}\equiv s_{i}s_{j}x_{j}\overline{\omega _{ij}}T^{-1},$ $\Psi
(u)=\exp \left( -%
{\frac12}%
u\right) $ to have proper contour conditions, \textit{T} is a
\textquotedblleft temperature\textquotedblright\ or stochasticity parameter,
and $S^{i}$ stands for \textit{S} after the change $s_{i}\rightarrow -s_{i}.$
(This formalism and its interpretation is described in detail by \citet{MarroBook}.)
We define the \textit{effective local fields} $%
h_{i}^{\text{eff}}=h_{i}^{\text{eff}}(S,T,q)$ via $\prod_{j}\varphi
_{ij}^{-}/\varphi _{ij}^{+}=\exp \left( -h_{i}^{\text{eff}}s_{i}/T\right) ,$
where $\varphi _{ij}^{\pm }\equiv q\exp \left( \pm \Xi _{j}v_{ij}\right)
+(1-q)\exp \left( \pm v_{ij}\right) $, with $v_{ij}=\frac{1}{2}\hat{a}
_{ij}u_{ij}.$ Effective weights $\omega _{ij}^{\text{eff}}$ then follow from 
$h_{i}^{\text{eff}}=\sum_{j}\omega _{ij}^{\text{eff}}s_{j}\hat{a} _{ij}$.
To obtain an analytical expression, we linearize around $\overline{\omega
_{ij}}=0$ (a good approximation when $M\ll N$), which yields 
\[
\omega _{ij}^{\text{eff}}=\left[ 1+q\left( \Xi _{j}-1\right) \right] 
\overline{\omega _{ij}}. 
\]%
In order to fix $\Xi _{j}$ here, we first introduce the overlap vector $%
\overrightarrow{m}=(m^{1},...m^{M}),$ with $m^{\nu }\equiv N^{-1}\sum_{i}\xi
_{i}^{\nu }s_{i},$ which measures the correlation between the current
configuration and each of the stored patterns, and the \textit{local} one $%
\overrightarrow{m_{j}}$ of components $m_{j}^{\nu }\equiv \langle k\rangle
^{-1}\sum_{l}\xi _{l}^{\nu }s_{l}\hat{a} _{jl}$, where $\langle k\rangle $
is the mean node connectivity, i.e., the average of $k_{i}=\sum_{j}\hat{a}
_{ij}.$ We then assume, for any $q\neq 0,$ that the relevant factor is $\Xi
_{j}=1+\zeta (h_{j}^{\nu })(\Phi -1)/q,$ with 
$$
\zeta (h_{j}^{\nu })=\frac{\chi ^{\alpha }}{1+M/N} \sum_{\nu
}|h_{j}^{\nu }|^{\alpha },
$$
where $\chi \equiv N/\langle k\rangle $ and $\alpha >0$ is a parameter. This
comes from the fact that the field at node $j$ can be written as a sum of
components from each pattern, namely, $h_{j}=\sum_{\nu }^{M}h_{j}^{\nu }$,
where 
\[
h_{j}^{\nu }=\xi _{j}^{\nu }N^{-1}\sum_{i}\hat{a} _{ij}\xi _{i}^{\nu
}s_{i}=\chi ^{-1}\xi _{j}^{\nu }m_{j}^{\nu }. 
\]%
Our choice for $\Xi _{j},$ which amounts to assuming that the
\textquotedblleft fatigue\textquotedblright\ at a given edge increases with
the field at the preceding node $j$ (and allows to recover the
fully--connected limit described by \citet{Marro_complex} if $\alpha =2$), finally
leads to%
\[
\omega _{ij}^{\text{eff}}=\left[ 1+(\Phi -1)\zeta _{j}(\overrightarrow{m_{j}}%
)\right] \overline{\omega _{ij}}. 
\]%
Varying $\Phi $ one sets the nature of the weights. That is, $0<\Phi <1$
corresponds to resistance (\textit{depression}) due to heavy local work,
while the edge facilitates -- i.e., tends to increase the effect of the signal
under the same situation -- for $\Phi >1.$ (The action of the edge is reversed
for negative $\Phi .)$ We performed Monte Carlo simulations using standard
parallel updating with the effective rates $\bar{c}\left( S\rightarrow
S^{i}\right) $ computed using the latter effective weights.

\section{Edge of Chaos}

It is possible to solve the single pattern case ($M=1$) under a mean-field
assumption, which is a good approximation for large enough connectivity.
That is, we may substitute the matrix $\hat{a} _{ij}$ by its mean value
over network realizations to obtain analytical results that are independent
of the underlying disorder. Imagine that each node hosts $k_{i}$ \textit{%
half--edges} according to a distribution $p(k),$ the total number of
half--edges in the network being $\langle k\rangle N$. Choose a node $i$ at
random and randomly join one of its half--edges to an available free
half--edge. The probability that this half--edge ends at node $j$ is $%
k_{j}/\left( \left\langle k\right\rangle N\right) .$ Once all the nodes have
been linked up, the expected value (as a quenched average over network
realizations) for the number of edges joining nodes \textit{i} and \textit{j}
is\footnote{Assuming one edge at most between any two nodes, $%
\hat{a} _{ij}=0,1,$ the value will be slightly smaller, but it is easy to
prove that this is also a good approximation if 
the network has a {\it structural cut-off}: $k_i<\sqrt{\langle k\rangle N}$, $\forall i$.
} $E(\hat{a}_{ij})=k_{i}k_{j}/\left( \left\langle k\right\rangle
N\right)$. This expression, which can be seen as a definition of the so-called 
{\it configuration model} for complex networks \citep{Newman_rev}, is valid for 
random networks with a given degree sequence (or, in practise, a given degree distribution) 
that have zero degree-degree correlations between neighbours \citep{Johnson_PRL}.
Using the notation $\eta _{i}\equiv \xi
_{i}s_{i}$, we have $m_{j}=\chi \langle \eta _{i}\hat{a} _{ij}\rangle _{i}=%
\frac{\chi }{N}\sum_{i}\eta _{i}\hat{a} _{ij}.$ Because node activity is
not statistically independent of connectivity \citep{Torres_influence}, we
must define a new set of overlap parameters, analogous to $m$ and $m_{j}.$
That is, $\mu _{n}\equiv \langle k_{i}^{n}\eta _{i}\rangle _{i}/\langle
k^{n}\rangle $ and the local versions $\mu _{n}^{j}\equiv \chi \langle
k_{i}^{n}\eta _{i}\hat{a} _{ij}\rangle _{i}/\langle k^{n}\rangle .$ After
using $\hat{a} _{ij}=E(\hat{a} _{ij}),$ one obtains the relation $%
\mu _{n}^{i}=\langle k^{n+1}\rangle k_{i}\mu _{n+1}/(\langle k^{n}\rangle
\langle k\rangle ^{2}).$ Inserting this expression into the definition of $%
\mu _{n}$, and substituting $\langle s_{i}\rangle =\tanh
[T^{-1}h_{i}^{eff}(S)]$ (for large N), standard mean-field analysis
yields%
\[
\mu _{n}(t+1)=\frac{1}{\left\langle k^{n}\right\rangle }\left\langle
k^{n}\tanh M_{T,\Phi }(k,t)\right\rangle _{k}, 
\]%
where the last quantity is defined as%
\[
M_{T,\Phi }=\frac{k}{TN}\left[ \mu _{1}(t)+(\Phi -1)\frac{\langle k^{\alpha
+1}\rangle }{\langle k\rangle ^{\alpha +1}}\left\vert \mu _{1}(t)\right\vert
^{\alpha }\mu _{\alpha +1}(t)\right] . 
\]%
This is a two-dimensional map which is valid for any random topology of
distribution $p(k)$. Note that the macroscopic magnitude of interest is $\mu
_{0}=m\equiv |\overrightarrow{m}|.$

A main consequence of this is the existence of a critical temperature, $%
T_{c},$ under very general conditions.
More specifically, as $T$ is decreased, the overlap $m$
describes a second--order phase transition from a disordered or, say,
\textquotedblleft paramagnetic\textquotedblright\ phase to an ordered
(\textquotedblleft ferromagnetic\textquotedblright ) phase which exhibits
associative memory. The mean-field temperature
at which
this transition occurs is
$$
T_{c}=\frac{\langle k^{2}\rangle}{\left\langle k\right\rangle N}.
$$
On the other hand, the map reduces to 
$$
\mu _{n}\left( t+1\right) =\mbox{sign}%
\left\{ \mu _{n}\left( t\right) \left[ 1+(\Phi -1)
\frac{\langle k^{\alpha+1}\rangle}{\langle k\rangle ^{\alpha +1}}
\right] \right\}
$$
for $T=0.$ This
implies the existence at $\Phi =\Phi _{0},$ where%
$$
\Phi _{0}=1-\frac{\langle k\rangle ^{\alpha +1}}{\langle k^{\alpha +1}\rangle} , 
$$
of a transition as $\Phi $ is decreased from the ferromagnetic phase to a
new phase in which periodic hopping between the attractor and its negative
occurs. This is confirmed by the Monte Carlo simulations for $M>1$; that is,
the hopping is also among different attractors for finite $T.$ The
simulations also indicate that this transition washes out at low enough
finite temperature. Instead, Monte Carlo evolutions show that, for a certain
range of $\Phi $ values, the system activity then exhibits chaotic behaviour.

The transition from ferromagnetic to chaotic states is a main concern
hereafter. Our interest in this regime follows from several recent
observations concerning the relevance of chaotic activity in a network. In
particular, it has been shown that chaos might be responsible for certain
states of attention during brain activity \citep{TorresNEW,Torres_int}, and
that some network properties such as the computational capacity \citep{Bert}
and the dynamic range of sensitivity to stimuli \citep{Assis} may become
optimal at the 
{\it Edge of Chaos}
in a variety of settings.

We next note that the critical values $T_{c}$ and $\Phi _{0}$ only depend
on the moments of the generic distribution $p(k),$ and that the ratio $%
\langle k^{a}\rangle /\langle k\rangle ^{a},$ $a>1$, is a convenient way of
characterizing heterogeneity. We studied in detail two particular types of
connectivity distributions with easily tunable heterogeneity; that is,
networks with $\langle k\rangle N/2$ edges randomly distributed with $%
p\left( k\right)$ such that the heterogeneity depends on a single
parameter. Our first case is the bimodal distribution, $p(k)=\frac{1}{2}%
\delta (k-k_{1})+\frac{1}{2}\delta (k-k_{2})$ with parameter $\Delta
=(k_{2}-k_{1})/2=\langle k\rangle -k_{1}=k_{2}-\langle k\rangle $. Our
second case is the \textit{scale--free} distribution, $p(k)\sim k^{-\gamma
}, $ which does not have any characteristic size but $k$ is confined to the
limits, $k_{0}$ and $k_{m}\leq \min (k_{0}N^{\frac{1}{\gamma -1}},N-1)$ for
finite $N$.
Notice that the network in this case gets more
homogeneous as $\gamma $ is increased\footnote{The distribution is truncated and therefore not strictly
scale free for $\gamma <2$. However, nature shows examples for which $\gamma 
$ is slightly larger than 1, so we consider the whole range here.}, and that this kind
of distribution seems to be most relevant in nature \citep{Newman_rev,Boccaletti}. 
In particular, it seems important
to mention that the \textit{functional} topology of the human brain, as
defined by correlated activity between small clusters of neurons, has been
shown to correspond to this case with exponent $\gamma \simeq 2$ \citep{Eguiluz}.
(It has not yet been possible to ascertain the brain's \textit{structural} topology 
experimentally, but there is some evidence that
function reflects structure at least to some extent \citep{Zhou97}.
Furthermore, it has been suggested, based on indirect methods, that the
structural connectivity of cat and macaque brains, at the level of brain
areas, may indeed be scale free \citep{Kaiser_EJN} -- and in any case displays
significantly higher heterogeneity than that of, say, Erd\H{o}s--R\'{e}nyi
random graphs.)

We obtained the critical value of the fatigue, $\Phi _{c}\left( T\right) ,$
from Monte Carlo simulations at finite temperature $T.$ These indicate that
chaos never occurs for $T\gtrsim 0.35T_{c}.$ On the other hand, a detailed
comparison of the value $\Phi _{c}$ with $\Phi _{0}$ -- as obtained
analytically for $T=0$ -- indicates that $\Phi _{c}\simeq \Phi _{0}.$ 
\begin{figure}[t!]
\begin{center}
\includegraphics
[
width=10cm
]
{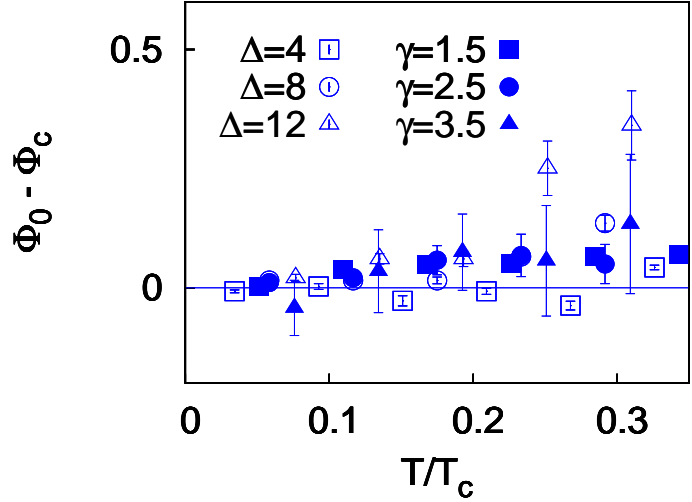} 
\end{center}
\caption{
The temperature dependence of the difference between the values for the fatigue at which the
ferromagnetic--periodic transition occurs, as obtained analytically for $T=0$
($\Phi _{0}$) and from MC simulations at finite $T$ ($\Phi _{c}$). The
critical temperature is calculated as $T_{c}=\langle k^{2}\rangle \left(
\left\langle k\right\rangle N\right) ^{-1}$ for each topology. Data are for
bimodal distributions with varying $\Delta $ and for scale--free topologies
with varying $\protect\gamma ,$ as indicated. Here, $\langle k\rangle =20$, $%
N=1600$ and $\protect\alpha =2.$ Standard deviations, represented as bars in
this graph, were shown to drop with $N^{-1/2}$ (not depicted).
}
\label{fig_Tc}
\end{figure}
\\
\linebreak
Figure \ref{fig_Tc} illustrates the \textquotedblleft
error\textquotedblright\ $\Phi _{0}-\Phi _{c}\left( T\right) $ for different
topologies. This shows that the approximation $\Phi _{c}\simeq \Phi _{0}$ is
quite good at low $T$ for any of the cases examined. Therefore, assuming the
critical values for the main parameters, $T_{c}$ and $\Phi _{0},$ as given
by our map, we conclude that the more heterogeneous the distribution of
connectivities of a network is, the lower the amount of fatigue, and the
higher the critical temperature, needed to destabilize the dynamics. As an
example of this interesting behaviour, consider a network with $\langle
k\rangle =\ln (N)$, and dynamics according to $\alpha =2$. If the
distribution were regular, the critical values would be $T_{c}=\ln (N)/N$
(which goes to zero in the thermodynamic limit) and $\Phi _{0}=0$. However,
a scale--free topology with the same number of edges and $\gamma =2$ would
yield $T_{c}=1$ and $\Phi _{0}=1-2(\ln N)^{3}/N^{2}$ (which goes to 1 as $%
N\rightarrow \infty ).$

\begin{figure}[t!]
\begin{center}
\includegraphics[
width=10cm
]{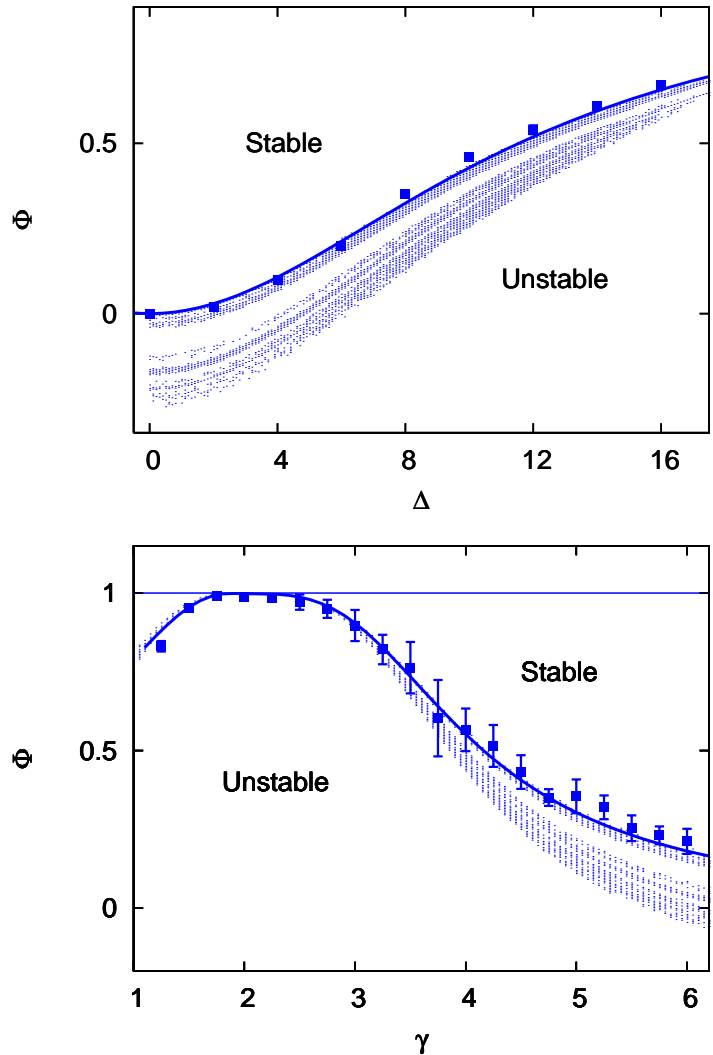} 
\end{center}
\caption{
The critical fatigue values $\Phi _{0}$ (solid lines) and $\Phi _{c}$ from MC
averages over 10 networks (symbols) with $T=2/N,$ $\langle k\rangle =20,$ $%
N=1600,$ $\protect\alpha =2$. The dots below the lines correspond to changes
of sign of the Lyapunov exponent as given by the iterated map, which
qualitatively agree with the other results. This is for bimodal and
scale--free topologies, as indicated.
}
\label{fig_phase}
\end{figure}
Figure \ref{fig_phase} illustrates, for two topologies, the phase diagram of the
ferromagnetic--chaotic transition. Most remarkable is the plateau observed
in the \textit{Edge-of-Chaos} or transition\ curve for scale--free
topologies around $\gamma \simeq 2$, for which very little fatigue, namely, $%
\Phi \lesssim 1$ which corresponds to slight \textit{depression}, is
required to achieve chaos. The limit $\gamma \rightarrow \infty $
corresponds to $\langle k\rangle $--regular graphs (equivalent to $\Delta =0$%
). If $\gamma $ is reduced, $k_{m}$ increases and $k_{0}$ decreases. The
network is truncated when $k_{m}=N$. It follows that a value of $\gamma $
exits at which $k_{0}$ cannot be smaller, so that $k_{m}$ must drop to
preserve $\langle k\rangle $. This explains the fall in $\Phi _{c}$ as $%
\gamma \rightarrow 1$.

Assuming that the \textquotedblleft ferromagnetic phase\textquotedblright\
here corresponds to a \textit{synchronous state}, our results are in
qualitative agreement with the ones obtained recently for coupled
oscillators \citep{Nishikawa, Zhou96}. As a matter of fact, the range of
coupling strengths which allow for stability of synchronous states in these
systems has been shown to depend on the spectral gap of the Laplacian matrix 
\citep{Barahona}, implying that the more heterogeneous a topology is, the
more easily activity can become unstable. It should be emphasized, however,
that the dynamics we are considering here does not come within the scope of
the formalism used to derive these results, since activity at node $i$
depends on the local field at node $j$.

\section{Network performance}
\begin{figure}[t!]
\begin{center}
\includegraphics[
width=10cm]
{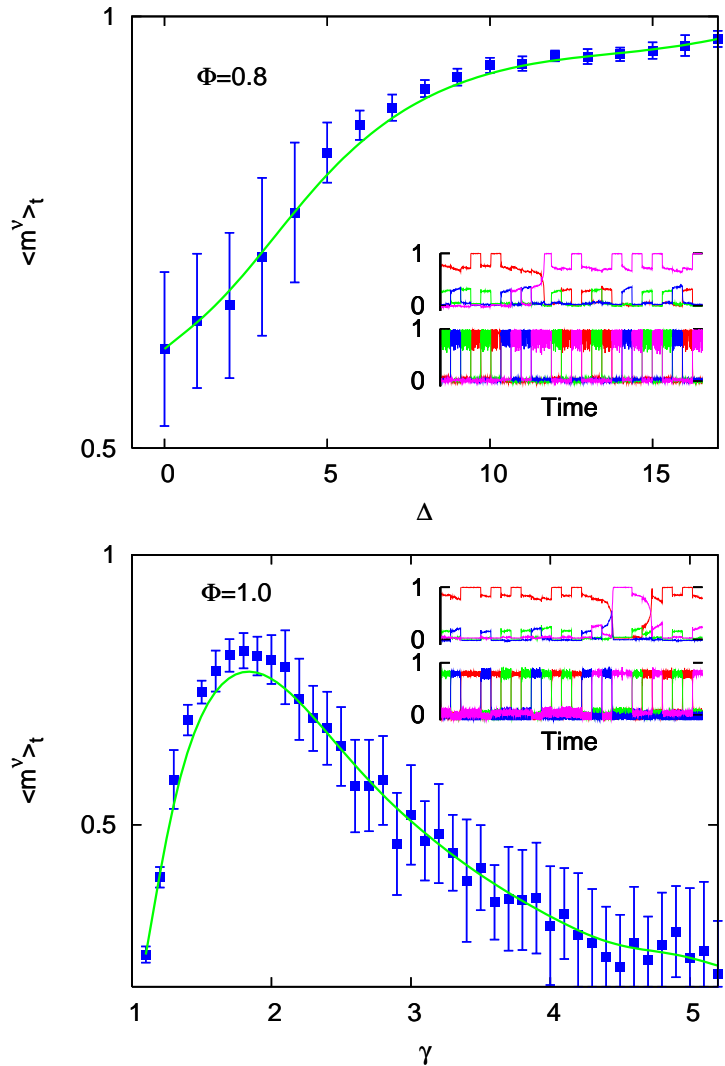} 
\end{center}
\caption{
Network
\textquotedblleft performance\textquotedblright\ (see the main text) against 
$\Delta$ for bimodal topologies (above) and against
$\protect\gamma $ for scale--free topologies (below). $\Phi=0.8$ for the first case and $\Phi =1$ in the second. Averages
over $20$ network realizations with stimulation every $50$ MC steps for $%
2000 $ MC steps, $\protect\delta =5$ and $M=4;$ other parameters as in Fig. 
\protect\ref{fig_phase}. Inset shows sections of typical time series of $m^{%
\protect\nu }$ for $\Delta=10$ (above) and $\protect\gamma =4$ (below); the corresponding stimulus for
pattern $\protect\nu $ is shown underneath.
}
\label{fig_stim}
\end{figure}
As a further illustration of our findings, we monitored the performance as a
function of topology during a simulation of pattern recognition. That is, we
\textquotedblleft showed\textquotedblright\ the system a pattern, say $\nu $
chosen at random from the set of $M$ previously stored, every certain number
of time steps. This was performed in practice by changing the field at each
node for one time step, namely, $h_{i}\rightarrow h_{i}+\delta \xi ^{\nu }$,
where $\delta $ measures the intensity of the input signal. Ideally, the
network should remain in this configuration until it is newly stimulated.
The performance may thus be estimated from a temporal average of the overlap
between the current state and the input pattern, $\langle m^{\nu }\rangle
_{time}$. This is observed to simply increase monotonically with $\Delta $
for the bimodal case. The scale--free case, however, as illustrated in
Fig. \ref{fig_stim}, shows how the task is better performed the closer to the Edge of
Chaos the network is. This is because the system is then easily destabilized
by the stimulus while being able to retrieve a pattern with accuracy. Figure %
\ref{fig_stim} also shows that the best performance for the scale--free
topology when $\Phi =1,$ i.e., in the absence of any fatigue, definitely
occurs around $\gamma =2.$%


\section{Discussion}

The model network we have studied is one of the simplest relevant situations one may
conceive. In particular, as emphasized above, we are
greatly simplyfieng
the elements at the nodes (neurons)
as binary variables. However, our assumption of dynamic connections
which depend on the local fields in such a simple scenario happens to show
that a close relation may exist between topological heterogeneity and
function, thus suggesting this may indeed be a relevant property for a
realistic network efficiently to perform certain high level tasks. In a
similar way to networks shown previously to be useful for pattern
recognition and family identification \citep{class}, our system retrieves
memory patterns with accuracy in spite of noise, and yet it is easily 
destabilized so as to change state in response to an input signal -- without
requiring excessive \textit{fatigue} for the purpose. There is a
relation between the amount $\Phi $ of fatigue and the value of $\gamma $
for which performance is maximized. One may argue that the plateau of
\textquotedblleft good\textquotedblright\ behaviour shown around $\gamma
\simeq 2$ for scale--free networks with $\Phi \lesssim 1$ (Fig. \ref%
{fig_phase}) is a possible justification for the supposed tendency of
certain systems in nature to evolve towards this topology. It may also 
prove useful for
implementing some artificial networks.%
%




%% file: prl/prl.tex



\chapter{Correlated networks and natural disassortativity}
\label{Chapter_PRL}


An intriguing feature of complex networks is the ubiquity of strong negative degree-degree correlations between neighbouring nodes -- the only exceptions being social systems, which tend to be {\it assortative} instead of {\it disassortative}. With the double purpose of addressing this mystery and uncovering the effects of correlations on network behaviour, we put forward a method which allows for the model-independent study of ensembles of correlated networks. We go on to show, by means of an information theory approach, that the expected value of correlations for a network at equilibrium (i.e., in the absence of specific correlating mechanisms) is not, as had been supposed, uncorrelated, but rahter disassortative. It turns out that the correlations of some networks are in excellent agreement with our predictions, while others, with known correlating or anticorrelating mechanisms, indeed appear to have been driven from their equilibrium points as expected. Therefore, our approach not only provides a parsimonious topological answer to a long-standing question, but also a neutral model against which to contrast experimental data to determine whether mechanisms must be sought to account for observed correlations. We go on to use our method, in Chapter \ref{Chapter_PRE}, to study the influence of assortativity on neural-network dynamics.





\section{Assortativity of networks}

Complex networks, whether natural or artificial, have non-trivial topologies
which are usually studied by analysing a variety of measures, such as the degree
distribution, clustering, average paths, modularity, etc.
\citep{Albert_rev,Dorogovtsev_book,Pastor-Satorras_book,Newman_rev,Boccaletti} The mechanisms which lead to a particular
structure and their relation to functional constraints are often not clear
and constitute the subject of much debate \citep{Newman_rev, Boccaletti}. When
nodes are endowed with some additional ``property,'' a feature known as
\textit{mixing} or \textit{assortativity} can arise, whereby edges are not
placed between nodes completely at random, but depending in some way on the property in
question. If similar (dissimilar) nodes tend to wire together, the network is
said to be \textit{assortative} (\textit{disassortative})
\citep{Newman_mixing_PRL, Newman_mixing_PRE}.

An interesting situation is when the property taken into account is the degree
of each node -- i.e., the number of neighbouring nodes connected to it. It
turns out that a high proportion of empirical networks -- whether biological,
technological, information-related or linguistic -- are disassortatively
arranged (high-degree nodes, or hubs, are preferentially linked to low-degree
neighbours, and viceversa) while social networks are usually assortative. Such
degree-degree correlations have
important consequences for
network characteristics such as connectedness and robustness
\citep{Newman_mixing_PRL, Newman_mixing_PRE}.

However, while assortativity in social networks can be explained
taking into account homophily \citep{Newman_mixing_PRL, Newman_mixing_PRE} or modularity
\citep{Newman_social}, the widespread prevalence and extent of
disassortative mixing in most other networks remains somewhat
mysterious. Maslov \textit{et al.} found that the restriction of having at most one
edge per pair of nodes induces some disassortative correlations
in heterogeneous networks \citep{Maslov},
and Park and Newman showed how this analogue of the Pauli exclusion principle leads to the edges following Fermi statistics \citep{Park_correlations} (see also \citep{Roma}).
However, this
restriction
is not sufficient to fully account for
empirical data. In general,
when one attempts to consider computationally
all the networks with the same distribution as a
given empirical one,
the mean assortativity is not necessarily zero
\citep{Zhao}. But since some ``randomization'' mechanisms induce positive correlations and others negative ones \citep{Farkas, Johnson_JSTAT}, it is not clear how the phase space can be properly sampled numerically.


In this chapter we develop a method for the study of correlated networks which is model-independent, and describe the main result of Ref. \citep{Johnson_PRL} -- namely, 
that there is a general reason, consistent
with empirical data, for the ``natural'' mixing of most networks to be
disassortative. Using an information-theory approach we find that the
configuration which can be expected to come about in the absence of specific additional
constraints turns out not to be, in general, uncorrelated. In fact,
for highly heterogeneous degree distributions such as those of the
ubiquitous scale-free networks, we show that the expected value of the
mixing is usually disassortative: there are simply more possible
disassortative configurations than assortative ones. This result
provides a simple topological answer to a long-standing question. Let
us caution that this does {\it not} imply that all scale-free networks
are disassortative, but only that,
in the absence of further information on the mechanisms behind their evolution, this is the neutral expectation.

\section{The entropy of network ensembles}

The topology of a network is entirely described by its adjacency
matrix $\hat{a}$; the element $\hat{a}_{ij}$ represents the number of
edges linking node $i$ to node $j$ (for undirected networks, $\hat{a}$
is symmetric). Among all the possible microscopically distinguishable
configurations a set of $L$ edges can adopt when distributed among $N$
nodes, it is often convenient to consider the set of configurations
which have certain features in common -- typically some macroscopic
magnitude, like the degree distribution. Such a set of configurations
defines an \textit{ensemble}. In a seminal series of papers Bianconi
has determined the partition functions of various ensembles of random
networks and derived their statistical-mechanics entropy
\citep{Bianconi_entropy, Bianconi_ensembles, Anand}. This allows the author to estimate the
probability that a random network with certain constraints has of
belonging to a particular ensemble, and thus assess the relative
importance of different magnitudes and help discern the mechanisms
responsible for a given real-world network. For instance,
she shows
that scale-free networks arise naturally when the total entropy
is restricted to a small finite value. Here we take a similar
approach: we
obtain the Shannon information entropy encoded in the
distribution of edges. As we shall see, both methods yield the same results
\citep{Jaynes, Anand}, but for our purposes the Shannon entropy is more tractable.

The Shannon entropy associated with a probability distribution $p_m$ is
$$
s=-\sum_{m}p_{m}\ln(p_{m}),
$$
where the sum extends over all possible outcomes
$m$. For a given pair of nodes $(i,j)$, $p_{m}$ can be considered to represent
the probability of there being $m$ edges between $i$ and $j$. For simplicity,
we shall focus here on networks such that
$\hat{a}_{ij}$ can only take values $0$ or $1$,
 although the method is applicable to any number of edges allowed. 
In this case, we have only two terms: $p_{1}=\hat{\epsilon}_{ij}$ and
$p_{0}=1-\hat{\epsilon}_{ij}$, where $\hat{\epsilon}_{ij}\equiv
E(\hat{a}_{ij})$ is the expected value of the element $\hat{a}_{ij}$ given
that the network belongs to the ensemble of interest. The entropy associated
with pair $(i,j)$ is then 
$$
s_{ij}=-\left[\hat{\epsilon}_{ij} \ln
 (\hat{\epsilon}_{ij})+(1-\hat{\epsilon}_{ij})\ln(1-\hat{\epsilon}_{ij})\right],
$$
while the total entropy of the network is $S=\sum_{ij}^{N}s_{ij}$: 
\begin{equation}
 S=-\sum_{ij}^{N}\left[\hat{\epsilon}_{ij} \ln (\hat{\epsilon}_{ij})
  +(1-\hat{\epsilon}_{ij})\ln(1-\hat{\epsilon}_{ij})\right].
\label{eq_s_exact}
\end{equation}
Since we have not imposed symmetry of the adjacency matrix,
this expression is in general valid for directed networks. For
undirected networks, however, the sum is only over $i\leq j$, with the
consequent reduction in entropy.

For the sake of illustration, we shall estimate the entropy of the Internet at
the autonomous system (AS) level and compare it with the values obtained in
\citep{Bianconi_entropy, Bianconi_ensembles, Anand} assuming the network belongs to two different
ensembles: the fully random graph, or Erd\H{o}s-R\'{e}nyi (ER) ensemble, and the
\textit{configuration} ensemble with a scale-free degree distribution ($p(k)\sim k^{-\gamma}$)
\citep{Newman_rev} and structural cutoff, $k_{i}<\sqrt{\langle k\rangle N}$,
$\forall i$ \citep{Bianconi_entropy, Bianconi_ensembles, Anand} ($\langle k\rangle$ is the mean degree). In this example, we assume the network to
be sparse enough to expand the term $\ln(1-\hat{\epsilon}_{ij})$ in
Eq. (\ref{eq_s_exact}) and keep only linear terms. This reduces
Eq. (\ref{eq_s_exact}) to
$$
S_{sparse}\simeq-\sum_{ij}^{N}\hat{\epsilon}_{ij}
[\ln(\hat{\epsilon}_{ij})-1]+O(\hat{\epsilon}_{ij}^{2}).
$$
In the ER ensemble, each of $N$ nodes has an equal probability of receiving each of $\frac{1}{2}\langle k\rangle N$ undirected edges.
So, writing $\hat{\epsilon}_{ij}^{ER}=\langle k\rangle/N$, we have
$$
S_{ER}=-\frac{1}{2}\langle k\rangle N\left[\ln\left(\langle k\rangle/N\right)-1\right].
$$
The configuration ensemble, which imposes a given
degree sequence $(k_{1},... k_{N})$, is defined via the expected value of the
adjacency matrix \citep{Newman_rev,Johnson_EPL}:
$$
\hat{\epsilon}_{ij}^{c}=k_{i}k_{j}/(\langle k\rangle
N).
$$ This value 
leads to
$$
S_{c}=\langle k\rangle N[\ln(\langle k\rangle
N)+1]- 2N\langle k \ln k\rangle,
$$
where $\langle \cdot \rangle \equiv N^{-1}\sum_{i}(\cdot)$ stands for
an average over nodes.
\begin{figure}
[t!]
\begin{center}
\includegraphics
[
width=10cm,
height=8cm
]
{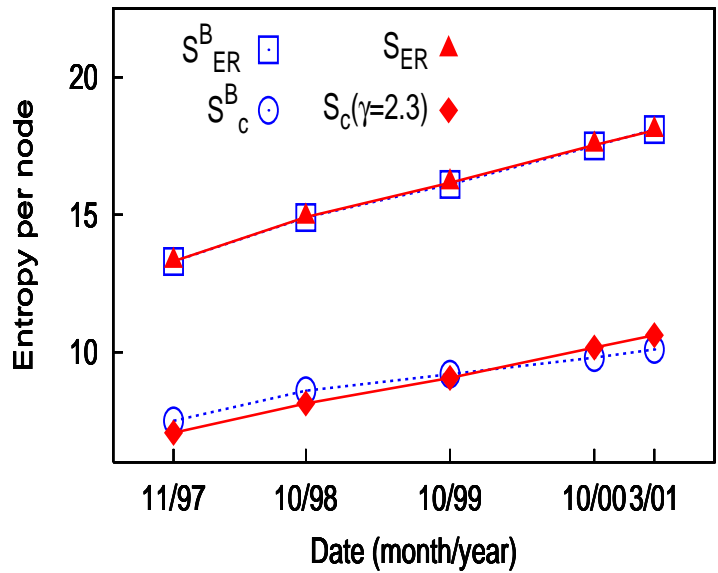}
\caption{
Evolution of the Internet at the AS
  level. Empty (blue) squares and circles:
entropy per node
  of randomized networks in the fully random and in the configuration ensembles,
  as obtained by Bianconi (hence the ``B'' superscription) \citep{Bianconi_entropy, Bianconi_ensembles, Anand}. Filled
  (red) triangles and diamonds:
Shannon entropy
for an ER network and a scale-free one with $\gamma=2.3$, respectively.
}%
\label{fig_int}%
\end{center}
\end{figure}

Fig. \ref{fig_int} displays the entropy per node obtained in
\citep{Bianconi_entropy, Bianconi_ensembles, Anand} for the first two levels of approximation
(ensembles) to the Internet at the AS level, first
taking into account only the numbers of nodes $N$ and edges $L=\frac{1}{2}\langle
k\rangle N$, and then also the degree sequence. Alongside these, we
plot the 
Shannon entropy
both for an ER random network,
(which coincides exactly with Bianconi's
expression), and for a scale-free network with $\gamma=2.3$
(the slight disparity arising from this exponent's changing a little with time).

\section{Entropic origin of disassortativity}

We shall now go on to analyse the effect of degree-degree correlations on the
entropy.
In the configuration ensemble, the expected value
of the mean degree of the neighbours of a given node is 
$$
k_{nn,i}=k_{i}^{-1}\sum_{j}\hat{\epsilon}_{ij}^{c}k_{j}=\frac{\langle
k^{2}\rangle}{\langle k\rangle},
$$
which is independent of $k_{i}$. However, as
mentioned above, real networks often display degree-degree correlations, with the result
that $k_{nn,i}=k_{nn}(k_{i})$. If $k_{nn}(k)$ increases (decreases)
with $k$, the network is assortative (disassortative). A measure of
this phenomenon is Pearson's coefficient applied to the
edges \citep{Newman_rev, Newman_mixing_PRL, Newman_mixing_PRE, Boccaletti}:
$$ r= \frac{[
k_{l}k'_{l}]-[ k_{l}]^{2}}{[ k_{l}^{2}]-[ k_{l}]^{2}},
$$
where
$k_{l}$ and $k'_{l}$ are the degrees of each of the two nodes
belonging to edge $l$, and $[\cdot]\equiv(\langle k\rangle
N)^{-1}\sum_{l}(\cdot)$ is an average over edges. Writing
$\sum_{l}(\cdot)=\sum_{ij}\hat{a}_{ij}(\cdot)$, $r$ can be expressed
as
\begin{equation}
  r=\frac{\langle k\rangle \langle k^{2} k_{nn}(k)\rangle - 
    \langle k^{2}\rangle^{2} }{\langle k\rangle \langle k^{3}\rangle 
    - \langle k^{2}\rangle^{2}}.
  \label{eq_r_gen}
\end{equation}
\begin{figure}
[t!]
\begin{center}
\hspace{-0.8cm}\includegraphics
[width=13cm,
height=7cm
]
{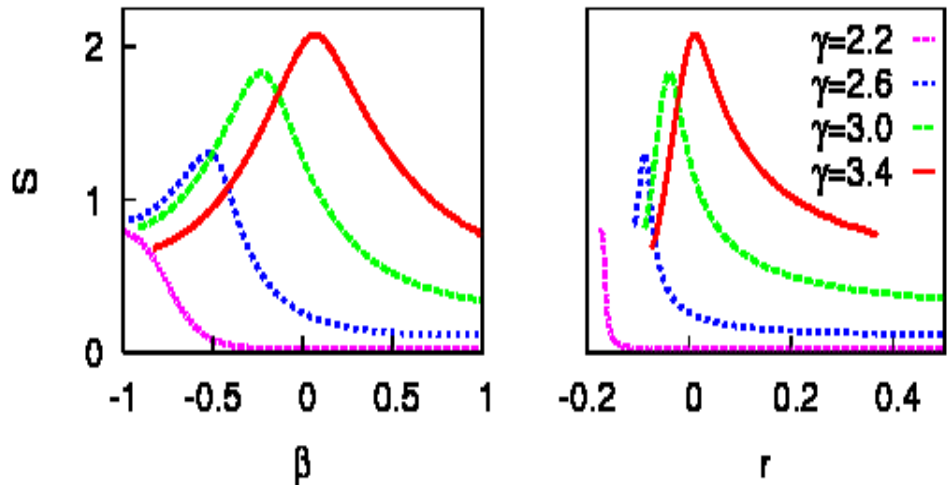}
\caption{
Shannon entropy of correlated scale-free networks
against parameter $\beta$ (left panel) and
  against Pearson's coefficient $r$ (right panel), for various
  values of $\gamma$ (increasing from bottom to top). $\langle
  k\rangle=10$, $N=10^{4}$.  }%
\label{fig_ent}%
\end{center}
\end{figure}
The ensemble of all networks with a given degree sequence $(k_{1},...k_{N})$
contains a subset for all members of which $k_{nn}(k)$ is constant (the configuration ensemble), but also subsets displaying other functions $k_{nn}(k)$. We can identify each one of these subsets (regions of phase space) with an
expected adjacency matrix $\hat{\epsilon}$ which simultaneously satisfies the following conditions: 
\begin{eqnarray*}
&\mbox{{\bf i)}   }&\sum_{j}k_{j}\hat{\epsilon}_{ij}=k_{i}k_{nn}(k_{i})\mbox{, }\forall i\mbox{, and}\\
\\
&\mbox{{\bf ii)}   }&\sum_{j}\hat{\epsilon}_{ij}=k_{i}\mbox{, }\forall i\mbox{ (for consistency).}
\end{eqnarray*}
An ansatz which fulfils these requirements is any matrix of the form
\begin{equation}
  \hat{\epsilon}_{ij}=\frac{k_{i}k_{j}}{\langle k\rangle N}
  +\int d\nu \frac{f(\nu)}{N}\left[\frac{(k_{i}k_{j})^{\nu}}
    {\langle k^{\nu}\rangle}-k_{i}^{\nu}-k_{j}^{\nu}+\langle k^{\nu}\rangle  \right],
\label{eq_epsi_gen}
\end{equation}
where 
$\nu\in\mathbb{R}$ and the function $f(\nu)$
is in general arbitrary,
although depending on the degree sequence
it shall here be restricted to values
which maintain $\hat{\epsilon}_{ij}\in [0,1]$, $\forall i,j$. This ansatz yields
\begin{eqnarray}
  k_{nn}(k)=\frac{\langle k^{2}\rangle}{\langle k\rangle}
  +\int d\nu f(\nu)\sigma_{\nu+1}\left[\frac{k^{\nu-1}}
    {\langle k^{\nu}\rangle}-\frac{1}{k} \right]
\label{eq_knn_gen}
\end{eqnarray}
(the first term being the result for the
configuration ensemble), where $\sigma_{b+1}\equiv \langle k^{b+1}\rangle -\langle k\rangle
\langle k^{b}\rangle$.
In practice, one could adjust Eq. (\ref{eq_knn_gen}) to fit any
given function $k_{nn}(k)$ and then wire up a network with the desired correlations: it suffices to throw random numbers
according to Eq. (\ref{eq_epsi_gen}) with
$f(\nu)$ as
obtained from
the fit to Eq. (\ref{eq_knn_gen})\footnote{Although, as with the configuration ensemble,
  it is not always possible to wire a network according to a
  given $\hat{\epsilon}$.}. To prove the uniqueness of a matrix $\hat{\epsilon}$ obtained in this way (i.e., that it is the only one compatible with a given $k_{nn}(k)$) assume that there exists another valid matrix $\hat{\epsilon}'\neq\hat{\epsilon}$. Writting $\hat{\epsilon}_{ij}'-\hat{\epsilon}_{ij}\equiv h(k_{i},k_{j})=h_{ij}$, then ${\bf i)}$ implies that $\sum_{j}k_{j}h_{ij}=0$, $\forall i$, while ${\bf ii)}$ means that $\sum_{j}h_{ij}=0$, $\forall i$. It follows that $h_{ij}=0$, $\forall i,j$.
\\
\linebreak

In many empirical networks, $k_{nn}(k)$ has the form $k_{nn}(k)=A+B
k^{\beta}$, with $A,B>0$ \citep{Boccaletti, Pastor-Satorras} -- the
mixing being assortative (disassortative) if $\beta$ is positive
(negative). Such a case is fitted by Eq. (\ref{eq_knn_gen}) if
$$
f(\nu)=C\left[\delta(\nu-\beta-1)\frac{\sigma_{2}}{\sigma_{\beta+2}}-\delta(\nu-1)\right],
$$
with $C$ a positive constant, since this choice yields
\begin{equation}
  k_{nn}(k)=\frac{\langle k^{2}\rangle}{\langle k\rangle}
  +C\sigma_{2}\left[\frac{k^{\beta}}{\langle
      k^{\beta+1}\rangle}-\frac{1}{\langle k\rangle} \right].
\label{eq_knn_simple}
\end{equation}
After plugging Eq. (\ref{eq_knn_simple}) into Eq. (\ref{eq_r_gen}), one obtains:
\begin{equation}
  r=\frac{C\sigma_{2}}{\langle k^{\beta+1}\rangle}
  \left(\frac{\langle k\rangle \langle k^{\beta+2} \rangle - 
      \langle k^{2}\rangle\langle k^{\beta+1}\rangle }
{\langle k\rangle \langle k^{3}\rangle - \langle k^{2}\rangle^{2}}\right).
\label{eq_r_simple}
\end{equation}
Inserting Eq. (\ref{eq_epsi_gen}) in Eq. (\ref{eq_s_exact}), we can calculate the
entropy of 
correlated
networks
as a function of $\beta$ and $C$ -- or,
by using Eq. (\ref{eq_r_simple}),
as a function of $r$. Particularizing for scale-free networks, then given
$\langle k\rangle$, $N$ and $\gamma$, there is always a certain combination of
parameters $\beta$ and $C$ which maximizes the entropy; we shall call these
$\beta^{*}$ and $C^{*}$.  For $\gamma\lesssim 5/2$ this point corresponds
to $C^{*}=1$. For higher $\gamma$, the entropy can be slightly higher for
larger $C$.
However, for these
values of $\gamma$, the assortativity $r$ of the point of maximum entropy
obtained with $C=1$ differs very little from the one corresponding to
$\beta^{*}$ and $C^{*}$ (data not shown). Therefore, for the sake of
clarity but with very little loss of accuracy, in the following we shall
generically set $C=1$ and vary only $\beta$ in our search for the level
of assortativity, $r^{*}$, that maximizes the entropy given $\langle
k\rangle$, $N$ and $\gamma$.  Note that $C=1$ corresponds to removing the
linear term, proportional to $k_i k_j$, in Eq. (\ref{eq_epsi_gen}), and leaving
the leading non-linearity, $(k_i k_j)^{\beta+1}$, as the dominant one.
\begin{figure}
[t!]
\begin{center}
\includegraphics
[width=10cm]
{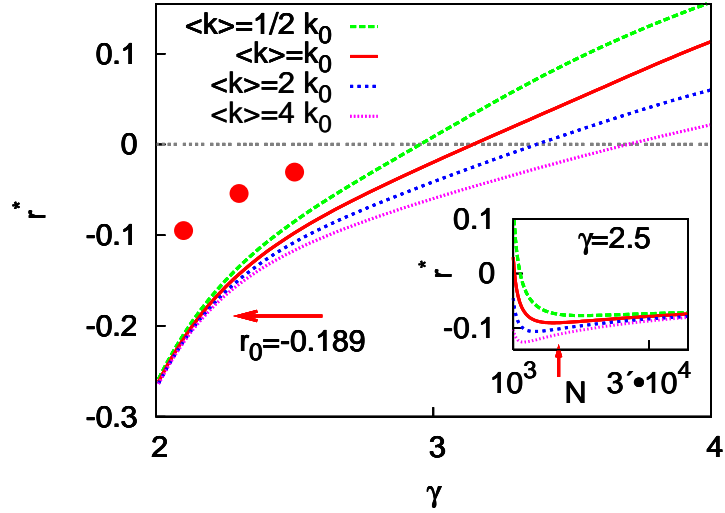}
\caption{
Lines from top to bottom: $r$ at which the entropy is maximized, $r^{*}$, against $\gamma$ for random scale-free networks with mean degrees $\langle k\rangle=\frac{1}{2}$, $1$, $2$ and $4$ times $k_{0}=5.981$, and $N=N_{0}=10697$ nodes ($k_{0}$ and $N_{0}$ correspond to the values for the Internet at the AS level in $2001$ \citep{Park_correlations}, which had $r=r_{0}=-0.189$). Symbols are the values obtained in \citep{Park_correlations} as those expected solely due to the one-edge-per-pair restriction (with $k_{0}$, $N_{0}$ and $\gamma=2.1$, $2.3$ and $2.5$). Inset: $r^{*}$ against $N$ for networks with fixed $\langle k\rangle/N$ (same values as the main panel) and $\gamma=2.5$; the arrow indicates $N=N_{0}$.
}%
\label{fig_max}%
\end{center}
\end{figure}

Fig. \ref{fig_ent} displays the entropy curves for various scale-free
networks, both as functions of $\beta$ and of $r$: {\it depending on
  the value of $\gamma$, the point of maximum entropy can be either
  assortative or disassortative}. This can be seen more clearly in
Fig. \ref{fig_max}, where $r^{*}$ is plotted against $\gamma$ for
scale-free networks with various mean degrees $\langle k\rangle$.
The values obtained by Park and Newman \citep{Park_correlations} as those resulting from the one-edge-per-pair restriction are also shown for comparison: notice that whereas this effect alone cannot account for the Internet's correlations for any $\gamma$, entropy considerations would suffice if $\gamma\simeq2.1$.
As shown in the inset, the results are robust in the large system-size limit.

\begin{figure}
[t!]
\begin{center}
\includegraphics
[width=10cm]
{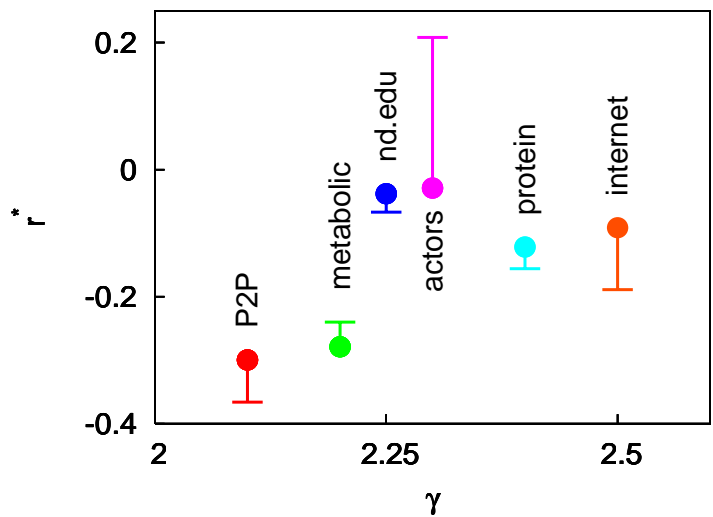}
\caption{
Level of assortativity that maximizes the
  entropy, $r^{*}$, for various real-world, scale-free networks, as
  predicted theoretically by Eq. (\ref{eq_s_exact}) (circles)
  and as directly measured (horizontal lines), against exponent $\gamma$.
}%
\label{fig_all}%
\end{center}
\end{figure}

Since most networks observed in the real world are highly
heterogeneous, with exponents in the range $\gamma\in (2,3)$, it is to
be expected that these should display a certain disassortativity --
the more so the lower $\gamma$ and the higher $\langle k\rangle$. In
Fig. \ref{fig_all} we test this prediction on a sample of empirical,
scale-free networks quoted in Newman's review \citep{Newman_rev} (p. 182). For each case, we found the value of $r$ that maximizes $S$
according to Eq. (\ref{eq_s_exact}), after inserting
Eq. (\ref{eq_epsi_gen}) with the quoted values of $\langle k\rangle$, $N$
and $\gamma$.
In this way, we obtained the expected
assortativity for six networks, representing: a peer-to-peer (P2P)
network, metabolic reactions, the nd.edu domain, actor collaborations,
protein interactions, and the Internet (see \citep{Newman_rev} and
references therein).  For the metabolic, Web domain and protein
networks, {\it the values predicted are in excellent agreement with
  the measured ones}; therefore, no specific anticorrelating mechanisms need to
be invoked to account for their disassortativity. In the other three
cases, however, the predictions are not accurate, so there must be
additional
correlating mechanisms at work. Indeed, it is known that small
routers tend to connect to large ones \citep{Pastor-Satorras}, so one
would expect the Internet to be more disassortative than predicted, as
is the case\footnote{However, as Fig. \ref{fig_max} shows, if the Internet exponent were the $\gamma=2.2\pm0.1$ reported in \citep{Pastor-Satorras} rather than $\gamma=2.5$, entropy would account more fully for these correlations.} -- an effect that is less pronounced but still detectable
in the more egalitarian P2P network. Finally, as is typical of social
networks, the actor graph is significantly more assortative than
predicted, probably due to the homophily mechanism whereby highly
connected, big-name actors tend to work together \citep{Newman_mixing_PRL, Newman_mixing_PRE}.

\section{To sum up...}

We have shown how
the ensemble of networks with a given degree sequence can be partitioned
into regions of equally correlated networks and found, using an information-theory approach, that
the largest (maximum entropy) region, for the case of scale-free networks, usually displays a certain disassortativity. Therefore, in the absence of knowledge regarding the specific evolutionary forces at work, this should be considered the most likely state.
Given the accuracy with which our approach can predict
the degree of assortativity of certain empirical networks {\it with no
  a priori information thereon}, we suggest this as a neutral model to
decide whether or not particular experimental data require specific
mechanisms to account for observed degree-degree
correlations.



%% file: pre/pre.tex

\chapter{Enhancing robustness to noise via assortativity} 
\label{Chapter_PRE}


As we saw in Chapter \ref{Chapter_EPL}, the performance of attractor neural networks 
depends
crucially on the heterogeneity of the underlying topology's degree distribution. We take this analysis a step further by examining the effect of degree-degree correlations -- or assortativity -- on neural-network behaviour.
In Chapter \ref{Chapter_PRL} we described a method
for studying correlated networks and dynamics thereon, both analytically and computationally, which is independent of how the topology may have evolved.
We now make use of this to
show how the robustness to noise is greatly enhanced in assortative (positively correlated) neural networks, especially if it is the hub neurons that store the information.


\section{Background}
\label{sec_introduction}

For a dozen years or so now, the study of complex systems has
been heavily influenced by
results from network science -- which one might regard as the fusion of graph theory with statistical physics \citep{Newman_rev, Boccaletti}. Phenomena as diverse as epidemics \citep{Watts}, cellular function \citep{Ojalvo}, power-grid failures \citep{Havlin_failures} or internet routing \citep{Boguna_hyperbolic}, among many others \citep{Arenas_rev}, depend crucially on the structure of the underlying network of interactions.
One of the earliest systems to have been described as a network was the brain, which is made up of a great many neurons connected to each other by synapses
\citep{Cajal, Amit, Abbot_from, Torres_rev}. Mathematically, the first neural networks combined the Ising model \citep{Baxter_exact} with the Hebb learning rule \citep{Hebb} to reproduce, very successfully, the storage and retrieval of information \citep{Amari, Hopfield, Amit_Hebb}. Neurons were simplified to binary variables (like Ising spins) representing firing or non-firing cells. By considering the trivial fully-connected topology, exact solutions could be reached, which at the time seemed more important than attempting to introduce biological realism. Subsequent work has tended to focus on considering richer dynamics for the cells rather than on the way in which these are interconnected \citep{Vogels, Tores_competition, Jorge_PLoS}. However, the topology of the brain -- whether at the level of neurons and synapses, cortical areas or functional connections -- is obviously far from trivial \citep{Amaral, Sporns, Eguiluz, Arenas_celegans, Sporns_09, Johnson_JSTAT}.

The number of neighbours a given node in a network has is called its degree, and much attention is paid to degree distributions since they tend to be highly heterogeneous for most real networks. In fact, they are often approximately scale-free (i.e., described by power laws) \citep{Newman_rev, Boccaletti, Peri, Barabasi_cell}. By including this topological feature in a Hopfield-like neural-network model, Torres {\it et al.} \citet{Torres_influence} found that degree heterogeneity increases the system's performance at high levels of noise, since the hubs (high degree nodes) are able to retain information at levels well above the usual critical noise. To prove this analytically, the authors considered the {\it configurational ensemble} of networks (the set of random networks with a given degree distribution but no degree-degree correlations) and showed that
Monte Carlo (MC) simulations were in good agreement with mean-field analysis, despite the approximation inherent to the latter technique when the network is not fully connected. A similar approach can also be used to show how heterogeneity may be advantageous for the performance of certain tasks in models with a richer dynamics \citep{Johnson_EPL}. It is worth mentioning that this influence of the degree distribution on dynamical behaviour is found in many other settings, such as the more general situation of systems of coupled oscillators \citep{Barahona}.

Another property of empirical networks that is quite ubiquitous is the existence of correlations between the degrees of nodes and those of their neighbours \citep{Pastor-Satorras, Newman_mixing_PRL, Newman_mixing_PRE}. If the average degree-degree correlation is positive the network is said to be {\it assortative}, while it is called {\it disassortative} if negatively correlated. Most heterogeneous networks are disassortative \citep{Newman_rev}, which,
as described in Chapter \ref{Chapter_PRL},
seems to be because this is in some sense their equilibrium (maximum entropy) state given the constraints imposed by the degree distribution \citep{Johnson_PRL}. However, there are probably often mechanisms at work which drive systems from equilibrium by inducing different correlations, as appears to be the case for most social networks, in which nodes (people) of a kind tend to group together. This feature, known as {\it assortativity} or {\it mixing by degree}, is also relevant for processes taking place on networks. For instance, assortative networks have lower percolation thresholds and are more robust to targeted attack \citep{Newman_mixing_PRE}, while disassortative ones make for more stable ecosystems and are -- at least according to the usual definition -- more synchronizable \citep{Brede}.

The approach usually taken when studying correlated networks computationally is to generate a network from the configuration ensemble and then introduce correlations (positive or negative) by some stochastic rewiring process \citep{Maslov}. A drawback of this method, however, is that results may well then depend on the details of this mechanism: there is no guarantee that one is correctly sampling the phase space of networks with given correlations. For analytical work, some kind of hidden variables from which the correlations originate are often considered
\citep{Caldarelli_fitness, Soderberg, Boguna, Fronczak} -- an assumption which can also be used to generate correlated networks computationally \citep{Boguna}. This can be a very powerful method for solving specific network models. However, it may not be appropriate if one wishes to consider all possible networks with given degree-degree correlations, independently of how these may have arisen. 
In this chapter,
we get round
the
problem by making use of
the method put forward by \citet{Johnson_PRL} (and described in Chapter \ref{Chapter_PRL}) whereby the ensemble of all networks with given correlations can be considered theoretically without recurring to hidden variables \citep{Sebas}. Furthermore, we show how this approach can be used computationally to generate random networks that are representative of the ensemble of interest (i.e., they are model-independent). In this way, we study the effect of correlations on a simple neural network model and find that assortativity increases performance in the face of noise -- particularly if
it is the hubs that are mainly responsible for storing information (and it is worth mentioning that there is experimental evidence suggestive of a main functional role played by hub neurons in the brain \citep{Morgan, Bonifazi}).
The
good agreement between the mean-field analysis and our MC simulations bears witness both to the robustness of the results as regards neural systems, and to the viability of using this method for studying dynamics on correlated networks.

\section{Preliminary considerations}

\subsection{Model neurons on networks}
\label{sec_model}

The attractor neural network model put forward by Hopfield \citep{Hopfield} consists of $N$ binary neurons, each with an activity given by the dynamic variable $s_{i}=\pm 1$. 
Every time step (MCS), each neuron is updated according to the stochastic transition probability
$
P(s_{i}\rightarrow \pm 1)=\frac{1}{2}\left[1\pm\tanh\left(h_{i}/T\right)\right]
$
(parallel dynamics),
where the field $h_{i}$ is the combined effect on $i$ of all its neighbours, $h_{i}=\sum_{j}\hat{w}_{ij}s_{j}$, and $T$ is a noise parameter we shall call {\it temperature}, but which represents any kind of random fluctuations in the environment. This is the same as the Ising model for magnetic systems, and the transition rule can be derived from a simple interaction energy such that aligned variables $s$ (spins) contribute less energy than if they were to take opposite values. However, this system can store $P$ given configurations ({\it memory patterns}) $\xi_{i}^{\nu}=\pm 1$ by having the interaction strengths ({\it synaptic weights}) set according to the Hebb rule \citep{Hebb}: $\hat{w}_{ij}\propto \sum_{\nu=1}^{P}\xi_{i}^{\nu}\xi_{j}^{\nu}$. In this way, each pattern becomes an attractor of the dynamics, and the system will evolve towards whichever one is closest to the initial state it is placed in. This mechanism is called {\it associative memory}, and is nowadays used routinely for tasks such as image identification. What is more, it has been established that something similar to the Hebb rule is implemented in nature via the processes of long-term potentiation and depression at the synapses \citep{Malenka,DeRoo,Ole1,Ole2}, and this phenomenon is indeed required for learning \citep{Gruart}.

To take into account the topology of the network, we shall consider the weights to be of the form $\hat{w}_{ij}=\hat{\omega}_{ij}\hat{a}_{ij}$, where the element $\hat{a}_{ij}$ of the adjacency matrix represents the number of directed edges (usually interpreted as synapses in a neural network) from node $j$ to node $i$, while $\hat{\omega}$ stores the patterns, as before: 
$$
\hat{\omega}_{ij}=\frac{1}{\langle k\rangle}\sum_{\nu=1}^{P}\xi_{i}^{\nu}\xi_{j}^{\nu}.
$$
For the sake of coherence with previous work, we shall assume $\hat{a}$ to be symmetric (i.e., the network is undirected), so each node is characterized by a single degree $k_{i}=\sum_{j}\hat{a}_{ij}$. However, all results are easily extended to directed networks -- in which nodes have both an {\it in} degree, $k_{i}^{\mbox{in}}=\sum_{j}\hat{a}_{ij}$, and an {\it out} degree, $k_{i}^{\mbox{out}}=\sum_{j}\hat{a}_{ji}$ -- by bearing in mind it is only a neuron's pre-synaptic neighbours that influence its behaviour. The mean degree of the network is $\langle k\rangle$, where the angles stand for an average over nodes\footnote{In directed networks the mean {\it in} degree and the mean {\it out} degree necessarily coincide, whatever the forms of the {\it in} and {\it out} distributions.}: $\langle \cdot\rangle\equiv N^{-1}\sum_{i}(\cdot)$.


\subsection{Network ensembles}
\label{sec_ensembles}

When one wishes to consider a set of networks which are randomly wired while respecting certain constraints -- that is, an {\it ensemble} -- it is usually useful to define the expected value of the adjacency matrix\footnote{As in statistical physics, one can consider the {\it microcanonical} ensemble, in which each element (network) satisfies the constraints exactly, or the {\it canonical} ensemble, where the constraints are satisfied on average \citep{Bianconi_ensembles}. Throughout this work, we shall refer to canonical ensembles.}, $E(\hat{a})\equiv\hat{\epsilon}$. The element $\hat{\epsilon}_{ij}$ of this matrix is the mean value of $\hat{a}_{ij}$ obtained by averaging over the ensemble. For instance, in the Erd\H{o}s-R\'{e}nyi (ER) ensemble all elements (outside the diagonal) take the value $\hat{\epsilon}_{ij}^{ER}=\langle k\rangle /N$, which is the probability that a given pair of nodes be connected by an edge. For studying networks with a given degree sequence, $(k_{1},...k_{N})$, it is common to assume the {\it configuration ensemble}, defined as
$$
\epsilon_{ij}^{conf}=\frac{k_{i}k_{j}}{\langle k\rangle N}
$$
This expression can usually be applied also when the constraint is a given degree distribution, $p(k)$, by integrating over $p(k_{i})$ and $p(k_{j})$ where appropriate. One way of deriving $\hat{\epsilon}^{conf}$ is to assume one has $k_{i}$ dangling half-edges at each node $i$; we then randomly choose pairs of half-edges and join them together until the network is wired up. Each time we do this, the probability that we join $i$ to $j$ is $k_{i}k_{j}/(\langle k\rangle N)^{2}$, and we must perform the operation $\langle k\rangle N$ times. Bianconi showed that this is also the solution for Barab\'asi-Albert evolved networks \citep{Bianconi_mean-field}. However, we should bear in mind that this result is only strictly valid for networks constructed in certain particular ways, such as in these examples. It is often implicitly assumed that were we to average over all random networks with a given degree distribution, the mean adjacency matrix obtained would be $\hat{\epsilon}^{conf}$. However, as we discussed in Chapter \ref{Chapter_PRL}, this is not in fact necessarily true \citep{Johnson_PRL}.


\begin{figure}[t!]
\begin{center}
\hspace*{-0.25cm}\includegraphics
[width=10.0cm]
{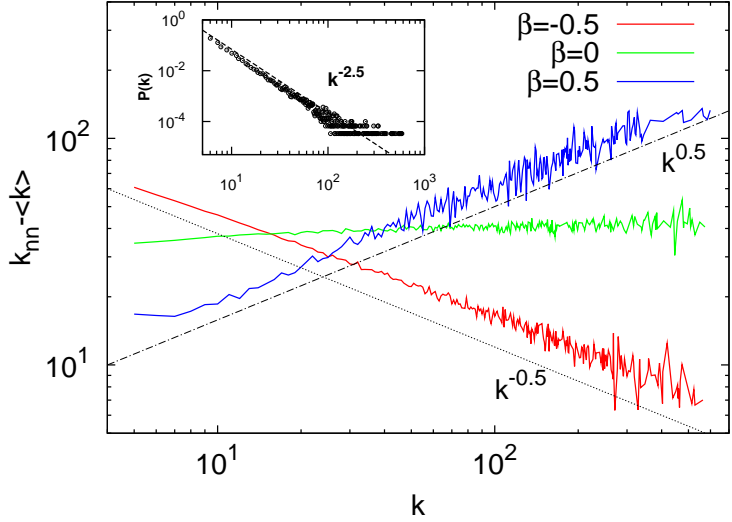}
\end{center}
\caption{
Mean-nearest-neighbour functions $\overline{k}_{nn}(k)$ for scale-free networks with $\beta=-0.5$ (disassortative), $0.0$ (neutral), and $0.5$ assortative, generated according to the algorithm described in Sec. \ref{sec_generating}. Inset: degree distribution (the same in all three cases). Other parameters are $\gamma=2.5$, $\langle k\rangle = 12.5$, $N=10^{4}$.}
\label{fig_knn}
\end{figure}


\subsection{Correlated networks}
\label{sec_correlated_nets}

In the configuration ensemble, the expected value
of the mean degree of the neighbours of a given node is $
\overline{k}_{nn,i}=k_{i}^{-1}\sum_{j}\hat{\epsilon}_{ij}^{conf}k_{j}=\langle
k^{2}\rangle/\langle k\rangle,
$ which is independent of $k_{i}$. However, as
mentioned above, real networks often display degree-degree correlations, with the result
that $\overline{k}_{nn,i}=\overline{k}_{nn}(k_{i})$. If $\overline{k}_{nn}(k)$ increases with $k$, the network is said to be assortative -- whereas it is disassortative if it decreases with $k$ (see Fig. \ref{fig_knn}). This is from the more general nomenclature (borrowed form sociology) in which sets are assortative if elements of a kind group together, or assort. In the case of degree-degree correlated networks, positive assortativity means that edges are more than randomly likely to occur between nodes of a similar degree.

The ensemble of all networks with a given degree sequence $(k_{1},...k_{N})$
contains a subset for all members of which $\overline{k}_{nn}(k)$ is constant (the configuration ensemble), but also subsets displaying other functions $\overline{k}_{nn}(k)$. We can identify each one of these subsets (regions of phase space) with an
expected adjacency matrix $\hat{\epsilon}$ which simultaneously satisfies the following conditions: ${\bf i)}$  $\sum_{j}k_{j}\hat{\epsilon}_{ij}=k_{i}\overline{k}_{nn}(k_{i})$, $\forall i$ (by definition of $\overline{k}_{nn}(k)$), and ${\bf ii)}$ $\sum_{j}\hat{\epsilon}_{ij}=k_{i}$, $\forall i$ (for consistency).
As we showed in Chapter \ref{Chapter_PRL}, the general solution to this problem is a matrix of the form
\begin{equation}
  \hat{\epsilon}_{ij}=\frac{k_{i}k_{j}}{\langle k\rangle N}
  +\int d\nu \frac{f(\nu)}{N}\left[\frac{(k_{i}k_{j})^{\nu}}
    {\langle k^{\nu}\rangle}-k_{i}^{\nu}-k_{j}^{\nu}+\langle k^{\nu}\rangle  \right],
\label{eq_epsi_PRE_gen_PRE}
\end{equation}
where $\nu\in\mathbb{R}$ and the function $f(\nu)$ is
determined by $\overline{k}_{nn}(k)$
\citep{Johnson_PRL}. (If the network were directed, then $k_{i}=k_{i}^{\mbox{in}}$ and $k_{j}=k_{j}^{\mbox{out}}$ in this expression.) 
This
yields
\begin{eqnarray}
  \overline{k}_{nn}(k)=\frac{\langle k^{2}\rangle}{\langle k\rangle}
  +\int d\nu f(\nu)\sigma_{\nu+1}\left[\frac{k^{\nu-1}}
    {\langle k^{\nu}\rangle}-\frac{1}{k} \right]
\label{eq_knn_gen_PRE}
\end{eqnarray}
(the first term being the result for the
configuration ensemble), where $\sigma_{b+1}\equiv \langle k^{b+1}\rangle -\langle k\rangle
\langle k^{b}\rangle$.
This means that $\hat{\epsilon}$ is not just one possible way of obtaining correlations according to $\overline{k}_{nn}(k)$; rather, there is a two-way mapping between $\hat{\epsilon}$ and $\overline{k}_{nn}(k)$: every network with this particular function $\overline{k}_{nn}(k)$ and no other ones are contained in the ensemble defined by $\hat{\epsilon}$. Thanks to this, if we are able to consider random networks drawn according to this matrix (whether we do this analytically or computationally; see Section \ref{sec_generating}), we can be confident that we are correctly taking account of the whole ensemble of interest. In other words, whatever the reasons behind the existence of degree-degree correlations in a given network, we can study the effects of these with only information on $p(k)$ and $\overline{k}_{nn}(k)$ by obtaining the associated matrix $\hat{\epsilon}$. This is not to say, of course, that all topological properties are captured in this way: a particular network may have other features -- such as higher order correlations, modularity, etc. -- the consideration of which would require concentrating on a sub-partition of those with the same $p(k)$ and $\overline{k}_{nn}(k)$. But this is not our purpose here.
\\
\linebreak

In many empirical networks, $\overline{k}_{nn}(k)$ has the form $\overline{k}_{nn}(k)=A+B
k^{\beta}$, with $A,B>0$ \citep{Boccaletti, Pastor-Satorras} -- the
mixing being assortative if $\beta$ is positive, and disassortative when negative.
Such a case is fitted by Eq. (\ref{eq_knn_gen_PRE}) if 
\begin{equation}
f(\nu)=C\left[\frac{\sigma_{2}}{\sigma_{\beta+2}}\delta(\nu-\beta-1)-\delta(\nu-1)\right],
\label{eq_f(nu)_PRE}
\end{equation}
with $C$ a positive constant, since this choice yields
\begin{equation}
  \overline{k}_{nn}(k)=\frac{\langle k^{2}\rangle}{\langle k\rangle}
  +C\sigma_{2}\left[\frac{k^{\beta}}{\langle
      k^{\beta+1}\rangle}-\frac{1}{\langle k\rangle} \right].
\label{eq_knn_simple_PRE}
\end{equation}


In Chapter \ref{Chapter_PRL} we discussed how the most likely configurations for networks with scale-free degree distributions ($p(k)\sim k^{-\gamma}$) and correlations given by Eq. (\ref{eq_knn_simple_PRE}) are generally disassortative. We also showed that the maximum entropy is usually obtained for values of $C$ close to one. Here, we shall use this result to justify concentrating on correlated networks with $C=1$, so that the only parameter we need to take into account is $\beta$. It is worth mentioning that Pastor-Satorras {\it et al.} originally suggested using this exponent as a way of quantifying correlations \citep{Pastor-Satorras}, since this seems to be the most relevant magnitude. Because $\beta$ does not depend directly on $p(k)$ (as $r$ does), and can be defined for networks of any size (whereas $r$, in very heterogeneous networks, always goes to zero for large $N$ due to its normalization \citep{Goltsev_zero}), we shall henceforth use $\beta$ as our assortativity parameter.

So, after plugging Eq. (\ref{eq_f(nu)_PRE}) into Eq. (\ref{eq_epsi_PRE_gen_PRE}), we find that the ensemble of networks exhibiting correlations given by Eq. (\ref{eq_knn_simple_PRE}) (and $C=1$) is defined by the mean adjacency matrix
\begin{eqnarray}
& \hat{\epsilon}_{ij} &  =  \frac{1}{N}[k_{i}+k_{j}-\langle k\rangle]
\nonumber
\\
& +& \frac{\sigma_{2}}{\sigma_{\beta+2}}\frac{1}{N}\left[\frac{(k_{i}k_{j})^{\beta+1}}{\langle k^{\beta+1}\rangle} -k_{i}^{\beta+1}-k_{j}^{\beta+1}+\langle k^{\beta+1}\rangle\right].
\label{eq_epsi_PRE}
\end{eqnarray}


\section{Analysis and results}

\subsection{Mean field}
\label{sec_asso_dyn}

Let us consider the single-pattern case ($P=1$, $\xi_{i}=\xi_{i}^{1}$). Substituting the adjacency matrix $\hat{a}$ for its expected value $\hat{\epsilon}$ (as given by Eq. (\ref{eq_epsi_PRE})) in the expression for the local field at $i$ -- which amounts to a mean-field approximation -- we have
\begin{eqnarray*}
h_{i} & = & \frac{1}{\langle k\rangle}\xi_{i}\left\{ \left[(k_{i}-\langle k\rangle)+\frac{\sigma_{2}}{\sigma_{\beta+2}}(\langle k^{\beta+1}\rangle-k_{i}^{\beta+1})\right]\mu_{0}\right.\\
\\
 & + & \left.\langle k\rangle\mu_{1}+\frac{\sigma_{2}}{\sigma_{\beta+2}}(k_{i}^{\beta}-\langle k^{\beta+1}\rangle)\mu_{\beta+1}\right\},
\end{eqnarray*}
where we have defined
$$
\mu_{\alpha}\equiv \frac{\langle k_{i}^{\alpha} \xi_{i}s_{i}\rangle}{\langle k^{\alpha}\rangle }
$$
for $\alpha=0,$ $1$, $\beta+1$. These order parameters measure the extent to which the system is able to recall information in spite of noise \citep{Johnson_EPL}. For the first order we have $\mu_{0}=m\equiv\langle \xi_{i}s_{i}\rangle$, the standard overlap measure in neural networks (analogous to magnetization in magnetic systems), which takes account of memory performance. However, $\mu_{1}$, for instance, weighs the sum with the degree of each node, with the result that it measures information per synapse instead of per neuron. Although the overlap $m$ is often assumed to represent, in some sense, the {\it mean firing rate} of neurological experiments, it is possible that $\mu_{1}$ is more closely related to the empirical measure, since the total electric potential in an area of tissue is likely to depend on the number of synapses transmitting action potentials. In any case, a comparison between the two order parameters is a good way of assessing to what extent the performance of neurons depends on their degree -- 
larger-degree model neurons can in general store information at higher temperatures than ones with smaller degree can \citep{Torres_influence}.

Substituting $s_{i}$ for its expected value according to the transition probability, $s_{i}\rightarrow \tanh(h_{i}/T)$, we have, for any $\alpha$,
\[
\langle k_{i}^{\alpha} \xi_{i}s_{i}\rangle=\langle k_{i}^{\alpha} \xi_{i}\tanh(h_{i}/T)\rangle;
\]
or, equivalently, 
the following 3-D map of closed
coupled equations for the macroscopic overlap observables $\mu_{0}$,
$\mu_{1}$ and $\mu_{\beta+1}$ -- which describes, in this mean-field approximation, the dynamics of the system:
\begin{eqnarray}
\mu_{0}(t+1) & = & \int p(k)\tanh[F(t)/(\langle k\rangle T)]\mbox{d}k
\nonumber\\
\nonumber\\
\mu_{1}(t+1) & = & \frac{1}{\langle k\rangle}\int p(k)k\tanh[F(t)/(\langle k\rangle T)]\mbox{d}k
\label{eq_3map}
\\
\nonumber\\
\mu_{\beta+1}(t+1) & = & \frac{1}{\langle k^{\beta+1}\rangle}\int p(k)k^{\beta+1}\tanh[F(t)/(\langle k\rangle T)]\mbox{d}k,
\nonumber
\end{eqnarray}
 with 
\begin{eqnarray*}
F(t) & \equiv & (k\mu_{0}(t)+\langle k\rangle\mu_{1}(t)-\langle k\rangle\mu_{0}(t))\\
\\
& + & \frac{\sigma_{2}}{\sigma_{\beta+2}}[k^{\beta+1}(\mu_{\beta+1}(t)-\mu_{0}(t))\\
\\
& + & \langle k^{\beta+1}\rangle(\mu_{0}(t)-\mu_{\beta+1}(t))].
\end{eqnarray*}
This can be easily computed for any degree distribution $p(k)$. Note that taking $\beta=0$ (the uncorrelated case) the system collapses to the 2-D map obtained by \citet{Torres_influence}, while it becomes the typical 1-D case for a homogeneous $p(k)$ -- say a fully-connected network \citep{Hopfield}. It is in principle possible to do similar mean-field analysis for any number $P$ of patterns, but the map would then be $3P$-dimensional, making the problem substantially more complex.

At a critical temperature $T_{c}$, the system will undergo the characteristic second order phase transition from a phase in which it exhibits memory (akin to ferromagnetism) to one in which it does not (paramagnetism). To obtain this critical temperature, we can expand the hyperbolic tangent in Eqs. (\ref{eq_3map}) around the trivial solution $(\mu_{0},\mu_{1},\mu_{\beta+1})\simeq(0,0,0)$ and, keeping only linear terms, write
\begin{eqnarray*}
\mu_{0} & = & \mu_{1}/T_{c}, \\
\\
\mu_{1} & = & \frac{1}{ \langle k\rangle ^{2}T_{c}}\left[ \langle k\rangle ^{2}\mu_{1}+\sigma_{2}\mu_{\beta+1}\right], \\
\\
\mu_{\beta+1} & = & \frac{1}{T_{c} \langle k\rangle  \langle k^{\beta+1}\rangle }\left[\sigma_{\beta+2}\mu_{0}\frac{}{}\right.
\\
\\
& + & \left. \frac{\sigma_{2}}{\sigma_{\beta+2}}\left( \langle k^{\beta+1}\rangle ^{2}- \langle k^{2(\beta+1)}\rangle \right)\mu_{0} \right.\\
\\
& + & \left.  \langle k\rangle  \langle k^{\beta+1}\rangle \mu_{1}-\frac{\sigma_{2}}{\sigma_{\beta+2}}\left( \langle k^{\beta+1}\rangle ^{2}- \langle k^{2(\beta+1)}\rangle \right)\mu_{\beta+1}\right].
\end{eqnarray*}
Defining
\begin{eqnarray*}
A         & \equiv & \frac{\sigma_{2}}{ \langle k\rangle ^{2}}, \\
\\
B         & \equiv & \frac{\sigma_{2}}{\sigma_{\beta+2}}  \frac{ \langle k^{2(\beta+1)}\rangle - \langle k^{\beta+1}\rangle ^{2}}{ \langle k\rangle  \langle k^{\beta+1}\rangle }, \\
\\
D & \equiv & \frac{\sigma_{\beta+2}}{ \langle k\rangle  \langle k^{\beta+1}\rangle },
\end{eqnarray*}
$T_{c}$ will be the solution to the third order polynomial equation:
\begin{equation}
T_{c}^{3}-(B+1)T_{c}^{2}+(B-A)T_{c}+A(B-D)=0.
\label{eq_PolyTc}
\end{equation}
Note that for neutral (i.e., uncorrelated) networks, $\beta=0$, and so $A=B=D$. We then have $T_{c}=\langle k^{2}\rangle/\langle k\rangle^{2}$, as expected \citep{Johnson_EPL}.

\subsection{Generating correlated networks}
\label{sec_generating}

Given a degree distribution $p(k)$, the ensemble of networks compatible with this constraint and with degree-degree correlations according to Eq. (\ref{eq_knn_simple_PRE}) (with some exponent $\beta$) is defined by the mean adjacency matrix $\hat{\epsilon}$ of Eq. (\ref{eq_epsi_PRE}) -- as described in Section \ref{sec_correlated_nets} and by \citet{Johnson_PRL}. Therefore, although there will generally be an enormous number of possible networks in this volume of phase space, we can sample them correctly simply by generating them according to $\hat{\epsilon}$. To do this, first we have to assign to each node a degree drawn from $p(k)$. If the elements of $\hat{\epsilon}$ were probabilities, it would suffice then to connect each pair of nodes $(i,j)$ with probability $\hat{\epsilon}_{ij}$ to generate a valid network. Strictly speaking, $\hat{\epsilon}$ is an expected value, which in certain cases can be greater than one. To get round this, we write a probability matrix $\hat{p}=\hat{\epsilon}/a$ with $a$ some value such that all elements of $\hat{p}$ are smaller than one. If we then take random pairs of nodes $(i,j)$ and, with probability $\hat{p}_{ij}$, place an edge between them, repeating the operation until $\frac{1}{2}\langle k\rangle N$ edges have been placed, the expected value of edges joining $i$ and $j$ will be $\hat{\epsilon}_{ij}$. This method is like the {\it hidden variable} technique
\citep{Boguna} in that edges are placed with a predefined probability (which is why the resulting ensemble is canonical). The difference lies in the fact that in the method here described correlations only depend on the degrees of nodes.

We are interested here in neural networks, in which a given pair of nodes can be joined by several synapses, so we shall not impose the restriction of so-called simple networks of allowing only one edge at most per pair. We shall, however, consider networks with a {\it structural cutoff}: $k_{i}< \sqrt{\langle k\rangle N}$, $\forall i$ \citep{Bianconi_entropy}. This ensures that, at least for $\beta\leq 0$, all elements of $\hat{\epsilon}$ are indeed smaller than one.



Because we can expect effects due to degree-degree correlations to be largest when $p(k)$ is very broad, and since most networks in nature and technology seem to exhibit approximately power-law degree distributions \citep{Newman_rev, Arenas_rev, Peri, Barabasi_cell}, we shall here test our general theoretical results against simulations of scale-free networks: $p(k)\sim k^{-\gamma}$.
This means that a network (or the region of phase space to which it belongs) is characterized by the set of parameters $\lbrace \langle k\rangle, N, \gamma, \beta \rbrace$.

\subsection{Assortativity and dynamics}
\label{sec_results}

\begin{figure}[t!]
\begin{center}
\hspace*{-0.2cm}\includegraphics
[width=10.0cm]
{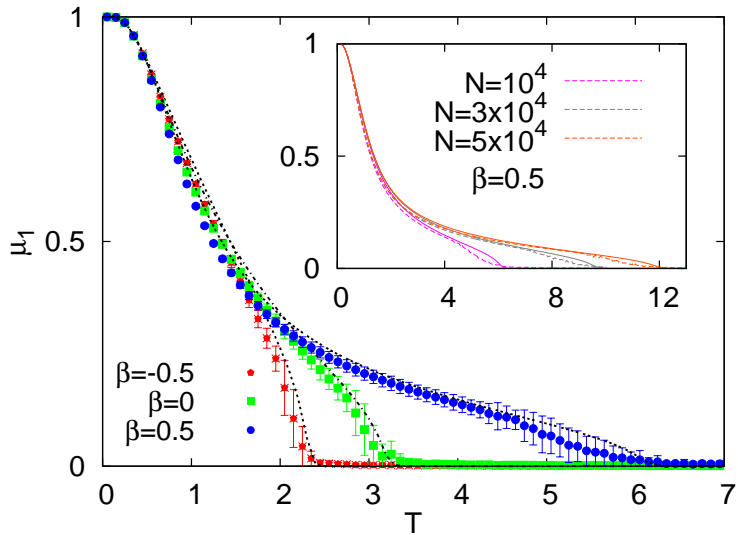}
\end{center}
\caption{ Stable stationary value of the weighted overlap $\mu_{1}$ against temperature $T$ for scale-free networks
with correlations according to $\overline{k}_{nn}\sim k^{\beta}$, for $\beta=-0.5$ (disassortative), $0.0$ (neutral), and $0.5$ (assortative). Symbols from MC simulations, with errorbars representing standard deviations, and lines from
Eqs. (\ref{eq_3map}). Other network  parameters as in Fig. \ref{fig_knn}. Inset: $\mu_{1}$ against $T$ for the assortative case ($\beta=0.5$) and different system sizes: $N=10^{4}$, $3\cdot 10^{4}$ and $5\cdot 10^{4}$.}
\label{fig_mu1}
\end{figure}

\begin{figure}[t!]
\begin{center}
\hspace*{-0.65cm}\includegraphics
[width=10.0cm]
{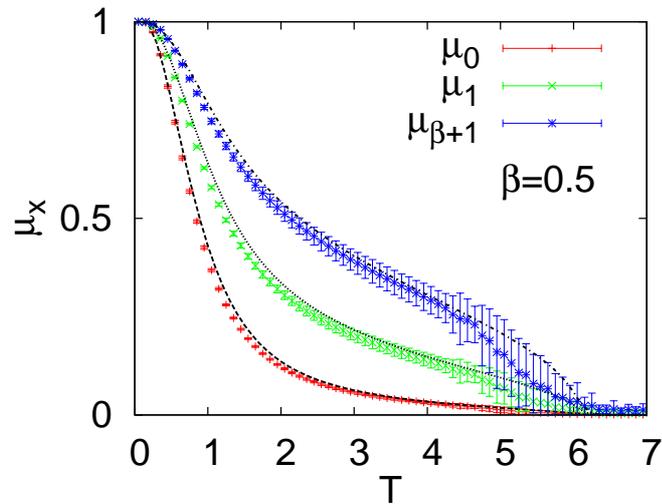}
\end{center}
\caption{ Stable stationary values of order parameters $\mu_{0}$, $\mu_{1}$ and $\mu_{\beta+1}$ against temperature $T$, for assortative networks according to $\beta=0.5$.
Symbols from MC simulations, with errorbars representing standard deviations, and lines from Eqs. (\ref{eq_3map}). Other parameters as in Fig. \ref{fig_knn}.
}
\label{fig_mub}
\end{figure}

\begin{figure}[t!]
\begin{center}
\hspace*{-0.65cm}\includegraphics
[width=10.0cm]
{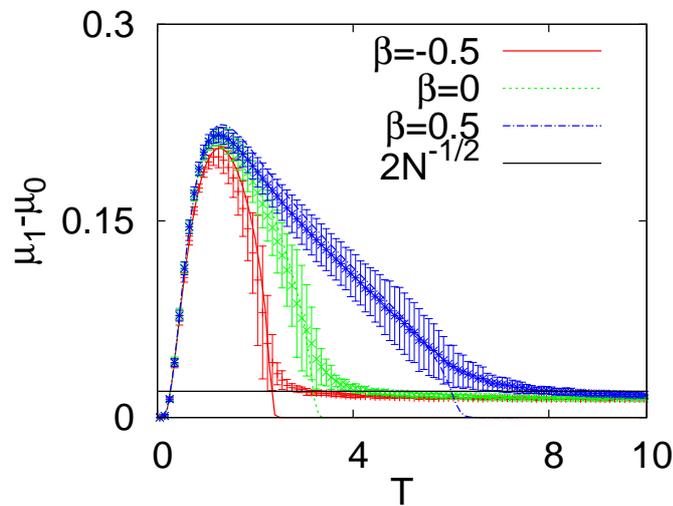}
\end{center}
\caption{
 Difference between the stationary values $\mu_{1}$ and $\mu_{0}$ for networks with $\beta=-0.5$ (disassortative), $0.0$ (neutral) and $0.5$ (assortative), against temperature.
Symbols from MC simulations, with errorbars representing standard deviations, and lines from Eqs. (\ref{eq_3map}). Line shows the expected level of fluctuations due to noise, $\sim N^{-\frac{1}{2}}$. Other parameters as in Fig. \ref{fig_knn}.
}
\label{fig_picos}
\end{figure}

In Fig. \ref{fig_mu1} we plot the stationary value of $\mu_{1}$ against the temperature $T$, as obtained from simulations and Eqs. (\ref{eq_3map}),
for disassortative, neutral and assortative networks. The three curves are similar at low temperatures, but as $T$ increases their behaviour becomes quite different. The disassortative network is the least robust to noise. However, the assortative one is capable of retaining some information at temperatures considerably higher than the critical value, $T_{c}=\langle k^{2}\rangle /\langle k\rangle$, of neutral networks. A comparison between $\mu_{1}$ and $\mu_{0}$ (see Fig. \ref{fig_mub}) shows that it is the high degree nodes that are mainly responsible for this difference in performance. This can be seen more clearly in Fig. \ref{fig_picos}, which displays the difference $\mu_{1}-\mu_{0}$ against $T$ for the same networks. It seems that, because in an assortative network a sub-graph of hubs will have more edges than in a disassortative one, it has a higher effective critical temperature. Therefore, even when most of the nodes are acting randomly, the set of nodes of sufficiently high degree nevertheless displays associative memory.


\begin{figure}[t!]
\begin{center}
\hspace*{0.25cm}
\epsfig{file=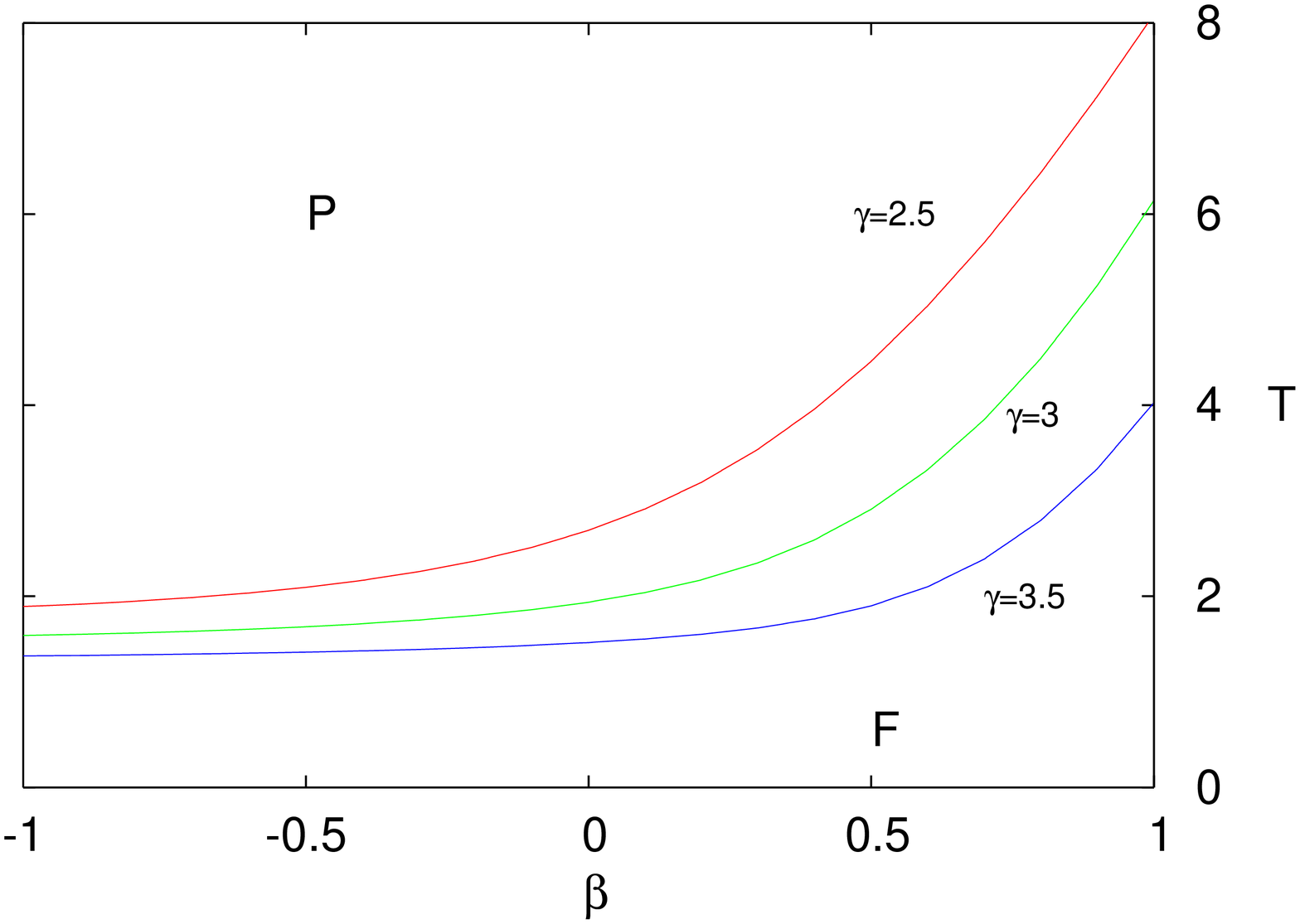,width=9.0cm}
\end{center}
\caption{ Phase diagrams for scale-free networks with $\gamma=2.5$, $3$, and $3.5$. Lines show the critical temperature $T_{c}$ marking the second-order transition from a memory (ferromagnetic) phase to a memoryless (paramagnetic) one, against the assortativity $\beta$, as
given by Eq. (\ref{eq_PolyTc}). Other parameters as in Fig. \ref{fig_knn}.
}
\label{fig_phase}
\end{figure}

The phase diagram if Fig. \ref{fig_phase} shows the critical temperature, $T_{c}$, as obtained from Eq. (\ref{eq_PolyTc}). In addition to the effect reported by \citet{Torres_influence} whereby the $T_{c}$ of scale-free networks grows with degree heterogeneity (decreasing $\gamma$), it also increases very significantly with positive degree-degree correlations (increasing $\beta$).

\begin{figure}[h!]
\begin{center}
\hspace*{-0.5cm}\includegraphics
[width=10.0cm]
{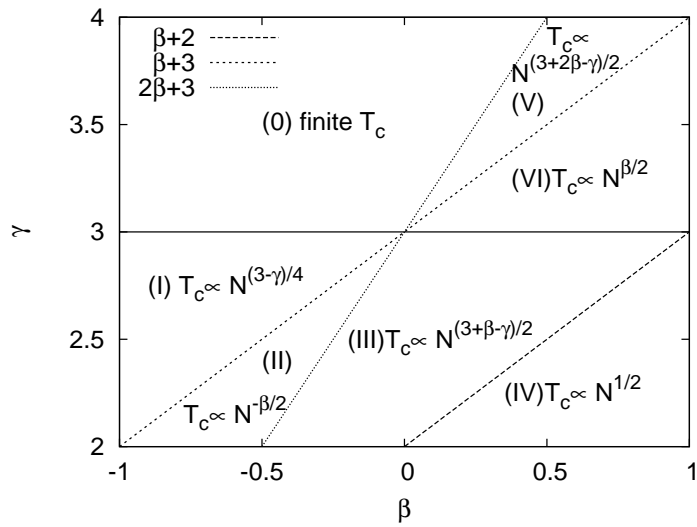}
\caption{ Parameter space $\beta-\gamma$ partitioned into the regions in which $b(\beta,\gamma)$ has the same functional form -- where $b$ is the scaling exponent of the critical temperature: $T_{c}\sim N^b$. Exponents obtained by taking the large $N$ limit in Eq. (\ref{eq_PolyTc}).}
\label{fig_scale}
\end{center}
\end{figure}

\begin{figure}[h!]
\begin{center}
\hspace*{-0.3cm}\includegraphics
[width=10.0cm]
{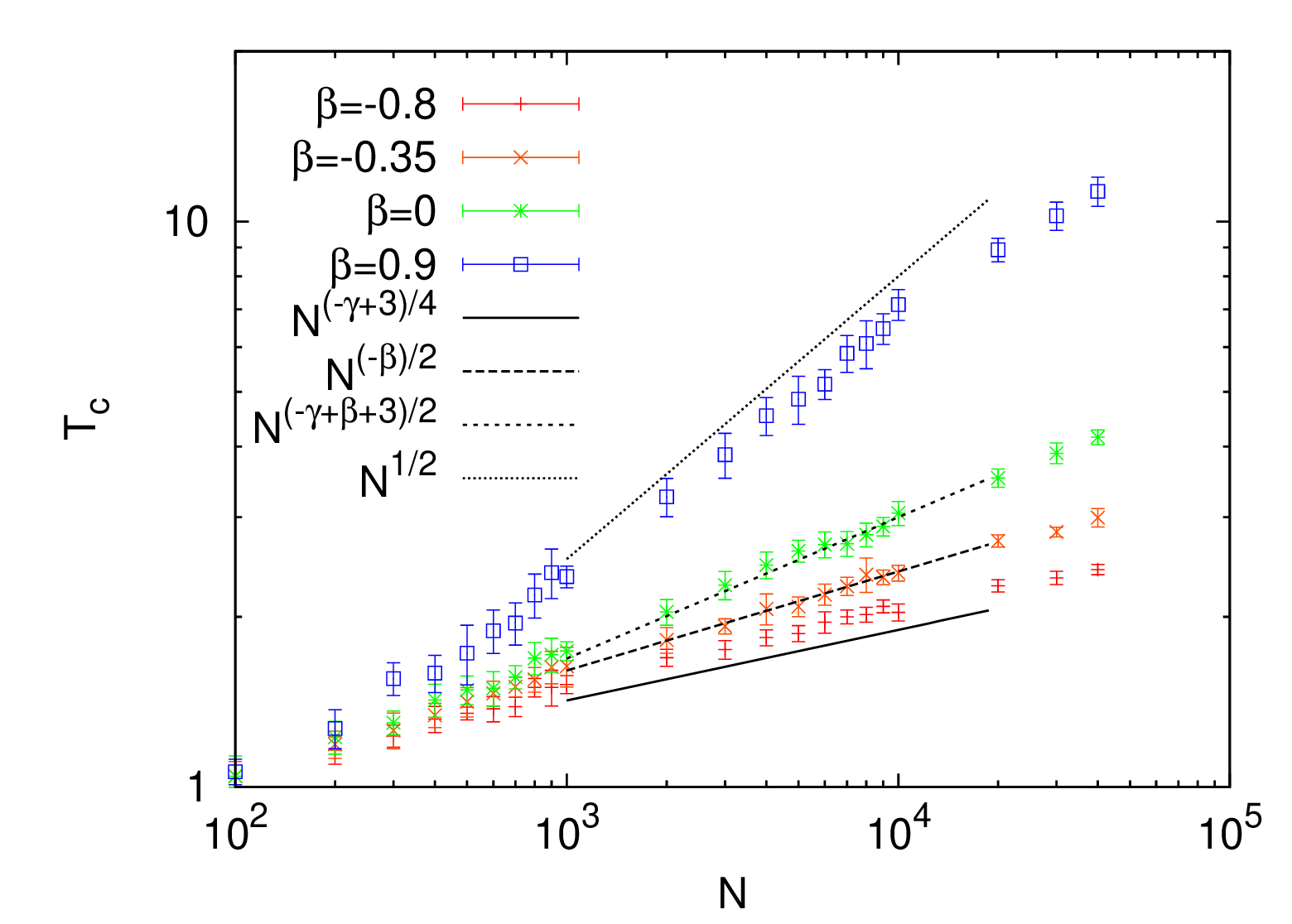}
\caption{
 Examples of how $T_{c}$ scales with $N$ for networks belonging to regions I, II, III and IV of Fig. \ref{fig_scale} ($\beta=-0.8$, $-0.35$, $0.0$ and $0.9$, respectively). Symbols from MC simulations, with errorbars representing standard deviations, and slopes from Eq. (\ref{eq_PolyTc}). All parameters -- except for $\beta$ and $N$ -- are as in Fig. \ref{fig_knn}.
}
\label{fig_4lines}
\end{center}
\end{figure}

At large values of $N$, the critical temperature scales as $T_{c}\sim N^b$, with $b\geq0$ a constant. However, because the moments of $k$ appearing in the coefficients of Eq. (\ref{eq_PolyTc}) can have different asymptotic behaviour depending on the values of $\gamma$ and $\beta$, the scaling exponent $b$ differs from one region to another in the space of these parameters. These are the seven regions shown in Fig. \ref{fig_scale}, along with the scaling behaviour exhibited by each one. This can be seen explicitely in Fig. \ref{fig_4lines}, where $T_{c}$, as obtained from MC simulations, is plotted against $N$ for cases in each of the regions with $\gamma<3$. In each case, the scaling is as given by Eq. (\ref{eq_PolyTc}) and shown in Fig. \ref{fig_scale}.
For the four regions with $\gamma<3$, from lowest to highest assortativity we have scaling exponents which are dependent on: only $\gamma$ (region I), only $\beta$ (region II), both $\gamma$ and $\beta$ (region III), and, perhaps most interestingly, neither of the two (region IV) -- with $T_{c}$ scaling, in the latter case, as $\sqrt{N}$. As for the more homogeneous $\gamma>3$ part, regions V and VI have a diverging critical temperature despite the fact that the second moment of $p(k)$ is finite, simply as a result of assortativity.

\begin{figure}[h!]
\begin{center}
\vspace{-1.8cm}
\hspace*{-0.3cm}\includegraphics
[width=14.1cm]
{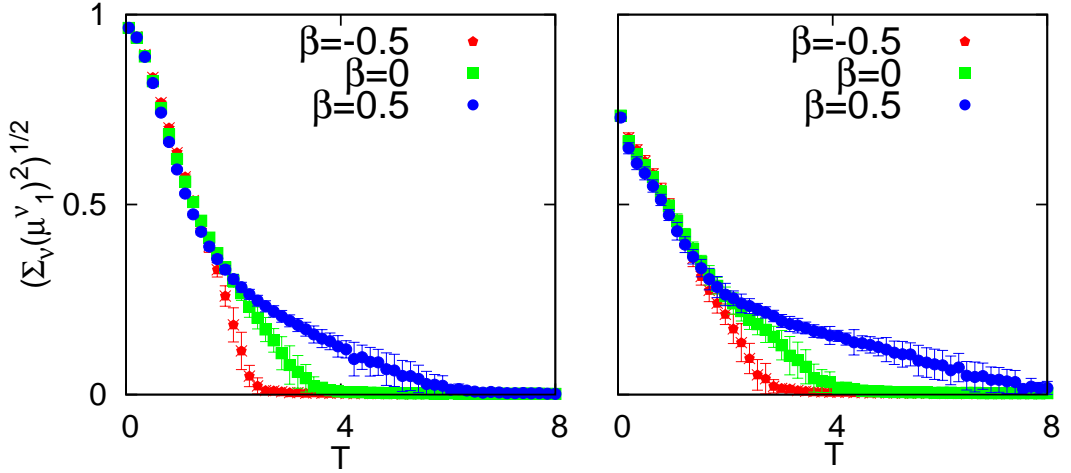}
\caption{
 Global order parameter $\zeta$ for assortative ($\beta=0.5$), neutral ($\beta=0.0$) and disassortative ($\beta=-0.5$) networks with $P=3$ (left panel) and $P=10$ (right panel) stored patterns. Symbols from MC simulations, with errorbars representing standard deviations. All parameters are as in Fig. \ref{fig_knn}.
}
\label{fig_P}
\end{center}
\end{figure}

The case in which more than one pattern are stored ($P>1$) can be explored numerically. Assuming there are $P$ uncorrelated patterns, we have an order parameter $\mu_{1}^{\nu}$ for each pattern $\nu$. A global measure of the degree to which there is memory can be captured by the parameter $\zeta$, where
$$
\zeta^2\equiv\frac{1}{1+P/N}\sum_{\nu=1}^{P}(\mu_{1}^{\nu})^{2}.
$$
Notice that the normalization factor is due to the fact that if one pattern is {\it condensed} -- i.e., $|\mu_{1}|\lesssim 1$ -- the others have $|\mu_{\nu}|\sim 1/\sqrt{N}$, $\nu=2,..P$, and so $\zeta\simeq 1$. Figure \ref{fig_P} shows how $\zeta$ decreases with $T$ in variously correlated networks for $P=3$ (left panel) and $P=10$ patterns (right panel). The behaviour is not qualitatively different from that observed for the single-pattern case in the main panel of Fig. \ref{fig_mu1}, suggesting that the influence of assortativity we report is robust as to the number of patterns stored, $P$.

\section{Discussion}
\label{sec_discussion}

We have shown that assortative networks of simple model neurons are able to exhibit associative memory in the presence of levels of noise such that uncorrelated (or disassortative) networks cannot. This may appear to be in contradiction with a recent result obtained using spectral graph analysis -- that synchronizability of a set of coupled oscillators is highest for disassortative networks \citep{Brede}. A synchronous state of model oscillators and a memory phase of model neurons are both sets of many simple dynamical elements coupled via a network in such a way that a macroscopically coherent situation is maintained \citep{Barahona}. Obviously both systems require the effective transmission of information among the elements. So why are opposite results as regards the influence of topology reported for each system? The answer is simple: whereas the definition of a synchronous state is that every single element oscillate at the same frequency, it is precisely when most elements are actually behaving randomly that the advantages to assortativity we report become apparent. In fact, it can be seen in Fig. \ref{fig_mu1} that at low temperatures disassortative networks perform the best, although the effect is small. This is reminiscent of percolation: at high densities of edges the giant component is larger in disassortative networks, but in assortative ones a non-vanishing fraction of nodes remain interconnected even at densities below the usual percolation threshold \citep{Newman_mixing_PRL,Newman_mixing_PRE}. Because in the case of targeted attacks it is this threshold which is taken as a measure of resilience, we say that assortative networks perform the best.
The relevance of partial synchronization and the important role of hubs have already been noted for systems of (weakly) coupled oscillators \citep{Arenas_paths, Pereira} -- for which, however, assortativity has not been expected to be of consequence \citep{Pereira}.
In general, the optimal network for good conditions (i.e., complete synchronization, high density of edges, low levels of noise) is not necessarily the one which performs the best in bad conditions (partial synchronization, low density of edges, high levels of noise). It seems that optimality --  whether in resilience or robustness -- should thus be defined for particular conditions.

We have used the technique suggested by \citet{Johnson_PRL} to study the effect of correlations on networks of model neurons, but many other systems of dynamical elements should be susceptible to a similar treatment. In fact, Ising spins \citep{Bianconi_mean-field}, Voter Model agents \citep{Eguiluz_voter}, or Boolean nodes \citep{Tiago}, for instance, are similar enough to binary neurons that we should expect similar results for these models. If a moral can be drawn, it is that persistence of partial synchrony, or coherence of a subset of highly connected dynamical elements, can sometimes be as relevant (or more so) as the possibility of every element behaving in the same way. In the case of real brain cells,
experiments suggest that hub neurons play key functional roles \citep{Morgan, Bonifazi}.
From this point of view, there may be a selective pressure for brain networks to become assortative -- although, admittedly, this organ engages in such complex behaviour that there must be many more functional constraints on its structure than just a high robustness to noise. Nevertheless, it would be interesting to investigate this aspect of biological systems experimentally. For this, it should be borne in mind that heterogeneous networks have a natural tendency to become disassortative, so it is against the expected value of correlations discussed by \citet{Johnson_PRL} that empirical data should be contrasted in order to look for meaningful deviations towards assortativity.
Similarly, it may be necessary to take into account the correlations that could emerge due to the spatial layout of neurons \citep{Kaiser_EJN, Johnson_CR}.
In any case, it would be in areas of the cortex specifically related to memory -- such as the temporal (long-term memory) \citep{Miyashita_temporal, Sakai_temporal} or prefrontal (short-term memory) \citep{Camperi_prefrontal, Compte} lobes -- that this effect might be relevant. A curious fact that would seem to support our hypothesis is that whereas the vast majority of non-social networks are disassortative \citep{Newman_rev}, one that appears actually to be strongly assortative is the functional network of the human cortex \citep{Eguiluz}.


%% file: cr/cr.tex

\chapter{Cluster Reverberation: A mechanism for robust short-term memory without synaptic learning} 
\label{Chapter_CR}


Short-term memory cannot in general be explained the way long-term memory can -- as a gradual modification of synaptic conductances -- since it takes place too quickly. Theories based on some  form of cellular bistability, however, do not seem to be able to account for the fact that noisy neurons can collectively store information in a robust manner. We show how a sufficiently clustered network of simple model neurons can be instantly induced into metastable states capable of retaining information for a short time. Cluster Reverberation, as we call it, could constitute a viable mechanism available to the brain for robust short-term memory with no need of synaptic learning. Relevant phenomena described by neurobiology and psychology, such as power-law statistics of forgetting avalanches, emerge naturally from this mechanism.



\section{Slow but sure, or fast and fleeting?}
\label{sec_intro}

Of all
brain phenomena, memory is probably one of the best understood \citep{Amit, Abbot_from, Torres_rev}. Consider a set of many neurons, defined as elements with two possible states (firing or not firing,
one or zero)
connected among each other in some way by synapses which carry a proportion of the current let off by a firing neuron to its neighbours; the probability that a given neuron has of firing at a certain time is then some function of the total current it has just received. Such a simplified model of the brain is able to store and retrieve information, in the form of patterns of activity (i.e., particular configurations of firing and non-firing neurons) when the synaptic conductances, or weights, have been appropriately set according to a learning rule \citep{Hebb}. Because each of the stored patterns becomes an attractor of the dynamics, the system will evolve towards whichever of the patterns most resembles the initial configuration. Artificial systems used for tasks such as pattern recognition and classification, as well as more realistic neural network models that take into account a variety of subcellular processes, all tend to rely on this basic mechanism, known as Associative Memory \citep{Amari, Hopfield}.

Synaptic conductances in animal brains have indeed been found to become strengthened or weakened during learning, via the biochemical processes of long-term potentiation (LTP) and depression (LTD) \citep{Malenka,Gruart, DeRoo, Ole1, Ole2}. Further support for the hypothesis that such a mechanism underlies long-term memory (LTM) comes from psychology, where it is being found more and more that so-called \textit{connectionist} models fit in well with observed
brain phenomena \citep{Marcus, Frank}. However, some memory processes take place on timescales of seconds or less
and in many instances cannot be accounted for by LTP and LTD \citep{Durstewitz}, since these require at least minutes to be effected \citep{Lee, Klintsova}.
For example, Sperling found that visual stimuli are recalled in great detail for up to about one second after exposure (iconic memory) \citep{Sperling}; similarly, acoustic information seems to linger for three or four seconds (echoic memory) \citep{Cowan}. 
In fact, it appears that the brain actually holds and continually updates a kind of buffer in which sensory information regarding its surroundings is maintained (sensory memory) \citep{Baddeley_Book}. This is easily observed by simply closing one's eyes and recalling what was last seen, or thinking about a sound after it has finished.
Another instance is the capability referred to as \textit{working} memory \citep{Durstewitz, Baddeley}: just as a computer requires RAM for its calculations despite having a hard drive for long term storage, the brain must continually store and delete information to perform almost any cognitive task.
To some extent, working memory could consist in somehow labelling or bringing forward previously stored concepts, like when one is asked to remember a particular sequence of digits or familiar shapes. But we are also able to manipulate -- if perhaps not quite so well -- shapes and symbols we have only just become acquainted with, too recently for them to have been learned synaptically.
We shall here use {\it short-term} memory (STM) to describe the brain's ability to store information on a timescale of seconds or less\footnote{We should mention that sensory memory is usually considered distinct from STM -- and probably has a different origin -- but we shall use ``short-term memory'' generically since the mechanism we propose in this paper could be relevant for either or both phenomena. On the other hand, the recent flurry of research in psychology and neuroscience on working memory has lead to this term sometimes being used to mean short-term memory; strictly speaking, however, working memory is generally considered to be an aspect of cognition which operates on information stored in STM.}.

Evidence that short-term memory is related to sensory information while long-term memory is more conceptual can again be found in psychology. For instance, a sequence of similar sounding letters is more difficult to retain for a short time than one of phonetically distinct ones, while this has no bearing on long-term memory, for which semantics seems to play the main role \citep{Conrad_1,Conrad_2}; and the way many of us think about certain concepts, such as chess, geometry or music, is apparently quite sensorial: we imagine positions, surfaces or notes as they would look or sound.
Most theories of short-term memory -- which almost always focus on working memory -- make use of some form of previously stored information (i.e., of synaptic learning) and so can account for the labelling tasks referred to above but not for the instant recall of novel information \citep{Wang, Barak, Roudi, Mongillo, Jorge}. Attempts to deal with the latter have been made by proposing mechanisms of \textit{cellular bistability}: neurons are assumed to retain the state they are placed in (such as firing or not firing) for some period of time thereafter \citep{Camperi, Fukai, Tarnow}.
Although there may indeed be subcellular processes leading to a certain bistability, the main problem with short-term memory depending exclusively on such a mechanism is that if each neuron must act independently of the rest the patterns will not be robust to random fluctuations \citep{Durstewitz} -- and the behaviour of individual neurons is known to be quite noisy \citep{Compte}. It is worth pointing out that one of the strengths of Associative Memory is that the behaviour of a given neuron depends on many neighbours and not just on itself, which means that robust global recall can emerge despite random fluctuations at an individual level. 

Something that, at least until recently, most neural network models have failed to take into account is the structure of the network -- its topology -- it often being assumed that synapses are placed  among the neurons completely at random, or even that all neurons are connected to all the rest (a mathematically convenient but unrealistic situation). Although relatively little is yet known about the architecture of the brain at the level of neurons and synapses, experiments have shown that it is heterogeneous (some neurons have very many more synapses than others), clustered (two neurons have a higher chance of being connected if they share neighbours than if not) and highly modular (there are groups, or modules, with neurons forming synapses preferentially to those in the same module) \citep{Sporns, Johnson_JSTAT}.
This chapter describes the main result of Ref. \citep{Johnson_CR} -- namely, 
that it suffices to use a more realistic topology, in particular one which is modular and/or clustered, for a randomly chosen pattern of activity the system is placed in to be metastable. This means that novel information can be instantly stored and retained for a short period of time in the absence of
both
synaptic learning
and
cellular bistability. The only requisite is that the patterns be coarse grained versions of the usual patterns -- that is, whereas it is
often assumed that each neuron 
in some way
represents one bit of information, we shall allocate a bit to a small group or neurons\footnote{This does not, of course, mean that memories are expected to be encoded as bitmaps. Just as with individual neurons, positions or orientations, say, could be represented by the activation of particular sets of clusters.} (four or five can be enough).

The mechanism, which we call Cluster Reverberation, is very simple. If neurons in a group are more highly connected to each other than to the rest of the network, either because they form a module or because the network is significantly clustered, they will tend to retain the activity of the group: when they are all initially firing, they each continue to receive many action potentials and so go on firing, whereas if they start off silent, there is not usually enough input current from the outside to set them off.
The fact that each neuron's state depends on its neighbours conferres to the mechanism a certain robustness in the face of random fluctuations. This robustness is particularly important for biological neurons, which as mentioned are quite noisy. Furthermore, not only does the limited duration of short-term memory states emerge naturally from this mechanism (even in the absence of interference from new stimuli) but this natural forgetting follows power-law statistics, as in experimental settings \citep{Wixted_1, Wixted_2, Sikstrom}.

The process is reminiscent both of block attractors in ordinary neural networks \citep{Dominguez}
and of domains in magnetic materials \citep{Hubert}, while Mu\~noz et al. have recently highlighted a similarity with Griffiths phases on networks \citep{Munoz}. It can also be interpreted as a multiscale phenomenon: the mesoscopic clusters take on the role usually played by individual neurons, yet make use of network properties.
Although the mechanism could also work in conjunction with other ones, such as synaptic learning or cellular bistability, we shall illustrate it by considering the simplest model which has the necessary ingredients: a set of binary neurons linked by synapses of uniform weight according to a topology whose modularity or clustering we shall tune. As with Associative Memory, this mechanism of Cluster Reverberation appears to be simple and robust enough not to be qualitatively affected by the complex subcellular processes incorporated into more realistic neuron models -- such as integrate-and-fire or Hodgkin-Huxley neurons. However, such refinements are probably needed to achieve graded persistent activity, since the mean frequency of each cluster could then be set to a certain value.


\section{The simplest neurons on modular networks}
\label{sec_model}

We consider a network of $N$
model neurons, with activities $s_{i}=\pm 1$.
The topology is given by the adjacency matrix $\hat{a}_{ij}=\lbrace 1,0\rbrace$, each element representing the existence or absence of a synapse from neuron $j$ to neuron $i$ ($\hat{a}$ need not be symmetric). In this kind of model, each edge usually has a \textit{synaptic weight} associated, $\omega_{ij}\in\mathbb{R}$; however, we shall here consider these to have all the same value: $\omega_{ij}=\omega$ $\forall i,j$. Neurons are updated in parallel (Little dynamics) at each time step, according to the stochastic transition rule
$$
P(s_{i}\rightarrow \pm1)=\pm \frac{1}{2} \tanh\left(\frac{h_{i}}{T}\right)+\frac{1}{2},
$$
where the \textit{field} of neuron $i$ is defined as
$$
h_{i}=\omega\sum_{j}^{N}\hat{a}_{ij}s_{j}
$$
and $T$ is a parameter we shall call \textit{temperature}.

First of all, we shall consider the network defined by $\hat{a}$ to be made up of $M$ distinct modules. To achieve this, we can first construct $M$ separate random directed networks, each with $n=N/M$ nodes and mean degree (mean number of neighbours) $\langle k\rangle$. Then we evaluate each edge and, with probability $\lambda$, eliminate it, to be substituted for another edge between the original post-synaptic neuron and a new pre-synaptic neuron chosen at random from among any of those in other modules\footnote{We do not allow self-edges (although these can occur in reality) since they can be regarded as a form of cellular bistability.}.
Note that this protocol does not alter the number of pre-synaptic neighbours of each node, $k^{in}_{i}=\sum_{j}\hat{a}_{ij}$ (although the number of post-synaptic neurons, $k^{out}_{i}=\sum_{j}\hat{a}_{ji}$, can vary).
The parameter $\lambda$ can be seen as a measure of \textit{modularity} of the partition considered, since it coincides with the expected value of the proportion of edges that link different modules. In particular, $\lambda=0$ defines a network of disconnected modules, while $\lambda=1-M^{-1}$ yields a random network in which this partition has no modularity. If $\lambda\in(1-M^{-1},1)$, the partition is less than randomly modular -- i.e., it is \textit{quasi-multipartite} (or multipartite if $\lambda=1$).

If the size of the modules is of the order of $\langle k\rangle$, the network will also be highly clustered. Taking into account that the network is directed, let us define the clustering coefficient $C_{i}$ as the probability, given that there is a synapse from neuron $i$ to a neuron $j$ and from another neuron $l$ to $i$, that there be a synapse from $j$ to $l$: that is, that there exist a feedback loop $i\rightarrow j\rightarrow l \rightarrow i$. Then, assuming $M\gg1$, the expected value of the clustering coefficient $C\equiv \langle C_{i}\rangle$ is
$$
 C \gtrsim \frac{\langle k\rangle-1}{n-1}(1-\lambda)^{3}.
$$

\section{Cluster Reverberation}
\label{sec_Cluster_Reverb}

\begin{figure}[t!]
\begin{center}
\includegraphics[
width=10cm]
{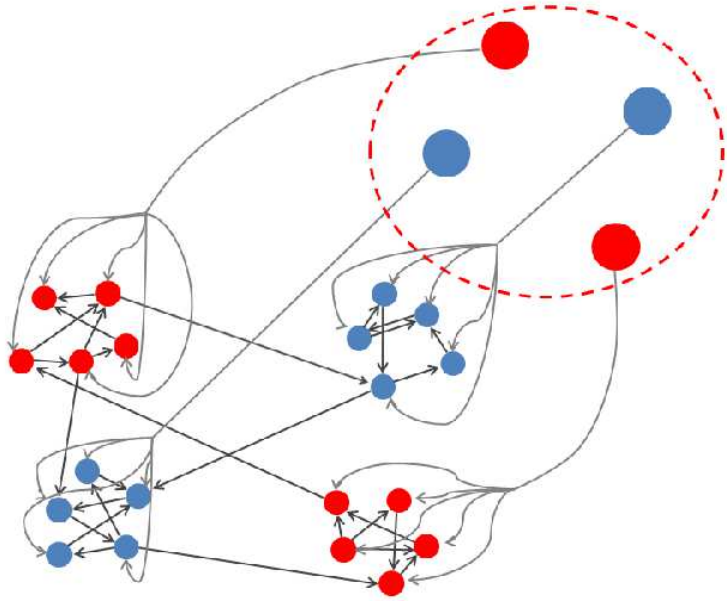}
\end{center}
\caption{Diagram of a modular network composed of four five-neuron clusters. The four circles enclosed by the dashed line represent the stimulus: each is connected to a particular module, which adopts the input state (red or blue) and retains it after the stimulus has disappeared via Cluster Reverberation.}
\label{fig_picture}
\end{figure}

A memory pattern, in the form of a given configuration of activities, $\lbrace \xi_{i} =\pm1\rbrace$, can be stored in this system with no need of prior learning. Imagine a pattern
such that the activities of all $n$ neurons found in any module are the same, i.e., $\xi_{i}=\xi_{\mu(i)}$, where the index $\mu(i)$ denotes the module that neuron $i$ belongs to. This can be thought of as a coarse graining of the standard idea of memory patterns, in which each neuron represents one bit of information. In our scheme, each module represents -- and stores -- one bit. The system can be induced into this configuration via the application of an appropriate \textit{stimulus} (see Fig. \ref{fig_picture}): the field of each neuron will be altered for just one time step according to
$$
h_{i}\rightarrow h_{i}+\delta \xi_{\mu(i)},\quad\forall i,
$$
where the factor $\delta$ is the intensity of the stimulus. This mechanism for dynamically storing information will work for values of parameters such that the system is sensitive to the stimulus, acquiring the desired configuration, yet also able to retain it for some interval of time thereafter.

The two main attractors of the system are $s_{i}=1$ $\forall i$ and $s_{i}=-1$ $\forall i$. These are the configurations of minimum energy (see the next section for a more detailed discussion on energy). However, the energy is locally minimised for any configuration in which $s_{i}=d_{\mu(i)}$ $\forall i$ with $d_{\mu}=\pm 1$; that is, configurations such that each module comprises either all active or all inactive neurons. These are the configurations that we shall use to store information. We define the mean activity\footnote{The mean activity in a neural network model is usually taken to represent the mean firing rate measured in experiments \citep{Torres_rev}.} of each module,
$$
m_{\mu}\equiv\frac{1}{n}\sum_{i\in \mu}^{n}s_{i},
$$
which is a mesoscopic variable, as well as the global mean activity,
$$
m\equiv\frac{1}{N}\sum_{i}^{N}s_{i}=\frac{1}{M}\sum_{\mu}^{M}m_{\mu}
$$
(these magnitudes change with time, but, where possible, we shall avoid writing the time dependence explicitely for clarity).
The extent to which the network, at a given time, retains the pattern $\lbrace\xi_{i}\rbrace$ with which it was stimulated is measured with the \textit{overlap} parameter
$$
m_{stim}\equiv\frac{1}{N}\sum_{i}^{N}\xi_{i}s_{i}=\frac{1}{M}\sum_{\mu}^{M}\xi_{\mu}m_{\mu}.
$$
Ideally, the system should be capable of reacting immediately to a stimulus by adopting the right configuration, yet also be able to retain it for long enough to use the information once the stimulus has disappeared. A measure of performance for such a task is therefore
$$
\eta\equiv\frac{1}{\tau}\sum_{t=t_{0}+1}^{t_{0}+\tau}m_{stim}(t),
$$
where $t_{0}$ is the time at which the stimulus is received and $\tau$ is the period of time we are interested in ($|\eta|\leq1$) \citep{Johnson_EPL}. If the intensity of the stimulus, $\delta$, is very large, then the system will always adopt the right pattern perfectly and $\eta$ will only depend on how well it can then retain it. In this case, the best network will be one
that is
made up of
unconnected modules. However, since the stimulus in a real brain can be expected to arrive via a relatively small number of axons, either from another part of the brain or directly from sensory cells, it is more realistic to assume that $\delta$ is of a similar order as the input a typical neuron receives from its neighbours, $\langle h\rangle \sim\omega \langle k\rangle$.

\begin{figure}[t!]
\begin{center}
\includegraphics[
width=10cm]{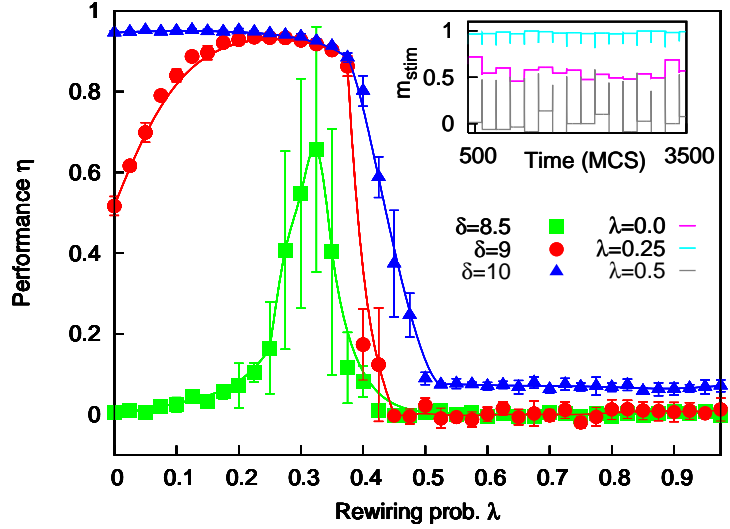}
\end{center}
\caption{Performance $\eta$ against $\lambda$ for networks of the sort described in the main text with $M=160$ modules of $n=10$ neurons, $\langle k\rangle=9$; patterns are shown with intensities $\delta=8.5$, $9$ and $10$, and
$T=0.02$
(lines -- splines -- are drawn as a guide to the eye). Inset: typical time series of $m_{stim}$ (i.e., the overlap with whichever pattern was last shown) for $\lambda=0.0$, $0.25$, and $0.5$, and $\delta=\langle k\rangle=9$.}
\label{fig_performance}
\end{figure}

Fig. \ref{fig_performance} shows the mean performance obtained when the network is repeatedly stimulated with different randomly generated patterns. For low enough values of the modularity $\lambda$ and stimuli of intensity $\delta \gtrsim \omega\langle k\rangle$, the system can capture and successfully retain any pattern it is ``shown'' for some period of time, even though this pattern was in no way previously learned. For less intense stimuli ($\delta<\omega\langle k\rangle$), performance is nonmonotonic with modularity: there exists an optimal value of $\lambda$ at which the system is sensitive to stimuli yet still able to retain new patterns quite well.

It is worth noting that performance can also break down due to thermal fluctuations. The two main attractors of the system ($s_{i}=1$ $\forall i$ and $s_{i}=-1$ $\forall i$) suffer the typical second order phase transition of the Hopfield model \citep{Hopfield}, from a 
memory phase (one in which $m=0$ is not stable and stable solutions $m\neq0$ exist) to
one with no memory (with $m=0$ the only stable solution), at the critical temperature \citep{Johnson_EPL}
$$
T_{c}=\omega\frac{\langle k_{in}^{2}\rangle}{\langle k\rangle}.
$$
(Note that, in a directed network, $\langle k_{in}\rangle=\langle k_{out}\rangle\equiv\langle k\rangle$, although the other moments can in general be different.)
The metastable states we are interested in, though, have a critical temperature
$$
T'_{c}=(1-\lambda)T_{c}
$$
(assuming that the mean activity of the network is $m\simeq0$). That is, the temperature at which the modules are no longer able to retain their individual activity is in general lower than that at which the
the solution $m=0$ for the whole network becomes stable.

\section{Energy and topology}
\label{sec_energy}

\begin{figure}[t!]
\begin{center}
\includegraphics[
width=10cm]{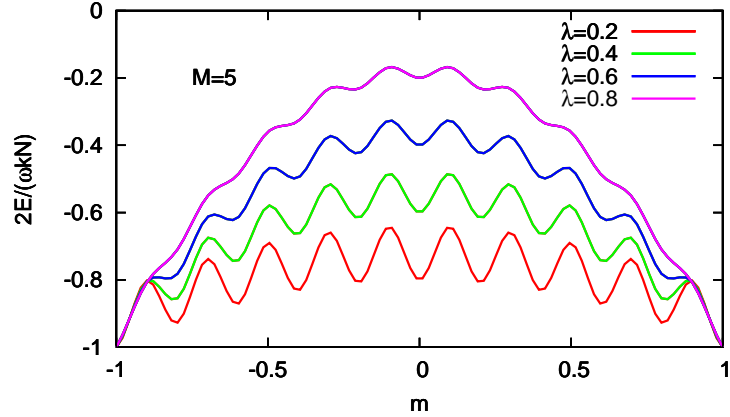}
\end{center}
\caption{
Configurational energy of a network composed of $M=20$ modules of $n=10$ neurons each, according to Eq. (\ref{E}), for various values of the rewiring probability $\lambda$. The minima correspond to situations such that all neurons within any given module have the same sign.
}
\label{fig_energy}
\end{figure}

Each pair of nodes contributes a configurational energy $e_{ij}=-\omega\frac{1}{2} (\hat{a}_{ij}+\hat{a}_{ji})s_{i}s_{j}$; that is, if there is an edge from $i$ to $j$ and  they have opposite activities, the energy is increased in $\frac{1}{2}\omega$, whereas it is decreased by the same amount if their activities are the same. Given a configuration, we can obtain its associated energy  by summing over all pairs. We shall be interested in configurations with $x$ neurons that have $s=+1$ (and $N-x$ with $s=-1$), chosen in such a way that one module at most, say $\mu$, has neurons in both states simultaneously. Therefore, $x=n\rho+z$, where $\rho$ is the number of modules with all their neurons in the positive state and $z$ is the number of neurons with positive sign in module $\mu$. We can write $m=(2x-1)/N$ and $m_{\mu}=(2z-1)/n$. The total configurational energy of the system will be $E=\sum_{ij}e_{ij}=\frac{1}{2}\omega(L_{\uparrow\downarrow}-\langle k\rangle N)$, where $L_{\uparrow\downarrow}$ is the number of edges linking nodes with opposite activities. By simply counting over edges, we can obtain the expected value of $L_{\uparrow\downarrow}$ (which amounts to a mean-field approximation because we are substituting the number of edges between two neurons for its expected value), yielding:
\begin{eqnarray}
\frac{E}{\omega\langle k\rangle}= (1-\lambda)\frac{z(n-z)}{n-1}
\nonumber
\\
+\frac{\lambda n}{N-n}\left\lbrace \rho[n-z+n(M-\rho-1)]+(M-\rho-1)(z+n\rho)\right\rbrace
-\frac{1}{2}N.
\label{E}
\end{eqnarray}
Fig. \ref{fig_energy} shows the mean-field configurational energy curves for various values of the modularity on a small modular network. The local minima (metastable states) are the configurations used to store patterns. It should be noted that the mapping $x\rightarrow m$ is highly degenerate: there are $C_{mM}^{M}$ patterns with mean activity $m$ that all have the same energy.

\section{Forgetting avalanches}
\label{sec_avalanches}

In obtaining the energy we have assumed that the number of synapses rewired from a given module is always $\nu=\langle k\rangle n\lambda$. However, since each edge is evaluated with probability $\lambda$, $\nu$ will in fact vary somewhat from one module to another, being approximately Poisson distributed with mean $\langle \nu\rangle=\langle k\rangle n\lambda$. The depth of the energy well corresponding to a given module is then, neglecting all but the first term in Eq. (\ref{E}) and approximating $n-1\simeq n$,
$$
\Delta E\simeq \frac{1}{4}\omega(n\langle k\rangle -\nu).
$$

\begin{figure}[t!]
\begin{center}
\hspace{-0.7cm}\includegraphics[
width=13.5cm]{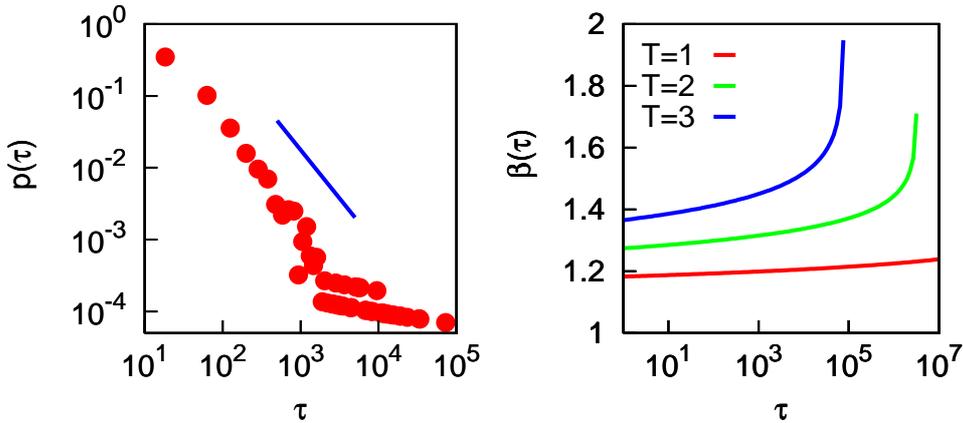}
\end{center}
\caption{
Left panel: distribution of escape times $\tau$, as defined in the main text, for $\lambda=0.25$ and $T=2$. Slope is for $\beta=1.35$. Other parameters as in Fig. \ref{fig_performance}. Symbols from MC simulations and line given by Eqs. (\ref{eq_Pt}) and (\ref{eq_beta}). Right panel: exponent $\beta$ of the quasi-power-law distribution $p(\tau)$ as given by Eq. (\ref{eq_beta}) for temperatures $T=1$ (red line), $T=2$ (green line) and $T=3$ (blue line). 
}
\label{fig_aval_r}
\end{figure}

The typical escape time $\tau$ from an energy well of depth $\Delta E$ at temperature $T$ is $\tau\sim e^{\Delta E/T}$ \citep{Levine}. Using Stirling's approximation in the Poissonian distribution of $\nu$ and expressing it in terms of $\tau$, we find that the escape times are distributed according to
\begin{equation}
P(\tau)\sim \left(1 -\frac{4T}{\omega n\langle k\rangle}
\ln\tau\right)^{-\frac{3}{2}}\tau^{-\beta(\tau)},
\label{eq_Pt}
\end{equation}
where
\begin{equation}
\beta(\tau)=1+\frac{4T}{\omega n\langle k\rangle}\left[1+\ln\left(\frac{\lambda n\langle k\rangle}{1-\frac{4T}{\omega n\langle k\rangle}\ln\tau}  \right) \right].
\label{eq_beta}
\end{equation}
Therefore, at low temperatures, $P(\tau)$ will behave approximately like a power-law.
The left panel of Fig. \ref{fig_aval_r} shows the distribution of time intervals between events in which the overlap $m_{\mu}$ of at least one module $\mu$ changes sign. The power-law-like behaviour is apparent, and justifies talking about \textit{forgetting avalanches} -- since there are cascades of many forgetting events interspersed with long periods of metastability.
This is very similar to the behaviour observed in other nonequilibrium settings in which power-law statistics arise from the convolution of exponentials \citep{Hurtado, Munoz}.

It is known from experimental psychology that forgetting in humans is indeed well described by power-laws \citep{Wixted_1, Wixted_2, Sikstrom}.
The right panel of Fig. \ref{fig_aval_r} shows the value of the exponent $\beta(\tau)$ as a function of $\tau$. Although for low temperatures it is almost constant over many decades of $\tau$ -- approximating a pure power-law -- for any finite $T$ there will always be a $\tau$ such that the denominator in the logarithm of Eq. (\ref{eq_beta}) approaches zero and $\beta$ diverges, signifying a truncation of the distribution.

\section{Clustered networks}
\label{sec_clustered_nets}

Although we have illustrated how the mechanism of Cluster Reverberation works on a modular network, it is not actually necessary for the topology to have this characteristic -- only for the patterns to be in some way ``coarse-grained,'' as described, and that each region of the network encoding one bit have a small enough parameter $\lambda$, defined as the proportion of synapses to other regions. 
For instance, for the famous Watts-Strogatz \textit{small-world} model \citep{Watts} -- a ring of $N$ nodes, each initially connected to its $k$ nearest neighbours before a proportion $p$ of the edges are randomly rewired -- we have $\lambda\simeq p$ (which is not surprising considering the resemblance between this model and the modular network used above). More precisely, the expected modularity of a randomly imposed box of $n$ neurons is
$$
\lambda=p-\frac{n-1}{N-1}p+\frac{1-p}{n}\left(\frac{k}{4}-\frac{1}{2}\right),
$$
the second term on the right accounting for the edges rewired to the same box, and the third to the edges not rewired but sufficiently close to the border to connect with a different box.

Perhaps a more realistic model of clustered network would be a random network 
embedded in $d$-dimensional Euclidean space. For this we shall use the scheme laid out by Rozenfeld \textit{et al.} \citep{Rozonfeld}, which consists simply in allocating each node to a site on a $d$-torus and then, given a particular degree sequence, placing edges to the nearest nodes possible -- thereby attempting to minimise total edge length\footnote{The authors also consider a cutoff distance, but we shall take this to be infinite here.}. For a scale-free degree sequence (i.e., a set $\lbrace k_{i}\rbrace$ drawn from a degree distribution $p(k)\sim k^{-\gamma}$) according to some exponent $\gamma$, then, as shown in \ref{Appendix_CR}, such a network has a modularity
\begin{equation}
\lambda\simeq\frac{1}{d(\gamma-2)-1}\left[d(\gamma-2)l^{-1}-l^{-d(\gamma-2)}\right],
\label{eq_lambda}
\end{equation}
where $l$ is the linear size of the boxes considered.

\begin{figure}[t!]
\begin{center}
\includegraphics[
width=10cm
]{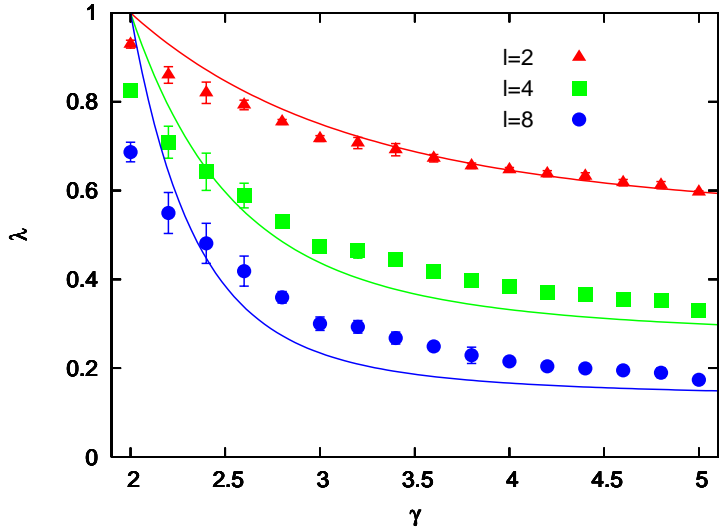}
\end{center}
\caption{
Proportion of outgoing edges, $\lambda$, from boxes of linear size $l$ against exponent $\gamma$ for scale-free networks embedded on $2D$ lattices. Lines from Eq. (\ref{eq_lambda})
and symbols from simulations with $\langle k\rangle=4$ and $N=1600$.
}
\label{fig_mod_new4}
\end{figure}

Fig. \ref{fig_mod_new4} compares this expression with the value obtained numerically after averaging over many network realizations, and shows that it is fairly good -- considering the approximations used for its derivation. It is interesting that even in this scenario, where the boxes of neurons which are to receive the same stimulus are chosen at random with no consideration for the underlying topology, these boxes need not have very many neurons for $\lambda$ to be quite low (as long as the degree distribution is not too heterogeneous).

\begin{figure}[t!]
\begin{center}
\includegraphics[
width=10cm]{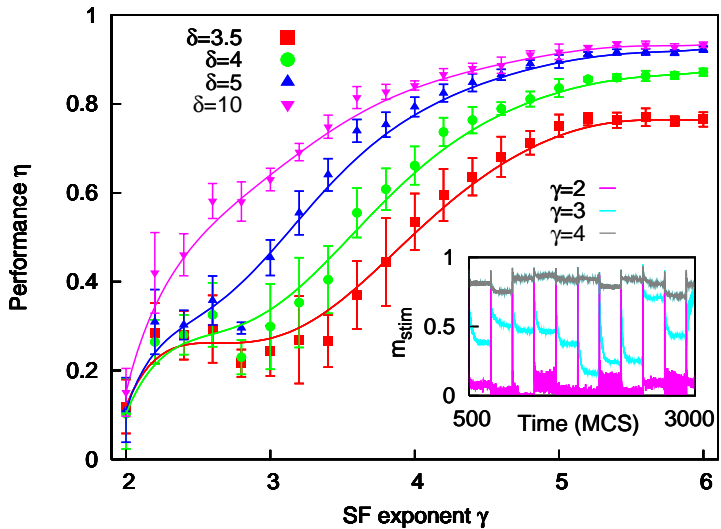}
\end{center}
\caption{Performance $\eta$ against exponent $\gamma$ for scale-free networks, embedded on a 2D lattice, with patterns of $M=16$ modules of $n=100$ neurons each, $\langle k\rangle=4$ and $N=1600$; patterns are shown with intensities $\delta=3.5$, $4$, $5$ and $10$, and $T=0.01$
(lines -- splines -- are drawn as a guide to the eye). Inset: typical time series for $\gamma=2$, $3$, and $4$, with $\delta=5$.}
\label{fig_serie_SF}
\end{figure}

Carrying out the same repeated stimulation test as on the modular networks in Fig. \ref{fig_performance}, we find a similar behaviour for the scale-free embedded networks. This is shown in Fig. \ref{fig_serie_SF}, where for high enough intensity of stimuli $\delta$ and scale-free exponent $\gamma$, performance can, as in the modular case, be $\eta\simeq1$.
We should point out that for good performance on these networks we require more neurons for each bit of information than on modular networks with the same $\lambda$ (in Fig. \ref{fig_serie_SF} we use $n=100$, as opposed to $n=10$ in Fig. \ref{fig_performance}).
However, that we should be able to obtain good results for such diverse network topologies underlines that the mechanism of Cluster Reverberation is robust and not dependent on some very specific architecture. In fact, we have recently shown that similar metastable memory states can also occur on networks which have random modularity and clustering, but a certain degree of \textit{assortativity}\footnote{The assortativity of a network is here understood to mean the extent to which the degrees of neighbouring nodes are correlated \citep{Johnson_PRL}.} \citep{Sebas}.


\section{Yes, but does it happen in the brain?}
\label{sec_discu}

As we have shown, Cluster Reverberation is a mechanism available to neural systems for robust short-term memory without synaptic learning. To the best of our knowledge, this is the first mechanism proposed which has these characteristics -- essential for, say,
sensory memory or certain working-memory tasks.
All that is needed is for the network topology to be highly clustered or modular, and for small groups of neurons to store one bit of information, as opposed to the conventional view which assumes one bit per neuron. Considering the enormous number of neurons in the brain, and the fact that real individual neurons are probably too noisy to store information reliably, these hypotheses do not seem farfetched. The mechanism is furthermore consistent both with what is known about the topology of the brain, and with experiments which have revealed power-law forgetting.


Since the purpose of this paper is only to describe the mechanism of Cluster Reverberation, we have made use of the simplest possible model neurons -- namely, binary neurons with static, uniform synapses -- for the sake of clarity and generality. However, there is no reason to believe that the mechanism would cease to function if more neuronal ingredients were to be incorporated. In fact, cellular bistability, for instance, would increase performance, and the two mechanisms could actually work in conjunction. Similarly, we have also used binary patterns to store information. But it is to be expected that patterns depending on any form of frequency coding, for instance, could also be maintained with more sophisticated neurons -- such that different modules could be set to different mean frequencies. 

Whether Cluster Reverberation would work for biological neural systems could be put to the test by growing such modular networks {\it in vitro}, stimulating appropriately, and observing the duration of the metastable states. {\it In vivo} recordings of neural activity during short-term memory tasks, together with a mapping of the underlying synaptic connections, could be used to ascertain whether the brain does indeed make use of this mechanism -- although for this it must be borne in mind that the neurons forming a module need not find themselves close together in metric space. We hope that experiments such as these will be carried out and eventually reveal something more about the basis of this puzzling emergent property of the brain's known as thought.






%% file: con/con.tex

\chapter{Concluding remarks} 
\label{Chapter_con}


``As long as the brain is a mystery, the universe will remain a mystery,'' claimed Santiago Ram\'on y Cajal. Our very essence seems to reside somehow in the workings of this organ, probably as a consequence of electro-chemical signalling that goes on among its hundred billion or so constituent neurons. Will this mystery ever be cleared up? We know of other objects that process information in highly sophisticated ways -- electronic computers. Faced with a sudden blue screen, one may be forgiven for calling these devices incomprehensible and capricious, even malevolent. But in fact most educated people understand, on some level at least, what mechanisms and physical processes are behind the complex behaviour displayed by computers, and do not consider the issue a mystery. This is not to suggest that the analogy between brain and computer should be taken any further than to illustrate how a great many elements, each executing some fairly simple and obvious operation, can ``cooperate'' to yield astonishingly complicated yet functional behaviour; and that one can grasp how this occurs without having to know every detail. But we have not yet reached this point as regards the brain. Much progress has been made concerning aspects of physiology, while once unassailable mental disorders such as phobias can now be easily cured by psychology. Yet as far as what mechanisms relate these two levels of description goes, perhaps all we can safely say for now is that synaptic plasticity is responsible for long-term memory. The origins of even some well-defined and much studied cognitive abilities -- such as probabilistic reasoning or short-term memory -- remain somewhat elusive, while the nature of consciousness, say, is still truly a mystery. However, if instead of developing computers ourselves we had been given them by an alien species, we could still hope one day to unravel the mysteries of their magic. In much the same manner, by searching for ways in which collections of neurons might perform tasks such as we know them to be capable of, we will some day understand not only how our stomachs digest and our hearts pump, but also how our brains think.

I cannot pretend that the work described here takes us more than, at best, a tiny step of the way along this path. The brain is, among other things, a network, and networks are a kind of mathematical object about which we now know much more than just a few years ago. In fact, they are a central element of what can arguably be called the most challenging frontier currently facing human understanding about the world -- the nature of complex systems. So, from among the innumerable aspects likely to shape and determine the way neurons cooperate, the research
presented here
focuses on the structure of the underlying network. First of all it looks at how this structure can develop. Chapter \ref{Chapter_JSTAT} addressed this by formalizing as a stochastic process a situation governed by probabilistic events like synaptic growth and death. Such simple individual behaviour was shown to be enough to explain many statistical features of real neural systems.
Furthermore, this Fokker-Planck description relating microscopic, stochastic actions to a macroscopic evolution of properties such as mean synaptic density, degree heterogeneity or assortativity may help to gain insights into the biochemical processes taking place.

The rest of the thesis is mostly devoted to how aspects of a neural network's topology might influence or even determine its ability to carry out certain tasks akin to those the brain undertakes. The fact that dynamical memory performance ensuing from synaptic depression is favoured by a highly heterogeneous degree distribution, laid out in Chapter \ref{Chapter_EPL}, may help to explain why the brain seems to display such a topology at several levels of description -- perhaps somehow maintaining itself close to a critical point. Similarly, the enhanced robustness to noise found for positively correlated networks in Chapter \ref{Chapter_PRE} suggests a functional advantage to a neural network being thus wired; a prediction also in agreement with some experimental findings.

As far as unearthing the mechanisms underpinning how neurons can perform cognitive tasks goes, though, perhaps the most interesting idea proposed is that of Cluster Reverberation, in Chapter \ref{Chapter_CR}, whereby thanks to modularity and/or clustering a neural network is able to store information instantly, without requiring biochemical changes in the synapses. Time will tell whether real neural systems do indeed harness this mechanism to perform certain short-term memory tasks.

A collateral but noteworthy aspect of this research is the potentiality for application elsewhere of some of the mathematical techniques developed. Most of all, the method for studying correlated networks and dynamics thereon put forward in Chapter \ref{Chapter_PRL} for use in Chapter \ref{Chapter_PRE} can be expected to find widespread use. The answer to the question of why most networks are disassortative given in Chapter \ref{Chapter_PRL}, or the relation between degree-degree correlations and nestedness described in Appendix \ref{Appendix_NEST} are examples of this.

Finally I must mention not just the answers I hope to have provided, or at least hinted at, to some unsolved problems, but also the questions that have been posed and challenges laid bare: Would a more detailed description of brain development still be possible with Fokker-Planck equations? Are these topological effects, found to be at work for the simplest neural models, indeed so relevant for real neurons? Can Cluster Reverberation be performed {\it in vitro}? The greatest function this thesis could perform would be to stimulate others to look into these or related issues in more depth than here. But I also hope it may serve to illustrate the sentiment, What matters how long the path to the final unravelling of the mysteries is, as long as the going is fun?

%% file: con_es/con_es.tex
\chapter{Conclusiones en espa\~nol} 
\label{Chapter_con_es}

``Mientras el cerebro sea un misterio, el universo continuar\'a siendo un misterio'', dijo una vez Santiago Ram\'on y Cajal. Parece que nuestra misma esencia reside de alguna manera en el funcionamiento de este \'organo, probablemente como consecuencia de las se\~nales electro-qu\'imicas entre sus aproximadamente cien mil millones de neuronas. ?`Se resolver\'a alg\'un d\'ia este misterio? Conocemos otros objetos capaces de procesar informaci\'on de manera altamente sofisticada: los ordenadores electr\'onicos. Confrontados con un pantallazo azul, se nos podr\'ia perdonar el tildar este tipo de aparatos de incomprensibles y caprichosos, por no decir mal\'evolos. Pero en realidad la mayor parte de la gente entiende, al menos en alg\'un nivel, cu\'ales son los mecanismos y procesos f\'isicos que subyacen el comportamiento complejo del que hacen gala los ordenadores, y no consideran que el tema sea un misterio. No es que la analog\'ia entre cerebro y ordenador deba ser llevado m\'as lejos que para ilustrar c\'omo muchos elementos, cada uno ejecutando alguna operaci\'on relativamente simple y obvia, pueden ``cooperar'' y mostrar un comportamiento colectivo asombrosamente complicado, pero funcional;
y que se puede comprender c\'omo ocurre esto sin necesidad de conocer hasta el \'ultimo detalle.
A\'un no hemos llegado  a poder responder  a esta pregunta en lo que respecta al cerebro. Hemos ampliado enormemente nuestro conocimiento de aspectos fisiol\'ogicos, y trastornos mentales anta\~no incurables, como las fobias, son f\'acilmente tratadas hoy en d\'ia por la psicolog\'ia. En cuanto a los mecanismos que relacionan estos dos niveles de descripci\'on, posiblemente lo \'unico que podamos decir a ciencia cierta es que la plasticidad sin\'aptica est\'a detr\'as de la memoria a largo plazo. Los or\'igenes incluso de algunas habilidades cognitivas bien definidas y extensamente estudiadas, como el razonamiento probabil\'istico o la memoria a corto plazo, est\'an a\'un por descifrar completamente; mientras que, por ejemplo, la naturaleza de la consciencia es verdaderamente a\'un un misterio. Sin embargo, si en lugar de haber desarrollado los ordenadores nosotros mismos los hubi\'esemos recibido de una especie alien\'igena, a\'un as\'i podr\'iamos esperar alg\'un d\'ia desenmara\~nar los misterios de su magia. Del mismo modo, buscando maneras de que conjuntos de neuronas puedan realizar el tipo de tareas de las que las sabemos capaces, alg\'un d\'ia entenderemos no s\'olo c\'omo nuestros est\'omagos digieren y nuestros corazones laten, sino tambi\'en c\'omo nuestros cerebros piensan.

Este trabajo, en el mejor de los casos, nos avanza un paso infinitesimal por este camino. El cerebro es, entre otras muchas cosas, una red, y las redes son objetos matem\'aticos sobre los que sabemos hoy mucho m\'as que hace tan s\'olo unos a\~nos. De hecho, son un elemento fundamental para uno de los mayores retos con los que se enfrenta actualmente el conocimiento humano: la naturaleza de los sistemas complejos. As\'i que, de entre los innumerables aspectos susceptibles de modificar y determinar c\'omo las neuronas cooperan, esta investigaci\'on se centra en la estructura de la red subyacente. Primero analiza c\'omo dicha estructura puede desarrollarse. El Cap\'itulo \ref{Chapter_JSTAT} enfoca esto formalizando mediante la teor\'ia de los procesos estoc\'asticos una situaci\'on gobernada por eventos probabil\'isticos tales como el crecimiento y la muerte sin\'apticas. Se demuestra que este tipo de comportamiento individual es suficiente para explicar muchas propiedades estad\'isticas de las redes de cerebros reales. Por otra parte, este marco te\'orico
puede ser reducido a una descripci\'on en t\'erminos de ecuaciones de Fokker-Planck, que relacionan acciones microsc\'opicas estoc\'asticas con la evoluci\'on macrosc\'pica de propiedades como la densidad sin\'aptica media, la heterogeneidad de la distribuci\'on de grados o la asortatividad, que quiz\'as nos permita extraer informaci\'on relevante acerca de los procesos bioqu\'imicos involucrados.

La mayor parte del resto de la tesis trata de c\'omo aspectos de la topolog\'ia de una red neuronal pueden influenciar o incluso determinar su habilidad para ejecutar ciertas tareas cognitivas como las que se describen en un cerebro o medio neuronal real. Por ejemplo, el hecho de que, en cuanto a la memoria din\'amica que emerge gracias a la depresi\'on sin\'aptica, el rendimiento es mayor para una distribuci\'on de grados altamente heterog\'enea, como demuestra el Cap\'itulo \ref{Chapter_EPL}, podr\'ia ayudar a explicar por qu\'e el cerebro parece mostrar una topolog\'ia de este tipo en varios niveles de descripci\'on, quiz\'as incluso manteni\'endo su actividad, de alguna manera todav\'ia no comprendida del todo, cerca de un punto cr\'itico. De igual modo, la mayor robustez durante los procesos cognitivos en presencia de ruido en el caso de redes con correlaciones
positivas como se ha descrito en el Cap\'itulo \ref{Chapter_PRE} sugiere que existe una ventaja funcional para una red neuronal en adoptar esta propiedad; una predicci\'on que tambi\'en encaja con algunos hallazgos experimentales.

En lo que se refiere a desentra\~nar los mecanismos que permiten a las neuronas realizar colectivamente tareas cognitivas, quiz\'as la idea m\'as interesante aqu\'i propuesta es la de {\it Cluster Reverberation} (Reverberaci\'on de Grupo), en el Cap\'itulo \ref{Chapter_CR}, seg\'un la cual, gracias a la modularidad y/o el grado de ``agrupamiento'', una red neuronal es capaz  de almacenar informaci\'on instant\'aneamente, sin requerir para ello cambios bioqu\'imicos de potenciaci\'on o depresi\'on a largo plazo en las sinapsis. El tiempo dir\'a si el cerebro aprovecha realmente este mecanismo para realizar ciertas tareas de memoria de corto plazo.

Un aspecto colateral pero digno de menci\'on de este trabajo es el de la potencialidad de algunas de las t\'ecnicas matem\'aticas desarrolladas de ser aplicadas para otras situaciones de inter\'es. Sobre todo, es de esperar que el m\'etodo para estudiar redes correlacionadas, y din\'amicas sobre ellas, propuesto en el Cap\'itulo \ref{Chapter_PRL} y utilizado en el Cap\'itulo \ref{Chapter_PRE}, sea \'util para una amplia gama de problemas. La respuesta, en el Cap\'itulo \ref{Chapter_PRL}, a la pregunta de por qu\'e la mayor\'ia de las redes son disasortativas, o la relaci\'on entre correlaciones entre los nodos y el ``anidamiento'' descrita en el Ap\'endice \ref{Appendix_NEST} son ejemplos de aplicaciones.

Finalmente, hay que mencionar no s\'olo las respuestas que se han intentado dar, o al menos sugerir, con esta tesis para algunos problemas sin resolver, sino tambi\'en las preguntas y los nuevos retos que han surgido: por ejemplo, ?`ser\'ia posible, tambi\'en con ecuaciones de Fokker-Planck, una descripci\'on m\'as detallada del desarrollo cerebral? ?`Son estos efectos topol\'ogicos, descritos para los modelos neuronales m\'as sencillos, realmente tan relevantes para neuronas de verdad? ?`Puede el mecanismo de {\em Cluster Reverberation} ocurrir {\it in vitro}? En definitiva, la mayor funci\'on
que pudiera cumplir esta tesis ser\'ia la de estimular a otra/os para que indaguen en estos y otros temas m\'as profundamente que aqu\'i. Pero quiz\'as tambi\'en sirva para ilustrar el siguiente sentimiento: ?`qu\'e m\'as da cu\'an largo sea el camino hacia el desenmara\~namiento \'ultimo de los misterios, siempre que el trayecto sea divertido?

%% file: rap/rap.tex
\chapter{Nonlinear preferential rewiring in fixed-size networks as a diffusion process} 
\label{Appendix_RAP}


We present an evolving network model in which the total numbers of nodes and
edges are conserved, but in which edges are continuously rewired according to
nonlinear preferential detachment and reattachment. Assuming power-law kernels
with exponents $\alpha$ and $\beta$, the stationary states the degree
distributions evolve towards exhibit a second order phase transition -- from
relatively homogeneous to highly heterogeneous (with the emergence of starlike
structures) at $\alpha=\beta$. Temporal evolution of the distribution in this
critical regime is shown to follow a nonlinear diffusion equation, arriving at
either pure or mixed power-laws, of exponents $-\alpha$ and $1-\alpha$.




Complex systems may often be described as a set of nodes with edges connecting some of them
-- the \textit{neighbours --} (see, for instance, Refs.\citep{Boccaletti,Arenas_rev,Marro_complex}).
The number of edges a particular node has is called its degree, $k$. The study of such large 
networks is usually made simpler by considering statistical properties, e.g.,
the degree distribution, $p(k)$ (probability of finding a node with a
particular degree). It turns out that a high proportion of real-world networks
follow power-law degree distributions, $p(k)\sim k^{-\gamma}$ -- referred to
as \textit{scale-free} due to their lack of a characteristic size. Also, many
of them have their edges placed among the nodes apparently in a random way
-- i.e., there is no correlation between the degree of a node and any other of
its properties, such as the degrees of its neighbours. Barab\'{a}si and Albert \citep{Barabasi} applied the mechanism of \textit{preferential attachment} to an evolving network model and showed how this resulted in the degree distributions becoming scale-free for long enough times. For this to work, attachment had to be linear -- i.e., the probability a node with degree $k$ has of receiving a new edge is $\pi(k)\sim k+q$. This results in scale-free stationary degree distributions with an exponent $\gamma=3-q$.

Preferential attachment seems to be behind the emergence of many real-world,
continuously growing networks. However, not all networks in which some nodes at times gain (or loose) new edges have a continuously growing number of nodes. For example, a given group
of people may form an evolving social network \citep{Kossinets} in which the
edges represent friendship. Preferential attachment may be relevant here -- the more people you know, the more likely it is that you will be introduced
to someone new -- but probabilities are not expected to depend linearly on
degree. For instance, there may be saturations (highly connected people might
become less accessible), threshold effects (hermits may be prone to antisocial
tendencies), and other non-linearities. The brain may also be a relevant case.
Once formed, the number of neurons does not seem to continually augment, and
yet its structural topology is dynamic \citep{Klintsova}. Synaptic growth and
dendritic arborization have been shown to increase with electric stimulation
\citep{Lee,DeRoo} -- and, in general, the more connected a neuron is, the more
current it receives from the sum of its neighbours.

Barab\'{a}si and Albert showed that both (linear) preferential attachment and
an ever-growing number of nodes are needed for scaling to emerge in their
model. In a fixed population, their mechanism would result in a
fully-connected network. However, this is not normally observed in real
systems. Rather, just as some new edges sprout, others disappear -- less used
synapses suffer atrophy, unstimulating friendships wither. Often, the numbers
of both nodes and edges remain roughly constant. The same authors did
therefore extend their model so as to include the effects of
\textit{preferential rewiring} (which could be applied to fixed-size
networks), although again probabilities depended linearly on node degree
\citep{Albert}. Another mechanism which (roughly) maintains constant the
numbers of nodes and edges is node fusing \citep{Thurner}, once more according
to linear probabilities. As to nonlinear preferential attachment, the
(growing) BA model was extended to take power-law probabilities into account
\citep{Krapivsky}, although the solutions are only scale free for the linear case.

In this note we present an evolving network model with preferential rewiring
according to nonlinear (power-law) probabilities. The number of nodes and
edges is conserved but the topology evolves, arriving eventually at a
macroscopically (nonequilibrium) stationary state -- as described by global
properties such as the degree distribution. Depending on the exponents chosen
for the rewiring probabilities, the final state can be either fairly
homogeneous, with a typical size, or highly heterogeneous, with the emergence
of starlike structures. In the critical case marking the transition between
these two regimes, the degree distribution is shown to follow a nonlinear
diffusion equation. This describes a tendency towards stationary states that are
characterized either by scale-free or by mixed scale-free distributions,
depending on parameters.

Our model consists of a random network with $N$ nodes of respective degree
$k_{i},$ $i=1,2,...,N,$ and $%
\frac12 N\left\langle k\right\rangle $ edges. Initially, the degrees have a given
distribution $p(k,t=0)$. At each time step, one node is chosen with a
probability which is a function of its degree, $\rho(k_{i})$. One of its edges
is then chosen randomly and removed from it, to be reconnected to another node
$j$ chosen according to a probability $\pi(k_{j})$. That is, an edge is broken
and another one is created, and the total number of edges, as well as the
total number of nodes, is conserved. The functions $\pi(k)$ and $\rho(k)$ are
arbitrary, but we shall explicitly illustrate here $\pi(k_{i})\sim
k_{i}^{\alpha}$ and $\rho(k_{i})\sim k_{i}^{\beta}$ that capture the essence
of a wide class of nonlinear monotonous response functions and are easy to
handle analytically.

The probabilities $\pi$ and $\rho$ a given node has, at each time step, of
increasing or decreasing its degree can be interpreted as transition
probabilities between states. The expected value of the increment in a given
$p(k,t)$ at each time step, $\Delta p(k,t)$, may then be written as%

\begin{align}
\frac{\partial p(k,t)}{\partial t}  &  =(k-1)^{\alpha}\,\bar{k}_{\alpha}%
^{-1}p(k-1,t)\nonumber\\
&  +(k+1)^{\beta}\,\bar{k}_{\beta}^{-1}\,p(k+1,t)\label{master}\\
&  -\left(  k^{\alpha}\,\bar{k}_{\alpha}^{-1}+k^{\beta}\,\bar{k}_{\beta}%
^{-1}\right)  \,p(k,t),\nonumber
\end{align}
where $\bar{k}_{a}=\bar{k}_{a}\left(  t\right)  =\sum_{k}k^{a}p(k,t).$
If it exists, any stationary solution must satisfy the condition
$p_{\text{st}}(k+1)\,(k+1)^{\beta}\,\bar{k}_{\alpha}^{\text{st}}=p_{\text{st}%
}(k)\,k^{\alpha}\,\bar{k}_{\beta}^{\text{st}}$ which, for $k\gg 1$, implies
that
\begin{equation}
\frac{\partial p_{\text{st}}(k)}{\partial k}=\left(  \frac{\bar{k}_{\alpha
}^{\text{st}}}{\bar{k}_{\beta}^{\text{st}}}\frac{k^{\alpha}}{(k+1)^{\beta}%
}-1\right)  p_{\text{st}}(k). \label{det_bal}%
\end{equation}
Therefore, the distribution will have an extremum at 
$k_{e}=\left(  \bar
{k}_{\beta}^{\text{st}}/\bar{k}_{\alpha}^{\text{st}}\right)  ^{\frac{1}%
{\alpha-\beta}}$ (where we have approximated $k_{e}\simeq k_{e}+1$). If
$\alpha<\beta$, this will be a maximum, signalling the peak of the
distribution. On the other hand, if $\alpha>\beta$, $k_{e}$ will correspond to
a minimum. Therefore, most of the distribution will be broken in two parts,
one for $k<k_{e}$ and another for $k>k_{e}$. The critical case for
$\alpha=\beta$ will correspond to a monotonously decreasing stationary
distribution, but such that $\mbox{lim}_{k\rightarrow\infty}\partial
p_{st}(k)/\partial k=0$. In fact, Eq. (\ref{master}) is for this situation
($\alpha=\beta$) the discretized version of a nonlinear diffusion equation,
\begin{equation}
\frac{\partial p(k,\tau)}{\partial\tau}=\frac{\partial^{2}}{\partial k^{2}%
}[k^{\alpha}p(k,\tau)], \label{diff}%
\end{equation}
after dynamically modifying the time scale according to $\tau=t/\bar
{k}_{\alpha}\left(  t\right)  $. Ignoring, for the moment, border effects, the
solutions of this equation are of the form
\begin{equation}
p_{\text{st}}(k)\sim Ak^{-\alpha}+Bk^{-\alpha+1}, \label{sol}%
\end{equation}
with $A$ and $B$ constants. If $\alpha>2$, then given $A$ we can always find a
$B$ which allows $p_{\text{st}}(k)$ to be normalized in the thermodynamic
limit \footnote{Although all moments of $k$ will diverge unless $B=0$.}. For example, if the lower limit is $k\geq1$, then
$B=(\alpha-2)\left[  1-A/(\alpha-1)\right]  $. However, if $1<\alpha\leq2$,
then only $A$ can remain non-zero, and $p_{\text{st}}(k)$ will be a pure power
law. For $\alpha\leq1$, both constants must tend to zero as $N\rightarrow
\infty$.%
\begin{figure}[t!]
\begin{center}
\epsfig{file=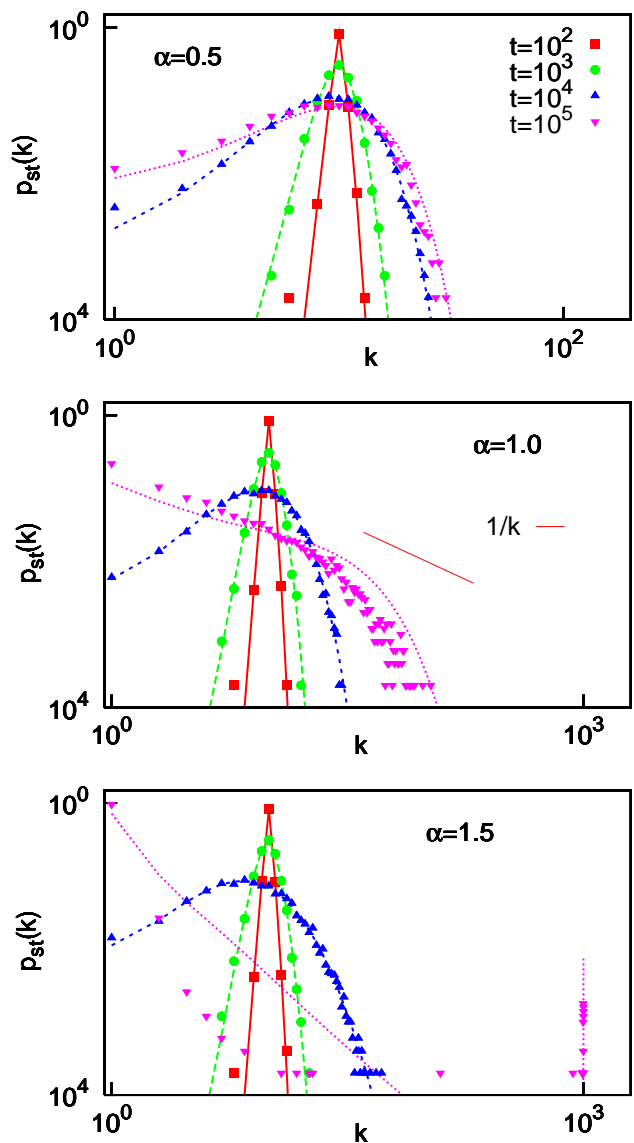,width=10.cm}
\caption{Degree distribution
$p(k,t)$ at four different stages of evolution: $t=10^{2}$ [(yellow) squares], $10^{3}$ [(blue) circles], $10^{4}$ [(red) triangles)] and $10^{5}$ MCS [(black) diamonds]. From top to bottom panels, subcritical ($\alpha=0.5$), critical ($\alpha=1$) and supercritical ($\alpha=1.5$) rewiring exponents. Symbols from MC simulations and corresponding solid lines from numerical integration of Eq. (\ref{master}). $\beta=1$, $\langle k\rangle=10$ and $N=1000$ in all cases.}%
\label{fig1}
\end{center}
\end{figure}
In finite networks, no node can have a degree larger than $N-1$ or lower than
$0$. In fact, one would usually wish to impose a minimum nonzero degree, e.g.
$k\geq1$. The temporal evolution of the degree distribution is illustrated in
Fig. \ref{fig1}. This shows the result of integrating Eq. (\ref{master}) for
$k\geq1,$ different times, $\beta=1,$ and three different values of $\alpha$,
along with the respective values obtained from Monte Carlo simulations.%

\begin{figure}[t!]
\begin{center}
\epsfig{file=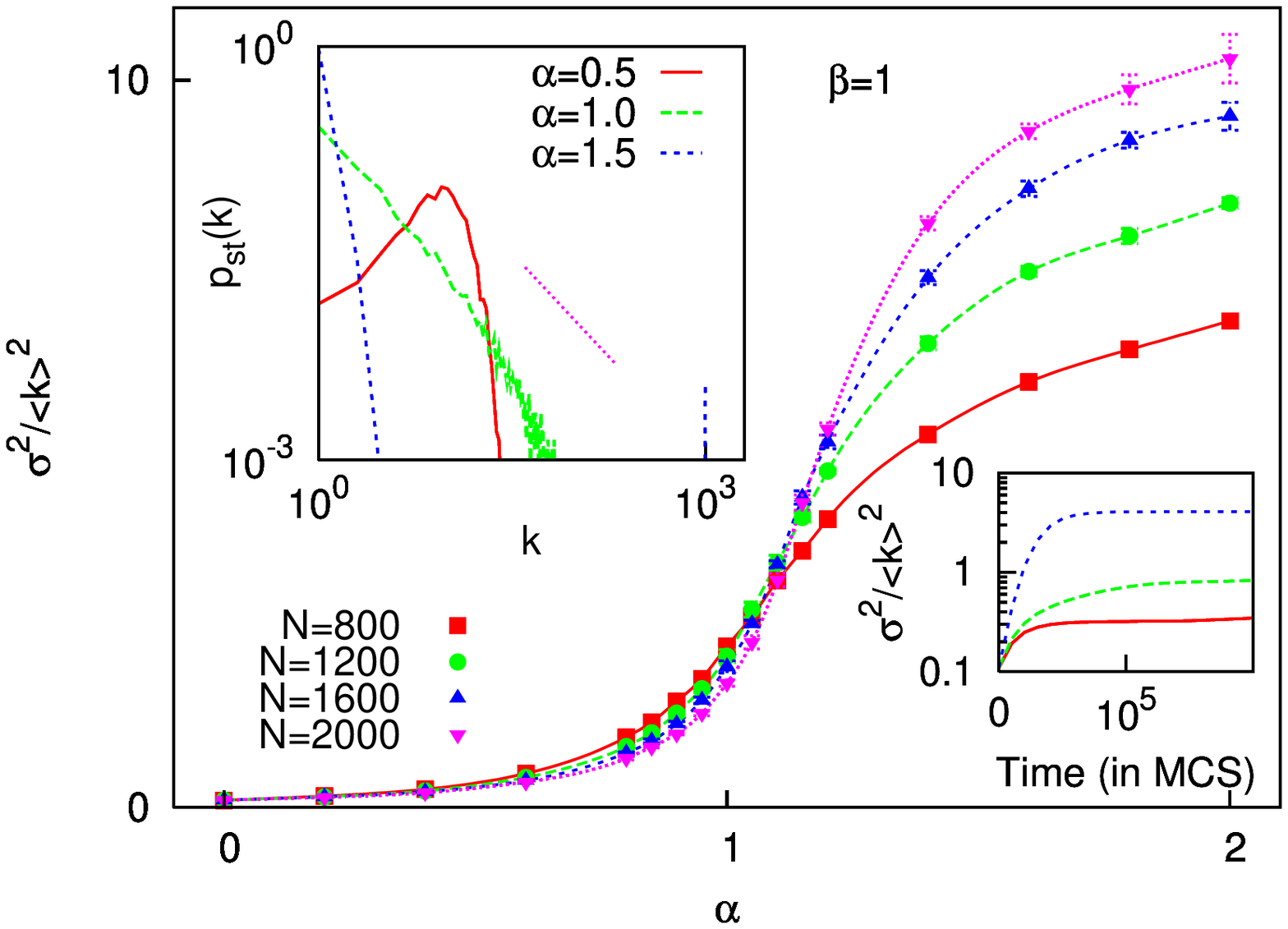,width=10.cm}
\caption{ Adjusted variance $\sigma^{2}/\langle k\rangle^{2}$ of the degree distribution after $2\times10^{5}$ MCS against $\alpha$, as obtained from MC simulations, for system sizes $N=800$ [(yellow) squares], $1200$ [(blue) circles], $1600$ [(red) triangles] and $2000$ [(black) diamonds]. Top left inset shows final degree distributions for $\alpha=0.5$ [light gray (blue)], $1$ [dark gray (red)] and $1.5$ (black), with $N=1000$. Bottom right inset shows typical time series of $\sigma^{2}/\langle k\rangle^{2}$ for the same three values of $\alpha$ and $N=1200$. In all cases, $\beta=1$ and $\langle k\rangle=10$.}%
\label{fig2}%
\end{center}
\end{figure}

The main result may be summarized as follows. For $\alpha<\beta$, the network
will evolve to have a characteristic size, centred around $\langle k\rangle$.
At the critical case $\alpha=\beta$, all sizes appear, according either to a
pure or a composite power law, as detailed above.

If we impose, say, $k\geq1$, then starlike structures will emerge, with a great many nodes connected to just a few
hubs \footnote{There is a finite-size effect not taken into account by the theory -- but relevant when $\alpha>\beta$ -- which provides a natural lower cutoff for $p_{st}(k)$: if there are, say, $m$ nodes which are connected to the whole network, then the minimum degree a node can have is $m$.}.

Figure \ref{fig2} illustrates the second order phase transition undergone by
the variance of the final (stationary) degree distribution, depending on the
exponent $\alpha$, where $\beta$ is set to unity. It should be mentioned that
this particular case, $\beta=1$, corresponds to edges being chosen at random
for disconnection, since the probability of a random edge belonging to node
$i$ is proportional to $k_{i}$.

This topological phase transition is similar to the ones that have been described in equilibrium network ensembles defined via an energy function, in the so-called \textit{synchronic} approach to network analysis \citep{Farkas,Park_equilibrium,Burda,Derenyi}. However, our (nonequilibrium) model does not come within the scope of this body of work, since the rewiring rates cannot, in general, be derived from a potential. Furthermore, we are here concerned with the time evolution rather than the stationary states, making our approach \textit{diachronic.} 

Summing up, in spite of its simplicity, our model captures the essence of many
real-world networks which evolve while leaving the total numbers of nodes and
edges roughly constant. The grade of heterogeneity of the stationary
distribution obtained is seen to depend crucially on the relation between the
exponents modelling the probabilities a node has of obtaining or loosing a new
edge. It is worth mentioning that the heterogeneity of the degree distribution
of a random network has been found to determine many relevant behaviours and
magnitudes such as its clustering coefficient and mean minimum path
\citep{Newman_rev}, critical values related to the dynamics of excitable
networks \citep{Johnson_EPL}, or the synchronizability for systems of coupled
oscillators (since this depends on the spectral gap of the Laplacian matrix)
\citep{Barahona}.

The above shows how scale-free distributions, with a range of
exponents, may emerge for nonlinear rewiring, although only in the critical
situation in which the probabilities of gaining or loosing edges are the same.
We believe that this non-trivial relation between the microscopic rewiring
actions (governed in our case by parameters $\alpha$ and $\beta$) and the emergent
macroscopic degree distributions could shed light on a class of biological,
social and communications networks.



%% file: cr/appendix_cr.tex
\chapter{Effective modularity of highly clustered networks} 
\label{Appendix_CR}


The number of nodes within a radius $r$ is $n(r)=A_{d}r^{d}$, with $A_{d}$ a constant. We shall therefore assume a node with degree $k$ to have edges to all nodes up to a distance $r(k)=(k/A_{d})^{1/d}$, and none beyond (note that this is not necessarily always feasible in practice). To estimate $\lambda$, we shall first calculate the probability that a randomly chosen edge have length $x$. The chance that the edge belong to a node with degree $k$ is $\pi(k)\sim kp(k)$ (where $p(k)$ is the degree distribution). The proportion of edges that have length $x$ among those belonging to a node with degree $k$ is $\nu(x|k)=dA_{d}x^{d-1}/k$ if $A_{d}x^{d}<k$, and $0$ otherwise.
Considering, for example, scale-free networks (as in Ref. \citep{Rozonfeld}), so that the degree distribution is $p(k)\sim k^{-\gamma}$ in some interval $k\in [k_{0},k_{max}]$ \citep{Barabasi}, and
integrating over $p(k)$, we have the distribution of lengths, 
$$
 P(x)=(Const.)\int_{\max(k_{0},Ax^{d})}^{k_{max}}\pi(k)\nu(k|x)dk=d(\gamma-2)x^{-[d(\gamma-2)+1]},
$$
where we have assumed, for simplicity, that the network is sufficiently sparse that $\max(k_{0},Ax^{d})=Ax^{d}$, $\forall x\geq1$, and where we have normalised for the interval $1\leq x<\infty$; strictly, $x\leq(k_{max}/A)^{1/d}$, but we shall also ignore this effect. Next we need the probability that an edge of length $x$ fall between two compartments of linear size $l$. This depends on the geometry of the situation as well as dimensionality; however, a first approximation which is independent of such considerations is
$$
P_{out}(x)=\min\left(1,\frac{x}{l}\right).
$$
We can now estimate the modularity
$\lambda$ as
$$
\lambda=\int_{1}^{\infty}P_{out}(x)P(x)dx=\frac{1}{d(\gamma-2)-1}\left[d(\gamma-2)l^{-1}-l^{-d(\gamma-2)}\right].
$$
Fig. \ref{fig_mod_new4} shows how $\lambda$ depends on $\gamma$ for $d=2$ and various box sizes.


%% file: nest/nest.tex
\chapter{Nestedness of networks}
\label{Appendix_NEST}


\label{Appendix_NEST}

The property of \textit{nestedness} has for some time aroused a fair amount of interest as regards ecological networks -- especially since a high nestedness in mutualistic systems has been shown to enhance biodiversity. However, because it is usually estimated with software, no analytical work has been done relating nestedness with other network characteristics, and consequently comparisons of experimental data with null-models can only be done computationally. We suggest a slightly refined version of the measure recently defined by Bastolla {\it et al.} and go on to study the effect of the degree distribution and degree correlations (assortativity). Our work provides a benchmark against which empirical networks can be contrasted. 

\section{Introduction}

\begin{figure}[t!]
\begin{center}
\includegraphics
[width=12.cm]
{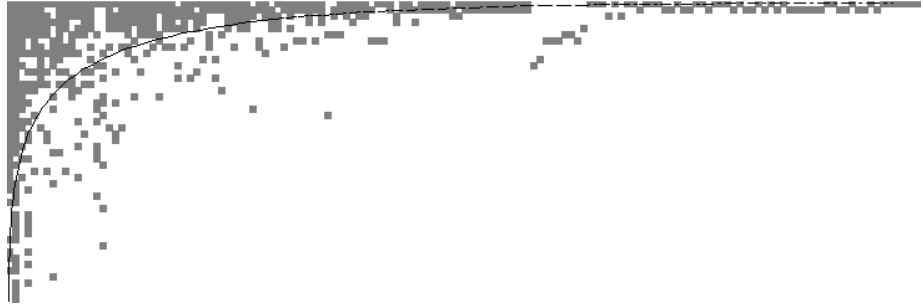}
\caption{
Maximally packed matrix representing a network of plants and islands off Perth \citep{Perth_plants} (because the network is bipartite, the adjacency matrix is composed of four blocks: two identical to this matrix, the other two composed of zeros). Data, image and line obtained from {\it NESTEDNESS CALCULATOR}, which returns a ``temperature'' of $T=0.69^{o}$ for this particular network.
}
\label{fig_Perth_plants}
\end{center}
\end{figure}

The intense study that complex networks have undergone over the past decade or so has shown how important topological features can be for properties of complex systems, such as dynamical behaviour, spreading of information, resilience to attacks, etc. \citep{Newman_rev, Boccaletti}. A paradigmatic case is that of ecosystems. The solution to May's paradox \citep{May} -- the fact that large ecosystems seem to be especially stable, when theory predicts the contrary -- is still not clear, but it is widely suspected that there is some structural feature of ecological networks which as yet eludes us. One aspect of such networks, which has been studied for some time by ecologists and may be related to this problem, is called {\it nestedness}. Loosly speaking, a network -- say of species and islands, linked whenever the former inhabit the latter -- is said to be highly nested if the species which exist on scarcely populated islands tend always to be found also on those islands inhabited by many different species. This can be most easily seen by graphically representing a matrix such that animals are columns and islands are files, with elements equal to one whenever two nodes are linked and zero if not. If, after ordering each kind of node by degree (number of neighbours), all the ones can be quite neatly packed into one corner, the network is considered highly nested. This is done in Fig. \ref{fig_Perth_plants} for a network of plants inhabiting islands off Perth. This rather vague concept is usually measured with software for the purpose. For Fig. \ref{fig_Perth_plants}, we have used {\it NESTEDNESS CALCULATOR}, which estimates a curve of equal density of ones and zeros, calculates how many ones and zeros are on the ``wrong'' side and by how much, and returns a number between $0$ and $100$ called ``temperature'' by analogy with some system such as a subliming solid. A low temperature indicates high nestedness. To determine how significantly nested a given network is, the usual procedure is to generate equivalent random networks computationally (with sone constraint such as the number of edges or the degree of each node being conserved) and estimate how likely it is that such a network be ``colder'' than that of the data.

Bastolla {\it et al.} \citep{basc_architecture} have recently shown how symbiotic interactions can reduce the effective competition between two species, say of insect, via common symbiotic hosts -- such as plants they pollinate. These authors define a measure to take into account the average number of shared partners in these {\it mutualistic} networks, and call it ``nestedness'' because it would seem to be related to the concept referred to above. They go on to show evidence of how the nestedness of empirical mutualistic networks is correlated with the biodiversity of the corresponding ecosystems. This beneficial effect ``enemy'' nodes can gain from sharing ``friendly'' partners is not confined to ecosystems. It is expected also to play a role, for instance, in financial networks or other economic systems \citep{financial}. The principle is simple. Say nodes A and B are in competition with each other. An increase in A will be to B's detriment, and viceversa; but if both A and B engage in a symbiotic relationship with node C, then A's thriving will stimulate C, which in turn will be helpful to B. Thus, the effective competition between A and B is reduced, and the whole system becomes more stable and capable of sustaining more nodes \citep{Virginia}.

In Ref. \citep{Johnson_NEST} we 
take up this idea of shared neighbours (though characterised with a slightly different measure, for reasons we shall explain in Section \ref{sec_def}) and study analytically the effect of other topological properties, such as the degree distribution and degree-degree correlations. This allows us to contrast empirical data with null-models and thus test for statistical significance with no need of computer randomisations. We also comment on how mutual-neighbour structure could develop in systems of interdependent networks (such as competition and symbiosis) so as to minimise the risk of a ``cascade of failures'' \citep{Havlin_failures}. Although we are not here concerned specifically with neural systems, a description of this work is included as an appendix since it serves as an example application of the method put forward in Ref. \citep{Johnson_PRL} and presented in Chapter \ref{Chapter_PRL}.

\section{Definition}
\label{sec_def}

Consider a network with $N$ nodes defined by the adjacency matrix $\hat{a}$: the element $\hat{a}_{ij}$ is equal to the number on links, or edges, from node $j$ to node $i$ (typically considered to be either one or zero). If $\hat{a}$ is symmetric, then the network is undirected and each node $i$ can be characterised by a degree 
$
k_{i}=\sum_{j}\hat{a}_{ij}
$. (If it is directed, $i$ has both an \textit{in} degree, $k_{i}^{\mbox{in}}=\sum_{j}\hat{a}_{ij}$, and an \textit{out} degree, 
$
k_{i}^{\mbox{out}}=\sum_{j}\hat{a}_{ji}
$; we shall focus here on undirected networks, although most of the results could be easily extended to directed ones.).

Bastolla \textit{et al.} \citep{basc_architecture} have shown that the effective competition between two species (say two species of insect) can be reduced if they have common neighbours with which they are in symbiosis (for instance, if they both pollinate the same plant). Therefore, in mutualistic networks (networks of symbiotic interactions) it is beneficial to the species at two nodes $i$ and $j$ for the number of shared symbiotic partners, $n_{ij}=\sum_{l}\hat{a}_{il}\hat{a}_{lj}=(\hat{a}^{2})_{ij}$, to be high. Going on this, and assuming the network is undirected, the authors suggest taking into account the following measure:
\begin{equation}
\eta_{B}=\frac{\sum_{i<j}\hat{n}_{ij}}{\sum_{i<j}\min(k_{i},k_{j})},
\label{eta_B}
\end{equation}
which they call \textit{nestedness} because it would seem to be highly correlated with the measures returned by nestedness software. Note that, although the authors were considering only bipartite graphs, this characteristic is not imposed in the above definition.
In this work, we shall take up the idea of the importance of $n_{ij}$, but use a slightly different measure of nestedness, for several reasons. One is that $\eta_{B}$ has a serious shortcoming. If we commute the sums\footnote{In an undirected network, $\sum_{i<j}=\frac{1}{2}\sum_{ij}$; we shall always sum over all $i$ and $j$, since it is easier to generalise to directed networks and often avoids writing factors 2.} in the numerator of Eq. (\ref{eta_B}), we find that the result only depends on the heterogeneity of the degree distribution:
$
\sum_{ij}\hat{n}_{ij}=\sum_{l}\sum_{i}\hat{a}_{il}\sum_{j}\hat{a}_{lj}=N\langle k^{2}\rangle.
$
Also, although the maximum value $\hat{n}_{ij}$ can take is $\min(k_{i},k_{j})$, this is not necessarily the best normalisation factor, since the expected number of paths of length 2 connecting nodes $i$ and $j$ depends on both $k_{i}$ and $k_{j}$ (as we show explicitely in Section \ref{sec_deg_dis}). Furthermore, it can sometimes be convenient to have a local measure of nestedness. For these reasons, we shall use
\begin{equation}
\eta_{ij}\equiv \frac{n_{ij}}{k_{i}k_{j}}= \frac{(\hat{a}^{2})_{ij}}{k_{i}k_{j}},
\label{eta_ij}
\end{equation}
which is defined for every pair of nodes $(i,j)$. This allows for the consideration of a nestedness per node, $\eta_{i}=N^{-1}\sum_{j}\eta_{ij}$, or of the global measure
\begin{equation}
\eta=\frac{1}{N^{2}}\sum_{ij}\eta_{ij}.
\label{eta_gen}
\end{equation}

\section{The effect of the degree distribution}
\label{sec_deg_dis}

Most networks have quite broad degree distributions $p(k)$, most notably the fairly ubiquitous scale-free networks, for which they follow power-laws, $p(k)\sim k^{-\gamma}$. Since this heterogeneity tends to have an importante influence on any network measure, it will be useful to take this effect into account analytically. As is standard, the null-model we shall use to do this is the \textit{configurational model} \citep{Newman_rev}: the set of random networks wired according to the constraints that a given degree sequence $(k_{1},...,k_{N})$ is respected, and also that there be no degree-degree correlations. The expected value of an element of the adjacency matrix for networks belonging to this ensemble is
\begin{equation}
\overline{\hat{a}_{ij}}\equiv\hat{\epsilon}_{ij}^{c}=\frac{k_{i}k_{j}}{\langle k\rangle N}.
\label{epsi_conf}
\end{equation}
We shall use a line, $\overline{(\cdot)}$, to represent expected values given certain constraints, and angles, $\langle \cdot\rangle$, for averages over nodes of a given network\footnote{In this case, for instance, the network considered for $\langle k\rangle$ is any of the members of the ensemble, since they all have the same mean degree by definition.}. For the case of the adjacency matrix, we use the notation $\hat{\epsilon}_{ij}^{c}=\overline{\hat{a}_{ij}}$ for clarity and coherence with previous work.
Plugging Eq. (\ref{epsi_conf}) into Eq. (\ref{eta_ij}), we have the expected value in the configuration ensemble,
\begin{equation}
\overline{\eta_{ij}}=\frac{\langle k^{2}\rangle}{\langle k\rangle^{2}N}\equiv \eta_{conf}.
\label{eta_conf}
\end{equation}
Since $\eta_{c}$ is independent of $i$ and $j$, it coincides with the expected value for the global measure, $\overline{\eta}=\eta_{conf}$ -- a fact that justifies the normalisation chosen in Eq. (\ref{eta_ij}). It is obvious from Eq. (\ref{eta_conf}) that degree heterogeneity will have an important effect on $\eta$. Therefore, if we are to capture aspects of network structure other than those directly induced by the degree distribution, it will in general be useful to consider the nestedness normalised to this expected value,
\begin{equation}
\tilde{\eta}\equiv\frac{\eta}{\eta_{conf}}=
\frac{\langle k\rangle^{2}}{\langle k^{2}\rangle N}\sum_{ij}\frac{(\hat{a}^{2})_{ij}}{k_{i}k_{j}}.
\label{eta_tilde}
\end{equation}

Although $\tilde{\eta}$ is unbounded, it has the advantage that it is equal to unity for any uncorrelated random network, independently of its degree heterogeneity, thereby making it possible to detect non-trivial structure in a given empirical network without the need for computational randomisations.

\section{Nestedness and assortativity}

In the configuration ensemble, the expected value
of the mean degree of the neighbours of a given node is $
\overline{k_{nn,i}}=k_{i}^{-1}\sum_{j}\hat{\epsilon}_{ij}^{c}k_{j}=\langle
k^{2}\rangle/\langle k\rangle,
$ which is independent of $k_{i}$. However,
real networks
usually display degree-degree correlations, with the result
that $\overline{k_{nn,i}}=\overline{k_{nn}}(k_{i})$. If $\overline{k_{nn}}(k)$ increases (decreases)
with $k$, the network is assortative (disassortative). A measure of
this phenomenon is Pearson's coefficient applied to the
edges \citep{Newman_rev, Newman_mixing_PRL, Newman_mixing_PRE, Boccaletti}: $ r= ([
k_{l}k'_{l}]-[ k_{l}]^{2})/([ k_{l}^{2}]-[ k_{l}]^{2}), $ where
$k_{l}$ and $k'_{l}$ are the degrees of each of the two nodes
belonging to edge $l$, and $[\cdot]\equiv(\langle k\rangle
N)^{-1}\sum_{l}(\cdot)$ is an average over edges. Writing
$\sum_{l}(\cdot)=\sum_{ij}\hat{a}_{ij}(\cdot)$, $r$ can be expressed
as
\begin{equation}
  r=\frac{\langle k\rangle \langle k^{2} \overline{k_{nn}}(k)\rangle - 
    \langle k^{2}\rangle^{2} }{\langle k\rangle \langle k^{3}\rangle 
    - \langle k^{2}\rangle^{2}}.
  \label{r_gen}
\end{equation}

The ensemble of all networks with a given degree sequence $(k_{1},...k_{N})$
contains a subset for all members of which $\overline{k_{nn}}(k)$ is constant (the configuration ensemble), but also subsets displaying other functions $\overline{k_{nn}}(k)$.

In Chapter \ref{Chapter_PRL} \citep{Johnson_PRL}
we showed that there is a one-to-one mapping between any mean-nearest-neighbour function $\overline{k_{nn}}(k)$ and its corresponding mean-adjacency-matrix $\hat{\epsilon}$, which is as follows: writing $\overline{k_{nn}}(k)$ as 
\begin{eqnarray}
  \overline{k_{nn}}(k)=\frac{\langle k^{2}\rangle}{\langle k\rangle}
  +\int d\nu f(\nu)\sigma_{\nu+1}\left[\frac{k^{\nu-1}}
    {\langle k^{\nu}\rangle}-\frac{1}{k} \right]
\label{knn_gen}
\end{eqnarray}
with $\sigma_{\nu+1}\equiv \langle k^{\nu+1}\rangle -\langle k\rangle
\langle k^{\nu}\rangle$
(which can always be done), the corresponding matrix $\hat{\epsilon}$ takes the form
\begin{equation}
  \hat{\epsilon}_{ij}=\frac{k_{i}k_{j}}{\langle k\rangle N}
  +\int d\nu \frac{f(\nu)}{N}\left[\frac{(k_{i}k_{j})^{\nu}}
    {\langle k^{\nu}\rangle}-k_{i}^{\nu}-k_{j}^{\nu}+\langle k^{\nu}\rangle  \right].
\label{epsi_gen}
\end{equation}

In many empirical networks, $\overline{k_{nn}}(k)$ has the form $\overline{k_{nn}}(k)=A+B
k^{\beta}$, with $A,B>0$ \citep{Boccaletti, Pastor-Satorras} -- the
mixing being assortative (disassortative) if $\beta$ is positive
(negative). Such a case is fitted by Eq. (\ref{knn_gen}) if $f(\nu)=C[\delta(\nu-\beta-1)\sigma_{2}/\sigma_{\beta+2}-\delta(\nu-1)]$, with $C$ a positive constant, since this choice yields
\begin{equation}
  \overline{k_{nn}}(k)=\frac{\langle k^{2}\rangle}{\langle k\rangle}
  +C\sigma_{2}\left[\frac{k^{\beta}}{\langle
      k^{\beta+1}\rangle}-\frac{1}{\langle k\rangle} \right].
\label{knn_simple}
\end{equation}
After plugging Eq. (\ref{knn_simple}) into Eq. (\ref{r_gen}), one obtains:
\begin{equation}
  r=\frac{C\sigma_{2}}{\langle k^{\beta+1}\rangle}
  \left(\frac{\langle k\rangle \langle k^{\beta+2} \rangle - 
      \langle k^{2}\rangle\langle k^{\beta+1}\rangle }
{\langle k\rangle \langle k^{3}\rangle - \langle k^{2}\rangle^{2}}\right).
\label{r_simple}
\end{equation}

It turns out that the configurations most likely to arise naturally (those with maximum entropy) usually have $C\simeq 1$ \citep{Johnson_PRL} (c.f. Chapter \ref{Chapter_PRL}). Therefore, and for the sake of analytical tractability, we shall 
do as in Chapter \ref{Chapter_PRE} and
consider this particular case\footnote{Note that $C=1$ corresponds to removing the
linear term, proportional to $k_i k_j$, in Eq. (\ref{epsi_gen}), and leaving
the leading non-linearity, $(k_i k_j)^{\beta+1}$, as the dominant one.} -- that is, we shall use
\begin{equation}
\hat{\epsilon}_{ij}= \frac{1}{N}\left\lbrace \frac{\sigma_{2}}{\sigma_{\beta+2}}
\left[ \frac{(k_{i}k_{i})^{\beta+1}}{\langle k^{\beta+1}\rangle}-k_{i}^{\beta+1}-k_{j}^{\beta+1}
+\langle k^{\beta+1}\rangle \right] +k_{i}+k_{j}-\langle k\rangle \right\rbrace.
\label{epsi_beta}
\end{equation}
Substituting the adjacency matrix for this expression in the definition of $\tilde{\eta}$ (Eq. (\ref{eta_tilde})), we obtain its expected value as a function of the remaining parameter $\beta$:
\begin{eqnarray}
\overline{\tilde{\eta}}_{\beta}
=
\nonumber
\\
\frac{\langle k\rangle^{2}}{\langle k^{2}\rangle}\left[1+(\sigma_{2}-\alpha_{\beta}^{2}\rho_{\beta})\left(2\frac{\langle k^{\beta}\rangle\langle k^{-1}\rangle}{\langle k^{\beta+1}\rangle}-\langle k^{-1}\rangle^{2}      \right)+\alpha_{\beta}^{2}\rho_{\beta}\left(\frac{\langle k^{\beta}\rangle}{\langle k^{\beta+1}\rangle}\right)^{2}\right],
\label{eta_mix}
\end{eqnarray}
where 
$
\alpha_{\beta}\equiv \sigma_{2}/\sigma_{\beta+2}
$
and $\rho_{\beta}\equiv\langle k^{2(\beta+1)}\rangle-\langle k\rangle^{2(\beta+1)}$. Note that 
$
\overline{\tilde{\eta}}_{0}=1
$.

\begin{figure}[t!]
\begin{center}
\includegraphics
[width=10.cm]
{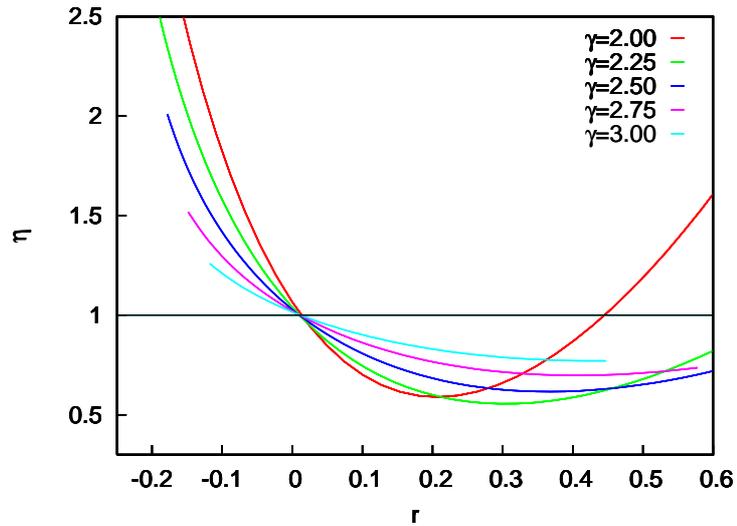}
\caption{
Nestedness against assortativity (as measured by Pearson's correlation coefficient) for scale-free networks as given by Eq. (\ref{eta_mix}). $\langle k\rangle =10$, $N=1000$.
}
\label{fig_num}
\end{center}
\end{figure}

\begin{figure}[t!]
\begin{center}
\includegraphics
[width=10.cm]
{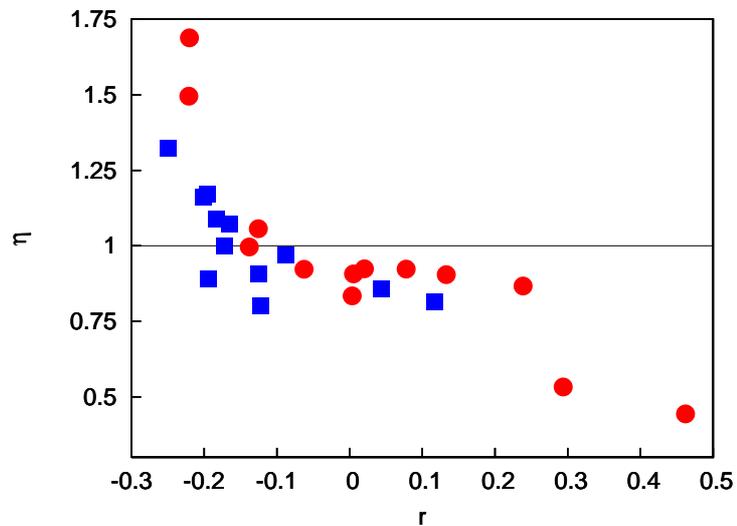}
\caption{
Nestedness against assortativity (as measured by Pearson's correlation coefficient) for data on a variety of networks. Blue squares are food webs (Table \ref{table_food_webs}) and red circles are networks of all other types (Table \ref{table_nidos}).
}
\label{fig_nidos}
\end{center}
\end{figure}

Fig. \ref{fig_num} shows the value of $\overline{\tilde{\eta}}_{\beta}$ given by Eq. (\ref{eta_mix}) against the assortativity $r$ for various scale-free networks. Nestedness is seen to grow very fast with increasing disassortativity (decreasing negative $r$), while in general slightly assortative networks are less nested than neutral ones. However, highly heterogeneous networks ($\gamma\rightarrow 2$) show an increase in $\overline{\tilde{\eta}}_{\beta}$ for large positive $r$. Fig. \ref{fig_nidos} shows a plot of nestedness against assortativity for the selection of empirical networks listed in Table \ref{table_nidos}. Although these networks are highly disparate as regards size, density, degree distribution, etc., it is apparent from the similarity to Fig. \ref{fig_num} that the main contribution to $\tilde{\eta}_{\beta}$ comes indeed from the assortativity.

\begin{table}
[th!]
\begin{center}
\label{table_food_webs}
\begin{tabular}{@{}llllll}
\\
Food web & $r$ & $\nu$ & $\langle k\rangle$ & $N$ & $\sigma/\langle k\rangle$\\
\hline
\hline
\\
Little Rock lake  &	-0.343   &	1.219   &	20.4   &	92   &	0.73 \\
Ythan Estuary (w/p)  & 	-0.249   &	1.323   &	8.9   &	82   &	0.93	 \\
Stony Stream & -0.201   &	1.163   &	14.7   &	109   &	0.75	 \\
Canton Creek & -0.196   &	1.171   &	13.5   &	102   &	0.69	 \\
Skipwith Pond & -0.194   &	0.891   &	14.2   &	25   &	0.37	 \\
El Verde & -0.183   &	1.088   &	18.4   &	155   &	0.88	 \\
Caribbean Reef (small)  & -0.172   &	1.000   &	19.7   &	50   &	0.49	 \\
St. Martin Island & -0.165   &	1.071   &	9.3   &	42   &	0.56 \\
UK Grassland  & -0.125   &	0.907   &	2.8   &	61   &	0.82	 \\
Chesapeake Bay & -0.123   &	0.801   &	4.1   &	31   &	0.60	 \\
NE US Shelf & -0.088   &	0.971   &	34.3   &	79   &	0.45	 \\
Coachella Valley & 0.043   &	0.857   &	14.6   &	29   &	0.41	 \\
St. Mark's Estuary  & 0.118   &	0.816   &	8.5   &	48   &	0.55	 \\
\end{tabular}
\caption{
Food webs appearing in Fig. \ref{fig_nidos} (listed from least to most assortative) : $r$ is the assortativity and $\nu$ the nestedness. The origins of all data cited in Ref. \citep{Dunne}, and kindly provided to us by Jennifer Dunne.
}
\end{center}
\end{table}

\begin{table}
[th!]
\begin{center}
\label{table_nidos}
\begin{tabular}{@{}lllllll}
\\
\cr Network & $r$ & $\nu$ & $\langle k\rangle$ & $N$ & $\sigma/\langle k\rangle$ & Ref. \\
\hline
\hline
\\
Political blogs	&	-0.221	&1.496 & 22.4 & 1490 & 1.62 & \citep{polblogs}\\
Metabolic	&	-0.220	&1.688 & 9.0 & 453 & 1.87 & \citep{metabolic}\\	
Political books	&	-0.138	&0.996 & 8.4 & 104 & 0.65 & \citep{polbooks}\\	
Adjectives and nouns		&	-0.125	&1.057 & 7.6 & 111 & 0.89 & \citep{Newman_finding}\\	
Dolphins	&	-0.063	&0.922 & 5.1 & 61 & 0.58 &\citep{dolphins}\\	
Power grid		&	0.003	&0.834 & 2.7 & 4940 & 0.67 & \citep{Watts}\\	
Neural		&	0.005	&0.907 & 5.9 & 306 & 0.81 & \citep{Watts}\\	
Jazz musicians		&	0.020	&0.924 & 27.6 & 198 & 0.63 & \citep{jazz}\\	
Email		&	0.078	&0.923 & 9.6 & 1133 & 0.97 & \citep{email}\\	
American football	&	0.133	&0.904 & 10.6 & 114 & 0.08 &\citep{Girvan}\\	
PGP		&	0.239	&0.867 & 4.6 & 10680 & 1.77 & \citep{pgp}\\		
High-energy arXiv		&	0.294	&0.533 & 3.8 & 8360 & 1.14 &\citep{hep-th}\\	
Net-science arXiv 	&	0.462	&0.443 & 3.45 & 1588 & 1.00 & \citep{Newman_finding}\\	
\end{tabular}
\caption{
Empirical networks appearing in Fig. \ref{fig_nidos} (listed from least to most assortative) : $r$ is the assortativity and $\nu$ the nestedness. All data available on the personal Web pages of \'Alex Arenas, Mark Newman and Duncan Watts.
}
\end{center}
\end{table}

\section{Bipartite networks}

Mutualistic networks are usually bipartite: two sets of nodes exist such that all edges are between nodes in one set and those of another. The ones considered in Ref. \citep{basc_architecture}, for instance, are composed of animals and plants which interact in symbiotic relations of feeding-pollination; these interactions only take place between animals and plants. Let us therefore consider a bipartite network and call the sets $\Gamma_{1}$ and $\Gamma_{2}$, with $n_{1}$ and $n_{2}$ nodes, respectively  ($n_{1}+n_{2}=N$). Using the notation $\langle \cdot \rangle_{i}$ for averages over set $\Gamma_{i}$, the total number of edges is $\langle k\rangle_{1}n_{2}=\langle k\rangle_{2}n_{1}=\frac{1}{2}\langle k\rangle N$. Assuming that the network is defined by the configuration ensemble, though with the additional constraint of being bipartite, the probability of node $l$ being connected to node $i$ is
$$
\hat{\epsilon}_{il}=2\frac{k_{i}k_{l}}{\langle k\rangle N}
$$
if they belong to different sets, and zero if they are in the same one. Proceeding as before, we find that the expected value of the nestedness for a bipartite network is
\begin{eqnarray}
\eta_{bip}=\frac{1}{N^{2}}\left[\sum_{i,j\in\Gamma_{1}}\frac{1}{k_{i}k_{j}}\sum_{l\in\Gamma_{2}}\frac{k_{i}k_{l}}{\langle k\rangle_{1}n_{2}}\frac{k_{l}k_{j}}{\langle k\rangle_{2}n_{1}}+
\sum_{i,j\in\Gamma_{2}}\frac{1}{k_{i}k_{j}}\sum_{l\in\Gamma_{1}}\frac{k_{i}k_{l}}{\langle k\rangle_{1}n_{2}}\frac{k_{l}k_{j}}{\langle k\rangle_{2}n_{1}}
 \right]=
\nonumber
\\
\frac{n_{1}\langle k^{2}\rangle_{2}+n_{2}\langle k^{2}\rangle_{1}}{\langle k\rangle_{1}\langle k\rangle_{2}(n_{1}+n_{2})^{2}}.
\quad\quad\quad
\end{eqnarray}
Interestingly, if $n_{1}=n_{2}$, the fact that the network is bipartite has no effect on the nestedness: $\eta_{bip}=\eta_{conf}$.

\section{Overlapping networks}

If the adjacency matrix $\hat{a}$ describes a mutualistic network, the benefit to its being nested resides in a counteraction of the competition matrix $\hat{c}$, which takes into account the extent to which one species is detrimental to another due to predation, sharing of resources, etc. From this point of view, it may be interesting to study to what extent matrices $\hat{c}$ and $\hat{a}^{2}$ overlap (note that both networks have the same nodes, but different edges). Presumably, if ecological networks are assembled in such a way that effective competition is minimised, this overlap should be higher than randomly expected. On the other hand, a certain degree of overlap may also arise from the fact that species interacting symbiotically with the same host are perhaps more than averagely likely to be phylogenetically close and/or phenotypically similar, leading (as Darwin noted) to a higher competition element.

In any case, a measure of this overlap is
\begin{equation}
r\equiv\frac{1}{\langle k\rangle_{c}N}\sum_{ij}\hat{c}_{ij}(\hat{a}^{2})_{ij},
\end{equation}
where $\langle \cdot \rangle_{c}$ represents an average over the competion network; similarly, $\langle \cdot \rangle_{a}$ will stand for an average over the mutualistic network. If the two networks are mutually uncorrelated\footnote{Note that we are saying nothing of the internal correlations that each network may display.} -- i.e., if the existence of an edge in one provides no information as to whether there is a corresponding one in the other -- we can write
\begin{equation}
r\simeq\frac{1}{\langle k\rangle_{c}N}\sum_{ij}\hat{c}_{ij}\frac{1}{N^{2}}\sum_{ij}(\hat{a}^{2})_{ij}\equiv r_{unc}.
\end{equation}
Using $\sum_{ij}(\hat{a}^{2})_{ij}=\langle k^{2}\rangle_{a}N$, and assuming that $\hat{c}$ is normalised so that $\sum_{ij}\hat{c}_{ij}=\langle k\rangle_{c}N$, we have\footnote{The competition matrix will in general be weighted, as could be the mutualistic one; we shall treat both as though they were not, but using weighted networks would only influence results by a normalisation factor.}
\begin{equation}
r_{unc}\simeq \frac{\langle k^{2}\rangle_{a}}{N},
\end{equation}
which only depends on the heterogeneity of the degree distribution of the mutualistic network. Again, it may be useful to consider the overlap normalised to this value,
\begin{equation}
\tilde{r}\equiv \frac{r}{r_{unc}}=\frac{1}{\langle k\rangle_{c}\langle k^{2}\rangle_{a}}\sum_{ij}\hat{c}_{ij}(\hat{a}^{2})_{ij}.
\end{equation}
This measure will equal unity when there is no statistical relation between the competition matrix and the mutualistic one, but can be expected to be greater if indeed such an overlap were contributing to a reduction in effective competition.

It has recently been shown that interconnected networks are prone to dangerous ``cascades of failures'' \citep{Havlin_failures}. It seems that the northen half of Italy was once left temporarily with no electric supply due to failures in the power-grid closing down dependent internet servers, which in turn further disrupted the grid, until many nodes of both networks were rendered dysfunctional. If two inter-dependent networks were to coincide perfectly ($r=1$), the resilience of the system to node removal would be the same as that of just one network; however, lower overlap leads to increased vulnerability to such cascades of failures. Since the extinction of a species can result in its host species also going extinct, such cascades of failures may be a threat to mutualistic systems. In such a case, it would seem that a high overlap $r$, as defined here, between the competition matrix and the mutualistic one would minimise this possibility. It would be interesting to test this experimentally.

\section{Discussion}

Whether or not the topological feature here described should be considered a measure of nestedness as it is usually understood in ecology is not clear. What is certain is that interactions between dynamical elements that are mediated by third parties, or common neighbours, can be relevant in a wide variety of settings. We have mentioned the paradigmatical case of ecosystems as well as financial and communications networks. But other examples spring easily to mind. For instance, two excitatory neighbouring neurons might have their mutual effect dampened  if they share inhibitory neighbours. Genetic networks are riddled with motifs such that switches activate or inactivate each other indirectly, via common neighbours. As we have shown, there are nontrivial relationships between nestedness, as it is here defined, and other topological features. If it turns out that this network property is indeed relevant for many complex systems, then we hope the null models we have laid out and analysed will prove useful in assessing its functional significance.


%% file: pub/pub.tex

\chapter{Publications derived from the thesis} 
\label{Appendix_PUB}

\section{Journals and book chapters (the most relevant ones marked with an asterisk)}


\begin{enumerate}
  \item 
* {\it Cluster Reverberation: A mechanism for robust short-term memory without synaptic learning}, S. Johnson, J. Marro,  and JJ. Torres,  submitted, arXiv:1007.3122

\item
* {\it Enhancing neural-network performance via assortativity}, S. de Franciscis, S. Johnson, and J.J. Torres, {\it Physical Review E} {\bf 83}, 036114 (2011)

\item
{\it Why are so many networks disassortative?} S. Johnson, J.J. Torres, J. Marro, and M.A. Mu\~noz, {\it AIP Conf. Proc.} {\bf 1332}, 249--50 (2011)

\item
{\it Shannon entropy and degree-degree correlations in complex networks}, S. Johnson, J.J. Torres, J. Marro, and M.A. Mu\~noz, ``Nonlinear Systems and Wavelet Analysis'', Ed. R. L\'opez-Ruiz, WSEAS Press, pp. 31--35 (2010)

\item
* {\it Entropic origin of disassortativity in complex networks}, S. Johnson, J.J. Torres, J. Marro, and M.A. Mu\~noz, {\it Physical Review Letters} {\bf 104}, 108702 (2010)

\item
* {\it Evolving networks and the development of neural systems}, S. Johnson, J. Marro, and J.J. Torres, {\it Journal of Statistical Mechanics} (2010) P03003

\item
{\it Excitable networks: Nonequilibrium criticality and optimum topology}, J.J. Torres, S. de Franciscis, S. Johnson, and J. Marro, {\it International Journal of Bifurcation and Chaos} {\bf 20}, 869--875 (2010)

\item
{\it Nonequilibrium behavior in neural networks: criticality and optimal performance}, J.J. Torres, S. Johnson, J.F. Mejias, S. de Franciscis, and J. Marro, ``Advances in Cognitive Neurodynamics (II)'' Eds. R. Wang and F. Gu, pp 597--603, Springer, 2011, ISBN: 978-90-481-9694-4, Proceedings of Second International Conference on Cognitive Neurodynamics (ICCN2009), Hangzhou 15-19 November 2009.

\item
{\it Development of neural network structure with biological mechanisms}, S. Johnson, J. Marro, J.F. Mejias, and J.J. Torres, {\it Lecture Notes in Computer Science} {\bf 5517}, 228--235 (2009)

\item
{\it Switching dynamics of neural systems in the presence of multiplicative colored noise}, J.F. Mejias, J.J. Torres, S. Johnson, and H.J. Kappen, {\it Lecture Notes in Computer Science} {\bf 5517}, 17--23 (2009)

\item
* {\it Nonlinear preferential rewiring in fixed-size networks as a diffusion process}, S. Johnson, J.J. Torres, and  J. Marro, {\it Physical Review E} {\bf 79}, 050104(R) (2009)

\item
* {\it Functional optimization in complex excitable networks}, S. Johnson, J.J. Torres, and J. Marro, {\it EPL} {\bf 83}, 46006 (2008)

\item
{\it Excitable networks: Non-equilibrium criticality and optimum topology}, J.J. Torres, S. de Franciscis, S. Johnson, and J. Marro, ``Modelling and Computation on Complex Networks and Related Topics'', Eds. Criado, Gonzalez-Viñas, Mancini and Romance. Proceedings of the conference "Net-Works 2008", 185--192, ISBN:978-84-691-3819-9.

\item
{\it Topology-induced instabilities in neural nets with activity-dependent synapses}, S. Johnson, J. Marro, and J. J. Torres, ``New Trends and Tools in Complex Networks'', Eds. Criado, Pello and Romance. Proceedings of the conference "Net-Works 2007", 59--71, ISBN:978-84-690-6890-8.

\end{enumerate}

\section{Abstracts}

\begin{enumerate}
 \item 
{\it Network topology and dynamical task performance}, S. Johnson, J. Marro, and J.J. Torres, {AIP Conf. Proc.} {\bf 1091}, 280 (2009)

\item
{\it Constructive chaos in excitable networks with tuneable topologies}, S. Johnson, J. Marro,  and J.J. Torres, XV Congreso de Física Estadística FisEs’08, 104 (2008)

\item
{\it The effect of topology on neural networks with unstable memories}, S. Johnson, J. Marro, and J.J. Torres, {\it AIP Conf. Proc.} {\bf 887} 261 (2006)

\item
{\it Relationship between the solar wind and the upper-frequency limit of Saturn Kilometric Radiation}, M.Y. Boudjada, P.H.M. Galopeau, H.O. Rucker, A. Lecacheux, W.S. Kurth, D.A. Gurnett, U. Taubencshuss, J.T. Steinberg, S. Johnson, and W. Vollerr, European Geosciences Union (2006)

\end{enumerate}